\documentclass[twocolumn,preprint]{aastex631}
\usepackage{graphicx}	% Including figure files
\usepackage{amsmath}	% Advanced maths commands
\usepackage{amssymb}	% Extra maths symbols
\usepackage{comment}

\let\splitbox\undefined
\usepackage{adjustbox}

\usepackage[T1]{fontenc}
\usepackage[font=small,labelfont=bf,tableposition=top]{caption}

\DeclareCaptionLabelFormat{andtable}{#1~#2  \&  \tablename~\thetable}
\newcommand{\bit}[1]{\ensuremath{\textit{\bfseries{#1}}}}
\def\be{\begin{equation}}
\def\ee{\end{equation}}
\def\ba{\begin{eqnarray}}
\def\ea{\end{eqnarray}}

\def\12{{1\over 2}}

\def\msun{M_\odot}

\def\ltgt{$\; \buildrel < \over \geq \;$}
\def\gtlt{\lower.5ex\hbox{\ltgt}}
\def\ltsima{$\; \buildrel < \over \sim \;$}
\def\simlt{\lower.5ex\hbox{\ltsima}}
\def\gtsima{$\; \buildrel > \over \sim \;$}
\def\simgt{\lower.5ex\hbox{\gtsima}}

\definecolor{arsenic}{rgb}{0.23, 0.27, 0.29}
\def\magenta{\textcolor{magenta}}

\definecolor{falured}{rgb}{0.5, 0.09, 0.09}
\definecolor{ao(english)}{rgb}{0.0, 0.5, 0.0}

\newcommand{\as}{^{\prime\prime}}

\begin{document}

\title{\bit{AstroSat}/UVIT Study of the Diffuse Ultraviolet Radiation in the Dwarf Galaxy Holmberg II}

\correspondingauthor{P. Shalima}
\email{shalima.p@manipal.edu}

\author[0009-0003-3435-8481]{Olag Pratim Bordoloi}
\affiliation{Department of Physics, Tezpur University, Napaam, India, 784028}

\author[0009-0001-4279-7041]{B. Ananthamoorthy}
\affiliation{Manipal Centre for Natural Sciences, Manipal Academy of Higher Education, Manipal, India, 576104}

\author[0000-0003-4528-9672]{P. Shalima}
\affiliation{Manipal Centre for Natural Sciences, Manipal Academy of Higher Education, Manipal, India, 576104}

\author[0000-0003-4893-6150]{Margarita Safonova}
\affiliation{Indian Institute of Astrophysics, Bengaluru, India, 560034}
 
\author[0000-0002-5123-972X]{Debbijoy Bhattacharya}
\affiliation{Manipal Centre for Natural Sciences, Manipal Academy of Higher Education, Manipal, India, 576104}

\author[0000-0002-3463-7339]{Yuri A. Shchekinov}
\affiliation{Raman Research Institute, Bengaluru, India, 560080}

\author[0000-0003-1715-0200]{Rupjyoti Gogoi}
\affiliation{Department of Physics, Tezpur University, Napaam, India, 784028}

\begin{abstract}
We present measurements of the diffuse ultraviolet (UV) emission in the dwarf irregular galaxy Holmberg II obtained with the Ultra Violet Imaging Telescope (UVIT) instrument onboard \textit {AstroSat}, India’s first multiwavelength space mission. With a spatial resolution of 1.2$\as$ - 1.6$\as$, these are the highest resolution UV observations of the galaxy to date. We find that diffuse emission accounts for $\sim $70.6\% ($\sim$58.1\%) of the total FUV (NUV) emission, respectively. In the FUV, this is reasonably close to the fraction reported for the SMC bar. We perform a UV–IR correlation study of the diffuse emission in this galaxy using infrared (IR) observations from \textit{Spitzer Space Telescope} and \textit{Herschel Space Observatory} for selected
locations, free of detectable bright point sources. The strongest positive correlation between FUV and IR is observed at 70 $\micron$ for high HI density (N(HI) $> 1 \times 10^{21}$ cm$^{-2}$) locations, indicating that warm dust grains dominate the IR emission, in agreement with earlier studies, while NUV is better correlated with 160 $\micron$ emission associated with cold dust grains. Low HI density regions (N(HI) $< 1 \times 10^{21}$ cm$^{-2}$), or cavities, do not show any significant UV–IR correlation except at 160 $\micron$, implying either the presence of colder dust grains in cavities being irradiated by the general radiation field, or insufficient amount of dust. The dust scattering contribution in high HI density regions, estimated using a single scattering model with foreground dust clouds with LMC reddening, gives best-fit albedo and asymmetry factor values of $\alpha = 0.2$ and $g = 0.5$, respectively, in reasonable agreement with the theoretical predictions for LMC dust. Our model-derived scattering optical depths in the FUV range from 0.02 to 0.12, implying the medium is optically thin. Therefore, in high HI density regions, dust scattering can be one of the sources of the observed diffuse UV emission, apart from possible contributions from H$_{2}$ fluorescence. However, the diffuse UV component in HI cavities can only be explained via other mechanisms, such as two-photon emission.
\end{abstract}

\section{Introduction} \label{sec:intro}

Holmberg II (Ho~II hereafter) is an Im-type irregular gas-rich dwarf galaxy residing in the M81-NGC 2403 group of galaxies at a distance of 3.39 Mpc \citep{karachentsev2002m}. It has a low metallicity of either 0.1 $Z_{\odot}$ or 0.3 $Z_{\odot}$, depending on the estimation method \citep{egorov2013emission}, based on its low gas-phase oxygen abundance \citep{pilyugin2014abundance}. These types of galaxies are thought to be very similar to the primordial galaxies of the early Universe and hence are crucial in understanding galaxy and star formation during that period. Ho~II has been mapped thoroughly over a large part of the electromagnetic spectrum through many surveys such as SINGS \citep{Kennicutt2003}, KINGFISH \citep{Kennicutt2011}, THINGS \citep{walter2008things}, etc. One of the most striking features of this galaxy is the  UV and H$\alpha$ bright central star-forming arc comprising giant star clusters with several young massive O-type stars, and the majority of the earlier studies have been dedicated to these star-forming regions. Dwarf irregulars like Ho~II are also characterized by the presence of a massive interstellar medium (ISM) dominated by neutral hydrogen (HI) \citep{karachentsev2019dwarf}. 
This galaxy contains numerous shells and cavities in its HI distribution, similar to M31 and M33, with sizes ranging from a few hundred parsecs to more than a kiloparsec \citep{puche1992holmberg}. \cite{hodge1994h} identified 82 HII regions in the galaxy. These bright HII regions form chain-like structures and constitute the central arc of the galaxy \citep{egorov2017complexes}. The distribution of these connected complexes of ongoing star formation, with a clear indication of triggering effects from HI supershell collisions, has been shown in \cite{egorov2017complexes}. The cavities were formed by stars born throughout the galaxy's evolution as well as by the explosive deaths of massive stars \citep{weisz2009does, puche1992holmberg}. 

Similar to other dwarf irregular galaxies, the dust in Ho~II is distributed in clumps in regions of high HI column density. This galaxy is also deficient in Polycyclic Aromatic Hydrocarbon (PAH) molecules, a conclusion inferred from \textit{Spitzer} IRS observations that showed a low PAH/continuum ratio \citep{walter2007dust}. The 8 $\micron$ emission, attributed to emission by PAHs is found only in regions coinciding with HII regions in Ho~II \citep{wiebe2014dust}. The mid-infrared (MIR) spectra of these kinds of galaxies are dominated by the warm dust emission from the Very Small Grains (VSG) \citep{madden2006ism}. The dust temperatures are found to be higher in Ho~II and other M81 dwarfs based on the observed high 70~$\micron$/160~$\micron$ ratio and this emission extends up to $8^{\prime}$ from the center and is correlated with HI emission \citep{walter2007dust}. This is unlike spiral galaxies where the average dust temperatures are lower and give rise to a peak at $\sim$$100 ~\micron$. This galaxy also lacks sufficient molecular emission to be detectable in the CO band \citep{kahre2018extinction}.  

Though most of the UV emission in Ho~II corresponds to the brightest star-forming regions associated with the central star-forming arc \citep{stewart2000star}, the rest of the galaxy, including the interiors of the HI cavities, are not free of UV emission. The HI cavities, which lack H$\alpha$ emission except at the boundaries, mostly contain FUV fainter stars of B or A spectral type. A significant amount of diffuse IR emission is also observed, which is mostly restricted to star-forming regions and is due to starlight being partly absorbed and re-emitted by dust. The dust-to-gas mass ratio is found to be 0.2 $\times$ 10$^{-3}$ \citep{walter2007dust}, implying the galaxy contains significantly more gas compared to dust. As mentioned earlier, the previous studies in the UV--IR range primarily focussed on the bright star-forming regions of Ho~II. With the availability of high-resolution UV observations from India's {\it AstroSat} mission, it has now become possible to map the diffuse UV emission in Ho~II and study its properties.

There have been several studies of the diffuse UV emission in our Galaxy, where various possible origins for this emission have been discussed extensively \citep{1999MmSAI..70..825H,bowyer,henry2014mystery, murthy2009observations}. {\it Voyager} observations of particular regions of the Milky Way ISM towards the North Galactic pole and at lower latitudes showed diffuse emission with the spectrum of a hot, UV-bright star \citep{Holberg1990}, which implied starlight scattered by dust. \citet{Schiminovich2001} and \citet{Sasseen1996} found the diffuse UV radiation to be well correlated with the IR sky background, once again indicating the role of dust in the observed diffuse UV emission. The diffuse emission produced by the scattering by dust of FUV photons from the hot stars is called Diffuse Galactic Light (DGL). Using data from the Galaxy Evolution Explorer ({\it GALEX}), \citet{murthy2014galex} generated an all-sky diffuse background map, where the emission was found to be following a cosecant distribution with galactic latitude. The diffuse map also showed the intensities to be asymmetric about the galactic plane. Although much of the diffuse FUV emission could be attributed to DGL, at high Galactic latitudes it possibly contained two more components: {\it a}) some diffuse background emission, attributed to the emission from other galaxies and intergalactic medium, part of the Extra-Galactic Background Light (EBL), and {\it b}) a component of unknown origin, referred to as ``Offset'' component \citep{hamden2013diffuse,akshaya2018diffuse}. There are several plausible origins of this offset component: two-photon decay continuum of the $2s$ state of H-atoms from warm ionized medium and low-velocity shocks, FUV line emission from hot ionized medium, two-photon emission from the interplanetary medium, exosphere and thermosphere of the Earth. These contributors can account for two-thirds of the offset component (for details of these processes, see \citet{kulkarni2022far}). In the galactic poles of the Milky Way, \citet{akshaya2018diffuse} found offsets of $230-290$ photons cm$^{-2}$ s$^{-1}$ sr$^{-1}$ \AA$^{-1}$ in FUV. Of the total radiation observed, approximately 120 photons cm$^{-2}$ s$^{-1}$ sr$^{-1}$ \AA$^{-1}$ has been attributed to the dust-scattered light which, by definition, can't contribute to the offsets as offsets are estimated from radiation coming from zero column density regions. Of the background components contributing to the offsets, 110 photons cm$^{-2}$ s$^{-1}$ sr$^{-1}$ \AA$^{-1}$ has been attributed to the extragalactic background, and another $120-180$ photons cm$^{-2}$ s$^{-1}$ sr$^{-1}$ \AA$^{-1}$ to the unidentified background. However, at low galactic latitudes, the diffuse emission is dominated by the scattered starlight from the dust grains \citep{jura1979observational}. Another contributor to the diffuse UV background can be the $H_{2}$ fluorescence observed by \citet{Witt1989} in IC 63, a nebula at about 500 pc distance, which contains a few embedded stars with non-negligible optical and UV depth. \citet{1999MmSAI..70..825H} suggested the integrated light from spiral galaxies as a possible contributor to the extragalactic component of the diffuse UV background, but its contribution was estimated to be insignificant (about 1/10 of the minimum observed signal). 

Another interesting source for the origin of diffuse FUV background was proposed by \citet{zhitnitsky2022mysterious}, based on the observations by \citet{henry2014mystery} and \citet{akshaya2018diffuse,akshaya2019components}. The author suggested that the diffuse FUV emission, not attributed to dust scattering, can originate from dark matter annihilation events within the Axion Quark Nugget (AQN) dark matter framework. When these AQNs enter the ISM, the annihilation process starts, thereby increasing the temperature of the nuggets. The estimated flux from the AQNs is argued to be consistent with the characteristics of the diffuse FUV radiation in the Milky Way \citep{zhitnitsky2022mysterious}. However, an estimate of the AQN contribution is model dependent, with uncertainty in several parameters, due to which we are unable to include it as part of the current analysis.

\citet{pradhan2010far} \& \citet{pradhan2011} studied the diffuse FUV emission from the Magellanic Clouds, which are also low-metallicity dwarf irregular galaxies like Ho~II. Using observations from  the \textit{FUSE} and \textit{UIT} satellites, they measured the FUV diffuse fractions to be in the range of 5\% to 20\% in the LMC in the \textit{FUSE} bands 1100~\AA~--~1150~\AA, which rises to $\sim$43\% in the \textit{UIT} band at 1615~\AA~(see Fig.~4 of \citet{pradhan2011}). For the SMC, they obtained values of 34\% -- 44\% in the 4 \textit{FUSE} bands ($905-1187$~\AA) and $\sim$63\% in the \textit{UIT} band at 1615~\AA. They also found that in both galaxies, relatively sparse regions have a higher diffuse fraction than the crowded regions. 

\citet{Calzetti} have shown that SED spectral features of the diffuse light in spiral galaxies differ from those of the stellar clusters, which excludes their common origin; unless either most clusters dissolve shortly in 7--10 Myr \citep{Tremonti}, or they lack massive stars \citep{Weidner}. In order to verify whether the diffuse light spectrum in Ho~II is consistent with the spectrum of the stellar population, the availability of UV spectra for diffuse regions is crucial. Unfortunately, there are no available diffuse spectral observations for Ho~II. FUSE did observe 6 regions in the galaxy, but all of them are either star clusters or HII regions. Nevertheless, high-resolution photometric observations in the UV can also be useful.

In this paper, we utilize the UV observations of Ho~II obtained by the UVIT instrument of India's \textit{AstroSat} mission to study the nature and origin of the diffuse UV emission in this galaxy. These are the most resolved UV observations of Ho~II to date, with a spatial resolution of $1.2''-1.6''$ \citep{Tandon2020}. Using this data, we construct the diffuse UV map of Ho~II and extract the total diffuse fraction in the galaxy. We also extract the diffuse UV intensities in selected locations of 5$''$ radius (5$''$ corresponds to $\sim$ 82 pc at the distance of Ho~II), to look for variations at small spatial scales. We correlate these observations with archival HI column densities from NRAO Very Large Array (VLA) 21-cm observations (The HI Nearby Galaxy Survey (THINGS)), and IR intensities from \textit{Spitzer Space Telescope} and \textit{Herschel Space Observatory} archival data to investigate the distribution of dust clouds, their contribution to the total diffuse UV emission in the galaxy and to look for other possible sources of diffuse UV emission. Further, we derive the dust optical properties and their contribution to the diffuse FUV emission for regions with dense HI using 3D dust radiative transfer modelling. In addition, diffuse fraction values for individual locations containing a UV-detected source are also extracted to look for location specific variations in the diffuse intensities. 

\section{Observations and Analysis of Data}
 
UV imaging observations were obtained with the UVIT instrument onboard the Indian space mission \textit{AstroSat}\footnote{For details about \textit{AstroSat}, visit \url{https://www.issdc.gov.in/astro.html}.} \citep{singh2014astrosat} in 3 epochs in 2016, 2 epochs in 2019, and in 3 epochs in 2020 \citep{Vinokurov2022}. 
In this work, we have used the UVIT data obtained in December 2016 in FUV filter F154W (mean $\lambda = 1541$\AA) and NUV filter N245M (mean $\lambda = 2447$\AA), as some of the IR data we considered for the correlation study was obtained on epochs close to that date.

The IR observations of Ho~II are obtained from the SINGS \citep{Kennicutt2003} and KINGFISH \citep{Kennicutt2011} surveys at eight wavelengths: 3.6, 4.5, 5.8, 8, 24, 70, 100 and 160 $\mu$m. 
The integrated neutral hydrogen N(HI) data for Ho~II were taken from the THINGS survey \citep{walter2008things}.

\subsection{The data}

\subsubsection{UV data: UVIT instrument and data reduction}

UVIT consists of two co-aligned Ritchey-Chretien telescopes with a 375 mm aperture each, one feeding the FUV detector ($1300-1800$~\AA), and the other feeding two detectors: near ultraviolet (NUV: $1800-3000$~\AA) and visible (VIS: $3200-5500$~\AA) through a dichroic filter, providing a field of view of $28'$. Each channel has a $512 \times$512 CMOS detector, which reads frames $\sim$29 times per second with an exposure time of $\sim 35$ ms per frame \citep{tandon2017orbit}. 

\begin{figure*}[h!]
\centering
\includegraphics[scale=0.33]{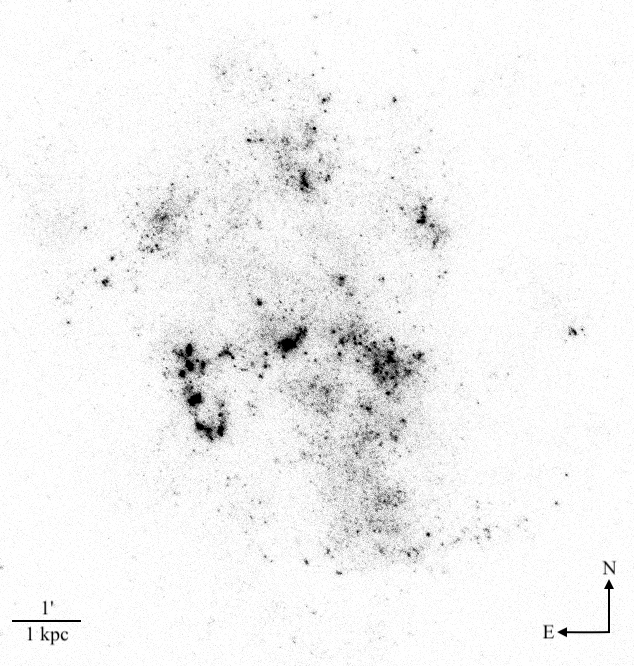}
\hskip 0.2in
\includegraphics[scale=0.33]{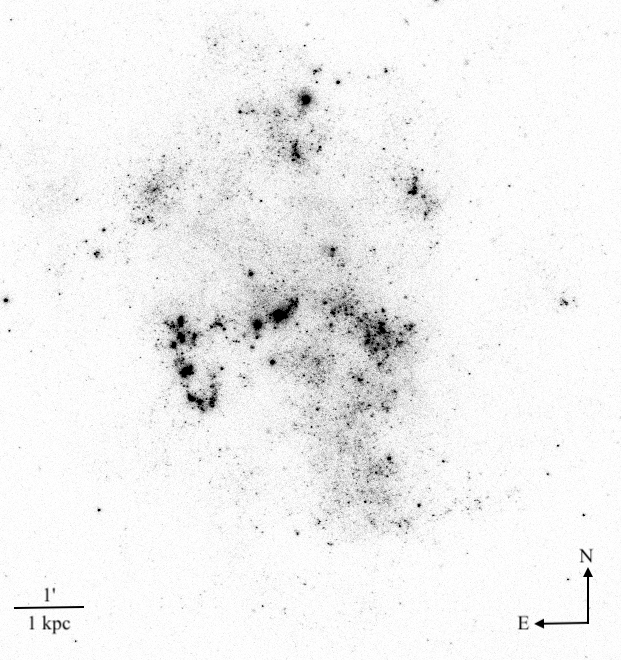}
\caption{FUV (\textit{left}) and NUV (\textit{right}) co-added images of Holmberg II observed by the UVIT instrument onboard \textit{AstroSat} at mean wavelengths of 1541~\AA~and 2447~\AA, respectively. These are the combined images with an integrated exposure time of $\sim$16.6 ks and $\sim$16.06 ks for FUV and NUV, respectively.}
\label{fig:astrosat_fuv}
\end{figure*}

The detectors provide images in the form of a list of centroids, computed to a precision of 1/32 of a pixel on the $512 \times 512$ CMOS detector, representing the detected photons in each frame \citep{tandon2017orbit}. Level 1 data, obtained from Indian Space Science Data Centre\footnote{\url{https://astrobrowse.issdc.gov.in/astro_archive/archive/Home.jsp}} (ISSDC), were converted to Level 2 using the CCDLAB pipeline \citep{Postma2017, Postma2021}. The pipeline applies various corrections, such as aspect, satellite drift, jitter, telescope distortion, flat field, etc., to provide science-ready counts and exposure maps. The details of the pipeline can be obtained from \citet{Postma2017,Postma2021}. Each of the pixels in the CMOS detector is mapped to 8$\times$8 subpixels in the final image to get a plate-scale of $\sim$0.416 per subpixel \citep{Tandon2020}. The final obtained Level~2 data had a full width at half maxima (FWHM) of $\sim 1.2^{\prime\prime}$ in FUV, and $\sim 1.05^{\prime \prime}$ in NUV. For this work, we considered all the UVIT observations available in the FUV BaF2 and NUV B13 filters, as detailed in Table~\ref{tab:observations}. For each filter, we combined multiple epoch observations into single images using CCDLAB. Additionally, we applied cosmic-ray correction by removing frames with counts above 4$\sigma$ of the median counts in the frame \citep{Postma2020} using CCDLAB. The combined images have a total exposure time of 16,644 seconds in the FUV (Fig.~\ref{fig:astrosat_fuv}, {\it Left}) and 16,061 seconds in the NUV B13 filter (Fig.~\ref{fig:astrosat_fuv}, {\it Right}). The conversion from counts per second (CPS) to physical flux units is according to the conversion factors provided in \citet{tandon2017orbit}.

\begin{table}[hb!]
    \centering
    \caption{Observational log}
   \begin{adjustbox}{width=0.5\textwidth, center}
         \begin{tabular}{ccc}
    \hline
        Observation date & Instrument & Total exposure (s) \\
        \hline
         21 November, 2016 & UVIT FUVF2 & 6687 \\
         & UVIT NUVF3 & 6454\\
         09 December, 2016 & UVIT FUVF2 & 9957 \\
         & UVIT NUVF3 & 9607\\
         \hline
         27 March, 2007 & IRAC 3.6 $\micron$ & 1286.4 \\
          & IRAC 4.5 $\micron$ & 1286.4 \\
             & IRAC 5.8 $\micron$ & 1286.4 \\
         & IRAC 8 $\micron$ & 1286.4 \\
         \hline
         14 October, 2004 & MIPS 24 $\micron$ & 146.8 \\
          & MIPS 70 $\micron$ & 83.8 \\
          & MIPS 160 $\micron$ & 16.76 \\
          \hline
         29 April, 2012 & PACS 100 $\micron$ & 12 scans$^1$\\
          \hline
$^1$\citet{Kennicutt2011}. &  &  \\         
    \end{tabular} 
    \end{adjustbox}
    \label{tab:observations}      
\end{table}

\begin{table*}
\centering
\caption{Diffuse FUV, NUV intensities, and neutral hydrogen column densities N(HI) for selected 33 locations having non-zero IR intensities at all considered wavelengths. The N(HI) values have been derived from the THINGS integrated HI map. Magenta colour indicates locations with N(HI) $<1\times$$10^{21}$ cm$^{-2}$. }
\begin{tabular}{cccccc}
\hline
Locations & \textit{l}   &  \textit{b} & FUV intensity & NUV intensity & N(HI)\\
      & (deg)  & (deg) & (photons cm$^{-2}$ s$^{-1}$ sr$^{-1}$ \AA$^{-1}$) & (photons cm$^{-2}$ s$^{-1}$ sr$^{-1}$ \AA$^{-1}$)  & ($10^{21}$ cm$^{-2}$) \\
\hline 
     1 & 144.2995 & 32.6691 & 3524.02 $\pm$ 195.43 & 2799.29 $\pm$ 103.08  & 1.60 $\pm$ 0.09 \\
     \magenta{2} &   \magenta{144.2058} & \magenta{32.7248} & \magenta{521.58 $\pm$ 105.50} & \magenta{419.11} $\pm$ \magenta{58.81} & \magenta{0.08} $\pm$ \magenta{0.008} \\
      3 &  144.2835 & 32.7029 & 6533.34 $\pm$ 255.21 & 5195.95 $\pm$ 133.38 & 1.48 $\pm$ 0.43 \\
      \magenta{4} &  \magenta{144.2829} & \magenta{32.7354} & \magenta{1010.03 $\pm$ 125.49} & \magenta{555.67 $\pm$ 62.18} & \magenta{0.97 $\pm$ 0.05} \\
      \magenta{5} &  \magenta{144.2714} & \magenta{32.6746} & \magenta{5021.40 $\pm$ 226.59} & \magenta{4901.48 $\pm$ 129.93} & \magenta{0.98 $\pm$ 0.11} \\
      6 &  144.2682 & 32.7303 & 3083.16 $\pm$ 186.10 & 2106.87 $\pm$ 92.62 & 1.04 $\pm$ 0.10 \\
      \magenta{7} &  \magenta{144.2128} & \magenta{32.6840} & \magenta{3899.69 $\pm$ 203.15} & \magenta{2457.21 $\pm$ 98.29} & \magenta{0.91 $\pm$ 0.11} \\
      \magenta{8} &  \magenta{144.3110} & \magenta{32.7007} & \magenta{3952.94 $\pm$ 204.34} & \magenta{2608.94 $\pm$ 100.12} & \magenta{0.36 $\pm$ 0.07} \\
      9 & 144.3063 & 32.6755 & 10922.56 $\pm$ 321.58 & 7897.64 $\pm$ 160.87 & 2.05 $\pm$ 0.21 \\
       10 & 144.2544 & 32.6734 & 3929.44 $\pm$ 205.28 & 3081.45 $\pm$ 107.11 & 1.44 $\pm$ 0.12 \\
       \magenta{11} & \magenta{144.3066} & \magenta{32.6670} & \magenta{2788.49 $\pm$ 177.58} & \magenta{1777.78 $\pm$ 86.90} & \magenta{0.96 $\pm$ 0.02} \\
      \magenta{12} &  \magenta{144.3100} & \magenta{32.6556} & \magenta{1505.13 $\pm$ 141.45} & \magenta{871.87 $\pm$ 69.63} & \magenta{0.40 $\pm$ 0.05} \\
      \magenta{13} &  \magenta{144.2417} & \magenta{32.6750} & \magenta{4876.30 $\pm$ 223.23} & \magenta{3274.28 $\pm$ 109.47} & \magenta{0.27 $\pm$ 0.03} \\
      14 &  144.2816 & 32.6657 & 3308.87 $\pm$ 190.74 & 2930.76 $\pm$ 104.94 & 1.63 $\pm$ 0.09\\
      15 &  144.2885 & 32.7289 & 4060.48 $\pm$ 207.78 & 2142.49 $\pm$ 93.16 & 2.74 $\pm$ 0.14 \\
       16 & 144.2847 & 32.6633 & 2864.80 $\pm$ 179.26 & 2344.61 $\pm$ 96.10 & 1.53 $\pm$ 0.08 \\
      \magenta{17} &  \magenta{144.3087} & \magenta{32.6598} & \magenta{1299.51 $\pm$ 135.44} & \magenta{776.81 $\pm$ 67.38} & \magenta{0.64 $\pm$ 0.12} \\
      \magenta{18} &  \magenta{144.2642} & \magenta{32.7385} & \magenta{1382.51 $\pm$ 137.36} & \magenta{779.92 $\pm$ 67.28} & \magenta{0.65 $\pm$ 0.10} \\
      19 &  144.2226 & 32.6552 & 1044.90 $\pm$ 126.64 & 585.59 $\pm$ 63.10 & 1.80 $\pm$ 0.12 \\
      \magenta{20} &  \magenta{144.2661} & \magenta{32.7527} & \magenta{451.27 $\pm$ 103.50} & \magenta{187.88 $\pm$ 52.70} & \magenta{0.78 $\pm$ 0.03} \\
       21 & 144.3336 & 32.6575 & 842.51 $\pm$ 119.50 & 435.68 $\pm$ 59.22 & 1.19 $\pm$ 0.09 \\
      22 &  144.2679 & 32.7278 & 6186.94 $\pm$ 249.07 & 3884.05 $\pm$ 118.01 & 1.42 $\pm$ 0.24 \\
       23 & 144.2625 & 32.6781 & 6031.80 $\pm$ 246.48 & 4926.88 $\pm$ 130.52 & 1.57 $\pm$ 0.08 \\
       \magenta{24} & \magenta{144.2465} & \magenta{32.6483} & \magenta{484.79 $\pm$ 104.94} & \magenta{373.81 $\pm$ 57.76} & \magenta{0.52 $\pm$ 0.04} \\
       25 & 144.2843 & 32.7338 & 1020.05 $\pm$ 125.75 & 571.37 $\pm$ 62.72 & 1.01 $\pm$ 0.04 \\
       26 & 144.2823 & 32.6637 & 2583.06 $\pm$ 172.47 & 2330.55 $\pm$ 96.08 & 1.54 $\pm$ 0.08 \\
       \magenta{27} & \magenta{144.2817} & \magenta{32.7376} & \magenta{656.15 $\pm$ 111.45} & \magenta{437.76 $\pm$ 59.30} & \magenta{0.92 $\pm$ 0.06} \\
       \magenta{28} & \magenta{144.2418} & \magenta{32.6507} & \magenta{705.38 $\pm$ 114.13} & \magenta{519.10 $\pm$ 61.29} & \magenta{0.53 $\pm$ 0.07} \\
       \magenta{29} & \magenta{144.2930} & \magenta{32.7406} & \magenta{790.26 $\pm$ 117.64} & \magenta{379.86 $\pm$ 57.89} & \magenta{0.93 $\pm$ 0.15} \\
       30 & 144.2723 & 32.6760 & 6651.67 $\pm$ 256.96 & 5633.73 $\pm$ 138.20 & 1.22 $\pm$ 0.16 \\
       \magenta{31} & \magenta{144.2476} & \magenta{32.6519} & \magenta{1314.31 $\pm$ 135.42} & \magenta{1100.58 $\pm$ 74.24} & \magenta{0.90 $\pm$ 0.13} \\
       32 & 144.2721 & 32.7334 & 1441.79 $\pm$ 139.82 & 883.12 $\pm$ 69.57 & 1.10 $\pm$ 0.06 \\
       33 & 144.2765 & 32.6735 & 6188.49 $\pm$ 249.34 & 4984.33 $\pm$ 131.09 & 1.13 $\pm$ 0.07 \\
     \hline
    \end{tabular}
    \label{tab:fuv_intensity}
\end{table*}

\begin{table*}
    \centering
\caption{IR intensities for 33 locations with non-zero values in all bands.}
\begin{adjustbox}{width=\textwidth, center}
\begin{tabular}{ccccccccc}
\hline
Location & \textit{l} & \textit{b}  & I$_{4.5 \micron}$ & I$_{5.8 \micron}$  & I$_{24 \micron}$ & I$_{70 \micron}$ & I$_{100 \micron}$ & I$_{160 \micron}$\\
  No      & (degrees) & (degrees)  & (MJy sr$^{-1}$) & (MJy sr$^{-1}$) &  (MJy sr$^{-1}$) & (MJy sr$^{-1}$) & (MJy sr$^{-1}$) & (MJy sr$^{-1}$) \\
       \hline
{1} &  {144.2995} &  {32.6691} & 0.0062 $\pm$ 0.0041 & 0.0063 $\pm$ 0.0219  & 0.0212 $\pm$ 0.0370 & 1.0850 $\pm$ 0.4762 & 2.0711 $\pm$ 0.1891 & 1.3365 $\pm$ 0.6228 \\
{2} &  {144.2058} &  {32.7248} & 0.0045 $\pm$ 0.0035 & 0.0097 $\pm$ 0.0228  & 0.0335 $\pm$ 0.0476 & 0.3675 $\pm$ 0.3499 & 0.0888 $\pm$ 0.4188 & 0.1896 $\pm$ 0.2158 \\

{3} & {144.2835} &  {32.7029} & 0.0131 $\pm$ 0.0072 & 0.0052 $\pm$ 0.0410  & 0.1527 $\pm$ 0.1028 & 4.1669 $\pm$ 0.9138 & 5.2371 $\pm$ 0.3456 & 5.3175 $\pm$ 1.1359 \\
{4} & {144.2829} &  {32.7354} & 0.0043 $\pm$ 0.0127 & 0.0245 $\pm$ 0.0210  & 0.0739 $\pm$ 0.0915 & 0.7040 $\pm$ 0.2525 & 2.1856 $\pm$ 0.4774 & 0.4082 $\pm$ 0.3509 \\
{5} & {144.2714} &  {32.6746} & 0.0052 $\pm$ 0.0061 & 0.0221 $\pm$ 0.0241  & 0.0152 $\pm$ 0.0263 & 1.4038 $\pm$ 0.4099 & 0.2419 $\pm$ 0.2397 & 2.3658 $\pm$ 0.3959 \\
{6} & {144.2682}&  {32.7303} & 0.0042 $\pm$ 0.0051 & 0.0009 $\pm$ 0.0343 & 0.0536 $\pm$ 0.0373 & 0.8714 $\pm$ 0.4047 & 0.1217 $\pm$ 0.4199 & 1.2483 $\pm$ 0.5298 \\

{7} & {144.2128} &  {32.6840} & 0.0089 $\pm$ 0.0114 & 0.0492 $\pm$ 0.0365  & 0.0013 $\pm$ 0.0364 & 0.0950 $\pm$ 0.3539 & 0.0938 $\pm$ 0.3735 & 1.1237 $\pm$ 0.1573 \\
{8} & {144.3110} & {32.7007} & 0.0070 $\pm$ 0.0065 & 0.0219 $\pm$ 0.0195  & 0.0382 $\pm$ 0.0332 & 0.0932 $\pm$ 0.1962 & 2.6554 $\pm$ 0.4565 & 0.2270 $\pm$ 0.2994 \\
{9} & {144.3063} &  {32.6755} & 0.0049 $\pm$ 0.0080 & 0.0117 $\pm$ 0.0273  & 0.0649 $\pm$ 0.0327 & 1.6511 $\pm$ 0.3393 & 1.8380 $\pm$ 0.4956 & 1.5806 $\pm$ 0.2642 \\

{10} &{144.2544} & {32.6734} & 0.0038 $\pm$ 0.0048 & 0.0111 $\pm$ 0.0222  & 0.0287 $\pm$ 0.0242 & 1.0782 $\pm$ 0.4586 & 0.4431 $\pm$ 0.5541 & 1.1277 $\pm$ 0.1548 \\

{11} & {144.3066} &  {32.6670} & 0.0029 $\pm$ 0.0055 & 0.0052 $\pm$ 0.0278  & 0.0135 $\pm$ 0.0385 & 0.2171 $\pm$ 0.3018 & 1.1570 $\pm$ 0.4715 & 0.8655 $\pm$ 0.6826 \\
{12} &{144.3100} & {32.6556} & 0.0030 $\pm$ 0.0049 & 0.0138 $\pm$ 0.0262  & 0.0061 $\pm$ 0.0289 & 0.0810 $\pm$ 0.2384 & 2.2887 $\pm$ 0.3257 & 0.7581 $\pm$ 0.0427 \\
{13} &{144.2417} & {32.6750} & 0.0115 $\pm$ 0.0522 & 0.0088 $\pm$ 0.0437  & 0.0135 $\pm$ 0.0402 & 0.3768 $\pm$ 0.1624 & 1.4094 $\pm$ 0.3236 & 0.5471 $\pm$ 0.5152 \\
{14} &{144.2816} &{32.6657} & 0.0049 $\pm$ 0.0051 & 0.0080 $\pm$ 0.0208  & 0.0202 $\pm$ 0.0351 & 1.2717 $\pm$ 0.2959 & 0.9758 $\pm$ 0.4028 & 1.0013 $\pm$ 0.2101 \\
{15} &{144.2885} &{32.7289} & 0.0011 $\pm$ 0.0036 & 0.0107 $\pm$ 0.0259  & 0.0525 $\pm$ 0.0558 & 2.1309 $\pm$ 0.4932 & 2.6377 $\pm$ 0.5618 & 1.5193 $\pm$ 0.7586 \\
{16} &{144.2847} & {32.6633} & 0.0041 $\pm$ 0.0025 & 0.0168 $\pm$ 0.0184  & 0.0008 $\pm$ 0.0293 & 1.4128 $\pm$ 0.1951 & 1.6775 $\pm$ 0.4181 & 1.1351 $\pm$ 0.0553 \\

{17} &{144.3087} &{32.6598} & 0.0009 $\pm$ 0.0036 & 0.0052 $\pm$ 0.0243  & 0.0053 $\pm$ 0.0276 & 0.4203 $\pm$ 0.2480 & 2.2430 $\pm$ 0.5503 & 0.2245 $\pm$ 0.2224 \\
{18} &{144.2642} &{32.7385} & 0.0042 $\pm$ 0.0044 & 0.0064 $\pm$ 0.0225  & 0.0137 $\pm$ 0.0344 & 0.3062 $\pm$ 0.1321 & 0.3756 $\pm$ 0.5151 & 0.3313 $\pm$ 0.2365 \\
{19} &{144.2226} &{32.6552} & 0.0088 $\pm$ 0.0097 & 0.0014 $\pm$ 0.0242 & 0.0233 $\pm$ 0.0331 & 0.4190 $\pm$ 0.3380 & 0.8689 $\pm$ 0.4349 & 0.6305 $\pm$ 0.2124 \\
{20} &{144.2661} &{32.7527} & 0.0114 $\pm$ 0.0492 & 0.0095 $\pm$ 0.0298  & 0.0435 $\pm$ 0.0449 & 0.6044 $\pm$ 0.1305 & 2.1473 $\pm$ 0.4956 & 0.0313 $\pm$ 0.1627 \\
{21} &{144.3336} &{32.6575} & 0.0026 $\pm$ 0.0051 & 0.0240 $\pm$ 0.0496  & 0.0063 $\pm$ 0.0357 & 0.2864 $\pm$ 0.2863 & 1.2062 $\pm$ 0.5060 & 0.0144 $\pm$ 0.1607 \\

{22} &{144.2679} &{32.7278} & 0.0055 $\pm$ 0.0089 & 0.0270 $\pm$ 0.0316  & 0.0799 $\pm$ 0.0432 & 1.7395 $\pm$ 0.6421 & 0.7082 $\pm$ 0.4474 & 1.3012 $\pm$ 0.5407 \\
{23} &{144.2625} &{32.6781} & 0.0054 $\pm$ 0.0029 & 0.0146 $\pm$ 0.0219  & 0.0246 $\pm$ 0.0266 & 1.4907 $\pm$ 0.2342 & 3.2091 $\pm$ 0.3913 & 2.4704 $\pm$ 0.4102 \\
{24} &{144.2465} &{32.6483} & 0.0055 $\pm$ 0.0258 & 0.0102 $\pm$ 0.0152  & 0.0194 $\pm$ 0.0414 & 0.2141 $\pm$ 0.4245 & 0.5041 $\pm$ 0.3707 & 0.7612 $\pm$ 0.5002 \\
{25} &{144.2843} &{32.7338} & 0.0019 $\pm$ 0.0173 & 0.0034 $\pm$ 0.0137  & 0.0754 $\pm$ 0.0408 & 0.7050 $\pm$ 0.2847 & 2.8483 $\pm$ 0.2218 & 0.1798 $\pm$ 0.3059 \\

{26} &{144.2823} &{32.6637} & 0.0046 $\pm$ 0.0045 & 0.0105 $\pm$ 0.0183  & 0.0116 $\pm$ 0.0282 & 1.3005 $\pm$ 0.2660 & 0.9547 $\pm$ 0.5276 & 0.9787 $\pm$ 0.1291 \\
{27} &{144.2817} &{32.7376} & 0.0020 $\pm$ 0.0011 & 0.0143 $\pm$ 0.0243  & 0.0502 $\pm$ 0.0561 & 0.7400 $\pm$ 0.3167 & 1.4890 $\pm$ 0.2596 & 0.9973 $\pm$ 0.3334 \\
{28} &{144.2418} &{32.6507} & 0.0021 $\pm$ 0.0051 & 0.0058 $\pm$ 0.0238  & 0.0062 $\pm$ 0.0381 & 0.3664 $\pm$ 0.4157 & 1.5994 $\pm$ 0.5095 & 0.4009 $\pm$ 0.6219 \\

{29} &{144.2930} &{32.7406} & 0.0032 $\pm$ 0.0052 & 0.0009 $\pm$ 0.0310  & 0.0271 $\pm$ 0.0345 & 0.2107 $\pm$ 0.2292 & 2.2268 $\pm$ 0.3958 & 0.6244 $\pm$ 0.0767 \\
{30} &{144.2723} &{32.6760} & 0.0065 $\pm$ 0.0097 & 0.0180 $\pm$ 0.0330  & 0.0151 $\pm$ 0.0347 & 2.3598 $\pm$ 0.5313 & 0.0340 $\pm$ 0.2841 & 3.1375 $\pm$ 0.4827 \\
{31} &{144.2476} &{32.6519} & 0.0070 $\pm$ 0.0548 & 0.0124 $\pm$ 0.0338  & 0.0815 $\pm$ 0.0557 & 1.3963 $\pm$ 0.4828 & 3.3328 $\pm$ 0.6623 & 1.6340 $\pm$ 0.5010 \\
{32} &{144.2721} &{32.7334} & 0.0010 $\pm$ 0.0037 & 0.0159 $\pm$ 0.0286  & 0.0264 $\pm$ 0.0390 & 0.9325 $\pm$ 0.3304 & 0.6293 $\pm$ 0.4523 & 1.6947 $\pm$ 0.4395 \\
{33} &{144.2765} &{32.6735} & 0.0061 $\pm$ 0.0040 & 0.0315 $\pm$ 0.0278  & 0.0440 $\pm$ 0.0317 & 2.1421 $\pm$ 0.4936 & 1.0896 $\pm$ 0.3669 & 2.8156 $\pm$ 0.5833 \\
        \hline
    \end{tabular}
\end{adjustbox}  
    \label{tab:IR_intensity_30}
\end{table*}

\subsubsection{IR data: Spitzer and Herschel}
\label{sec:ir_data}

We acquired the IR images of Ho~II from two surveys, specifically the SINGS survey \citep{Kennicutt2003} and the KINGFISH survey \citep{Kennicutt2011}. These images were retrieved from the NASA/IPAC Infrared Science Archive, at eight different wavelengths\footnote{For accessing SINGS and KINGFISH data, please visit \url{https://irsa.ipac.caltech.edu/data/SPITZER/SINGS/galaxies/hoii/} and \url{https://irsa.ipac.caltech.edu/data/Herschel/KINGFISH/galaxies/HoII/PACS/}, respectively.}. Specifically, the images at wavelengths 3.6, 4.5, 5.8 and 8 $\micron$ were obtained using the Infrared Array Camera (IRAC) instrument of the \textit{Spitzer Space Telescope}\footnote{For detailed information about the IRAC instrument, refer to \citet{Fazio2004}.}, whereas images at 24, 70 and 160 $\mu$m were acquired using the Multiband Imaging Photometer for Spitzer (MIPS) instrument\footnote{For details about the MIPS instrument, see \citet{Rieke2004}.}. Additionally, the 100 $\micron$ image was obtained using the Photodetector Array Camera and Spectrometer (PACS) instrument\footnote{See \citet{poglitsch2010} for details.} onboard the \textit{Herschel Space Observatory}. It is worth noting that a constant background level had already been subtracted from each of these images.

The images were further processed to correct for contamination from stars. To effectively account for stellar contamination, we adopted a methodology outlined by \citet{Helou2004}, which is based on the assumption that the emission at 3.6 $\micron$ traces the total stellar emission. Here, we applied specific scaling factors, namely 0.596, 0.399, 0.232, and 0.032, to scale down the 3.6 $\micron$ intensities for the 4.5, 5.8, 8 and 24 $\micron$ images, respectively. These scaled 3.6 $\micron$ intensities were then subtracted from the corresponding intensities in the convolved images at their respective wavelengths. This process effectively isolated what we consider to be as diffuse dust emission.

We did not apply any correction for stellar contamination to the 70, 100 and 160 $\micron$ images, because the influence of stellar contributions diminishes significantly as we move to longer wavelengths in the IR spectrum, rendering it negligible within this wavelength range. 

\begin{figure*}
    \centering
\includegraphics[height=8cm,width=15cm]{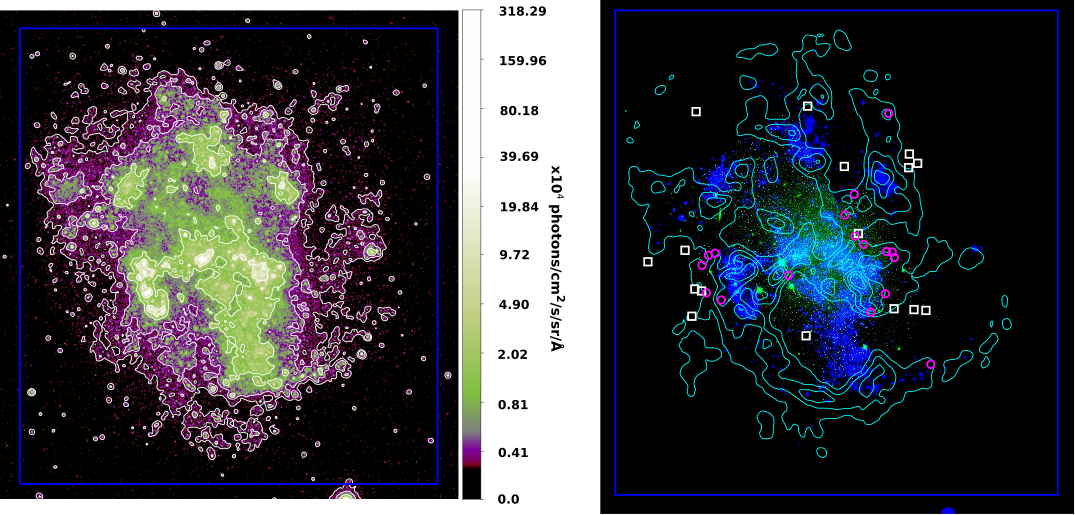}
\caption{{\it Left}: Smoothed FUV image of Ho~II with isophotes overplotted for five intensity levels. {\it Right}: Multiband image of Ho~II with FUV (F154W; {\it AstroSat}-UVIT) in blue, IR (3.6~$\mu$m; {\it Spitzer}) in green, and integrated HI map (THINGS) in teal contours. The HI contour corresponds to N(HI) = $1 \times 10^{21}$ cm$^{-2}$. Selected diffuse UV locations with N(HI) $>1 \times 10^{21}$ cm$^{-2}$ are indicated by purple circles, and locations with N(HI) $<1 \times 10^{21}$ cm$^{-2}$ are indicated by white squares. It is evident from the figure that the locations with N(HI) $<1 \times 10^{21}$ cm$^{-2}$ mostly lie in the cavities devoid of HI gas. The blue box in both panels indicates the analysis region.}
\label{fig:rgb}
\end{figure*}
 
\begin{figure*}
\centering
\includegraphics[height=6cm,width=6cm]{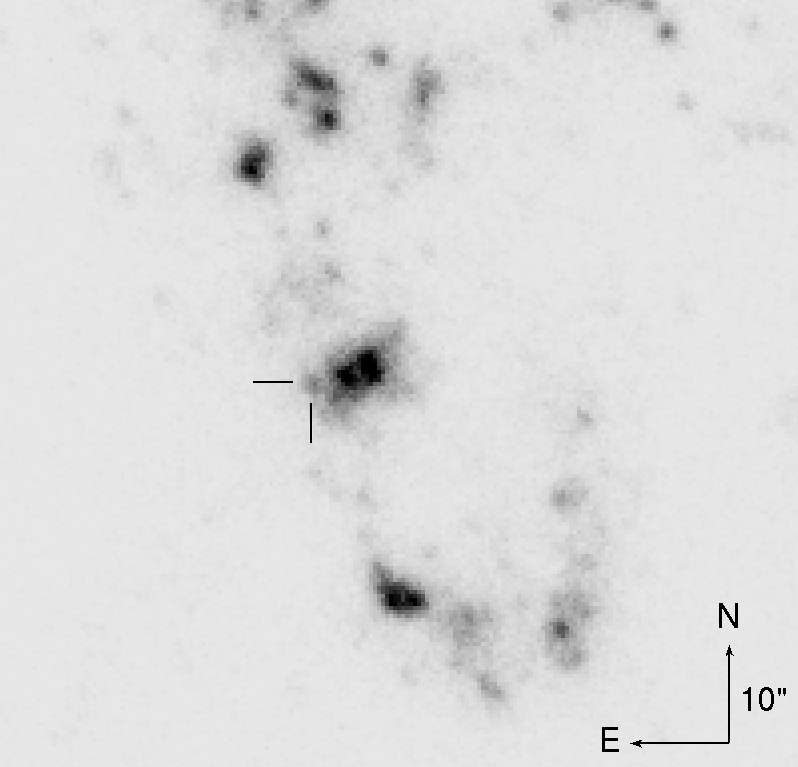}
\hskip 0.4in
\includegraphics[height=6cm,width=6cm]{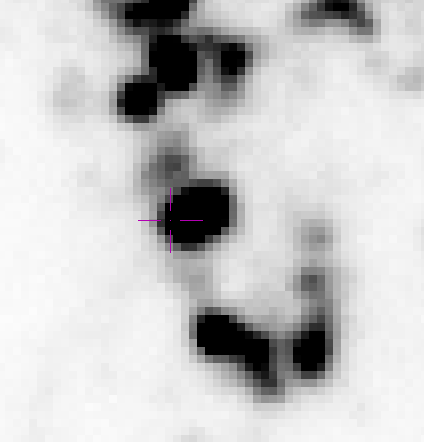}
\caption{\textit{Left}: UVIT FUV image of the star-forming region harboring the ULX source Ho~II X-1. \textit{Right}: GALEX FUV image of the same region in the same scale. The location of the ULX is shown with the reticle. These two images clearly highlight the improvement in resolution for UVIT compared to GALEX.}
\label{fig:astrosat_galex_comparison}
\end{figure*}
\section{Results and Discussions}

\subsection{Diffuse UV emission}
 
From the UV images of Holmberg~II (Fig.~\ref{fig:astrosat_fuv}), it is evident that the UV emission is patchy in nature with regions of recent massive star formation, like in the central arc, being the most prominent. In order to probe the FUV morphology of Ho~II, we have overplotted five FUV intensity contours on the UVIT image (shown in Fig.~\ref{fig:rgb}, {\it Left}), where the brightest level is at $3.49 \times 10^{6}$ photons cm$^{-2}$ s$^{-1}$ sr$^{-1}$ \AA$^{-1}$. These brightest values correspond to the centers of massive OB-associations, especially in the central star-forming arc, and a few other regions with young massive stars away from the center. Owing to the high spatial resolution of UVIT, we can clearly identify several FUV point sources that represent these clusters. The right panel of Fig.~\ref{fig:rgb} shows a multiband image of Ho~II with {\it AstroSat} FUV (F154W) in blue, {\it Spitzer} IR (3.6 $\mu$m) in green, and with overplotted THINGS integrated HI  contours. The FUV morphology is very different from the distribution of both the older stellar population traced by the near-IR observations and the HI distribution. 

The high resolution of UVIT images allows the study of diffuse UV emission, since most of the point sources can be identified and subtracted. Fig.~\ref{fig:astrosat_galex_comparison} shows a comparison between UVIT and GALEX\footnote{This GALEX image is available from \url{https://galex.stsci.edu/GR6/?page=downloadlist&tilenum=23043&type=coaddI}} (observation ID: GI3\_050003\_HolmbergII\_0001, observation date: 31 March, 2007) FUV images of a region in the central star-forming arc of the galaxy which harbors the ultraluminous X-ray (ULX) source Holmberg II X-1, the location of which is shown with the reticle on both images. The comparison clearly highlights the significant improvement in resolution in the UVIT images, where the ULX source can be identified as a distinct source, while it is blended in the GALEX image. 

\begin{figure*}[h]
\centering
\hskip -0.65in
\includegraphics[scale=0.6]{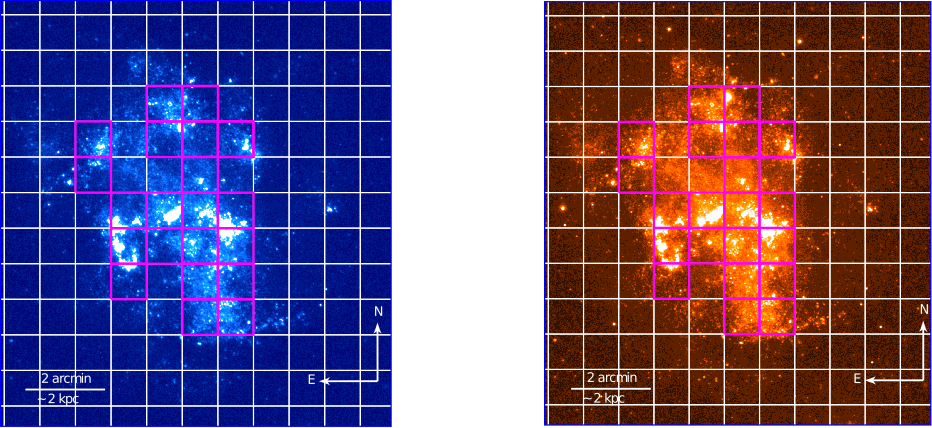}
\vskip 0.1in
\hskip 0.1in
\centering
\includegraphics[scale=0.6]{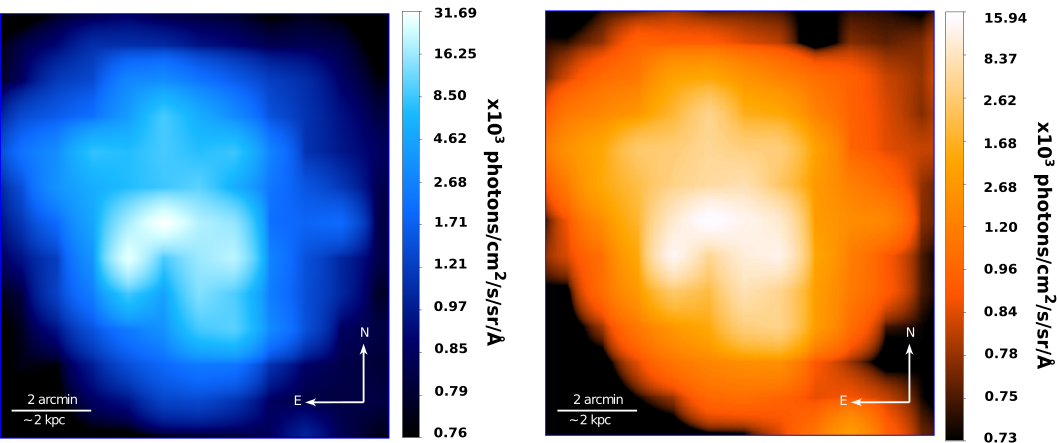}
\caption{Top panel: Regions used for deriving FUV (left) and NUV (right) diffuse maps from UVIT observations. The blocks of $128 \times 128$ pixels used for background calculation are marked by white boxes for non-crowded regions, and by purple boxes for crowded regions. Bottom panel: FUV (left) and NUV (right) diffuse maps of Ho~II. }
\label{fig:diffuse}
\end{figure*}
 
The background in the UVIT field includes contributions from Ho II and emissions external to Ho II. In this work, we used the term “diffuse emission” to represent the astrophysical diffuse emission and to distinguish it from the calculated background that is used to set up the threshold for source detection. Due to the low UV background counts, we considered the Poisson distribution for the calculation of the background \citep{Morrissey2007} as implemented in the GALEX pipeline. A rectangular region of size 1408$\times$1536 pixels ($\sim9.7\times10.6$ arcmin), nearly the size of the galaxy \cite[major axis 9.3 arcmin;][]{Nilson1973}, comprises the analysis region in this work. This region (indicated by the blue box in Fig.~\ref{fig:rgb}) was divided into blocks of 128$\times$128 pixels, within which the background is assumed to be constant (see later top panel of Fig.~4 and Sec.~3.1.1). The background is calculated iteratively by removing the pixels above the $3\sigma$ equivalent threshold, calculated from the Poisson distribution (confidence level of 99.73\%) until the mean converges in each block. The mean obtained value was considered as the background for the central pixel of each block. Background maps were bilinear interpolated to obtain the background map of the original image size. Sources were detected using SExtractor \citep{bertin1996}. Due to crowded regions, we set the deblending parameter (DEBLEND\_MINCONT) to 5e-7. Other parameters were fixed at values as provided by \citet{Ananthamoorthy2024}. Source fluxes were calculated within a 3-pixel aperture centered on each source and corrected for aperture effects using \citet{Tandon2020}. 

\begin{figure}[h]
\centering
\includegraphics[scale=0.13]{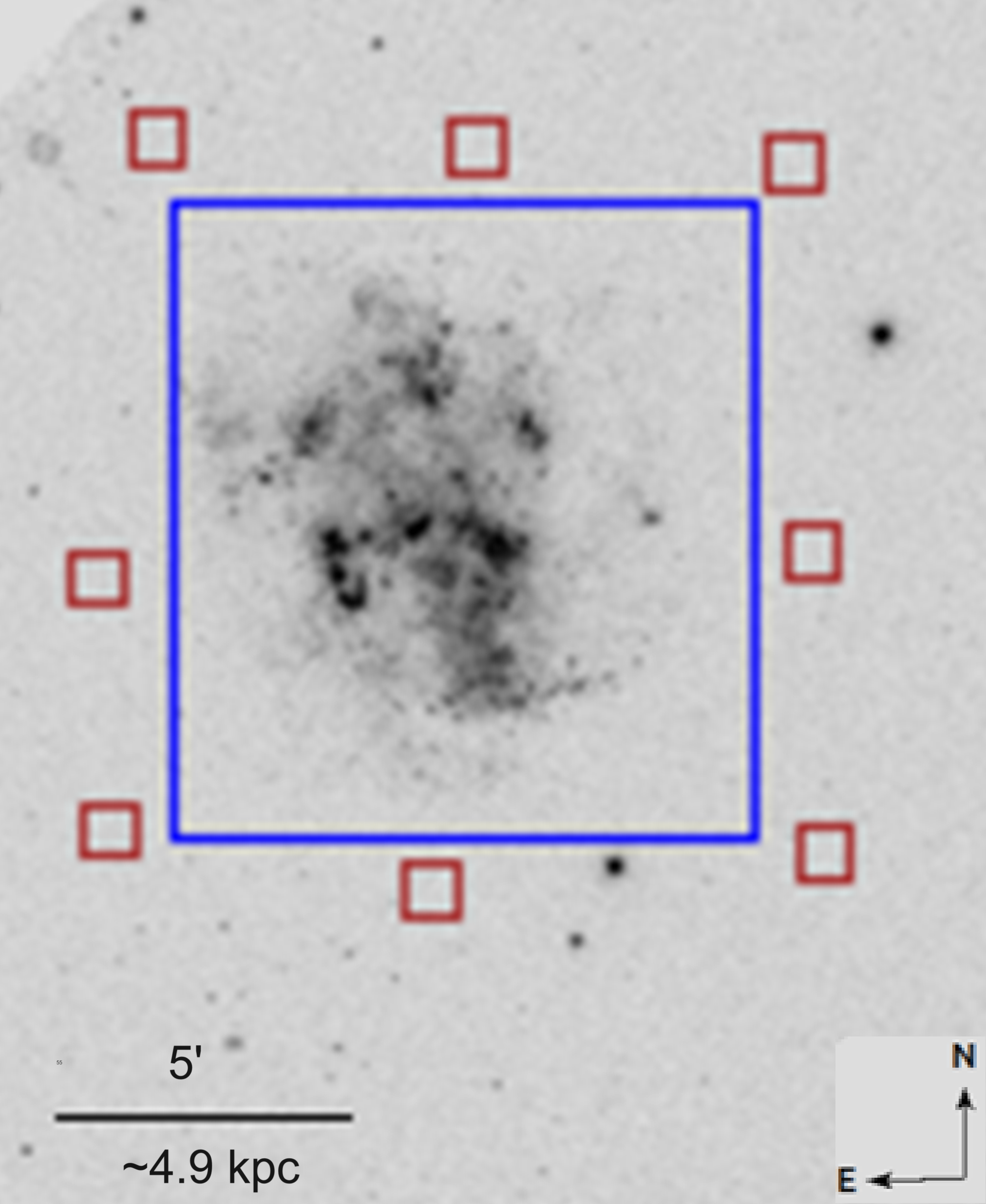}
\caption{The analysis region of Ho~II. The blue box corresponds to the region used for diffuse fraction calculation. Red boxes correspond to regions used for the removal of external background.}
\label{fig:outside}
\end{figure}

To obtain the diffuse UV map of Ho~II, we performed the following steps:
\begin{enumerate}
\item We generated the point spread function (PSF) from the isolated sources in the field using the \textit{psf} task in the image reduction and analysis facility (IRAF) \citep{iraf1986}.
\item The detected sources are removed from the field according to the PSF.

\item A constant emission in source-free regions calculated far outside of Ho II (indicated by red boxes in Fig.~\ref{fig:outside}) is subtracted to remove the diffuse emission of external origin, presented in Table~\ref{tab:external}.

\item The obtained diffuse map is averaged over $128\times128$ pixel blocks and linearly interpolated to the remaining pixels to obtain the final diffuse map from Ho II.
\end{enumerate}

\begin{table}[h]
\centering
\caption{Regions to calculate the background outside the galaxy (marked by red boxes in Fig.~\ref{fig:outside}).}
\begin{tabular}{cccc}
\hline
 $l$  &  $b$ & FUV background  & NUV background  \\
 (deg)  & (deg) & (CPS) & (CPS) \\
\hline
144.13329  & 32.74809 & 1.05562e-05 & 3.37928e-05 \\
144.15633  & 32.66062 & 1.07369e-05 & 3.30274e-05\\
144.18223  & 32.57405 & 1.03907e-05 & 3.49069e-05\\
144.30992  & 32.58952 & 1.11986e-05 & 3.39291e-05 \\
144.40942  & 32.60196 & 1.0715e-05  & 3.15944e-05\\
144.39718  & 32.71293 & 1.09996e-05 & 3.28439e-05\\
144.35667  & 32.79848 & 1.03039e-05 & 3.1771e-05 \\
144.27337  & 32.78856 & 1.11127e-05 & 3.53945e-05 \\
\hline
\label{tab:external}
\end{tabular}
\end{table}

\begin{figure*}[htb]
\centering
\includegraphics[scale=0.25]{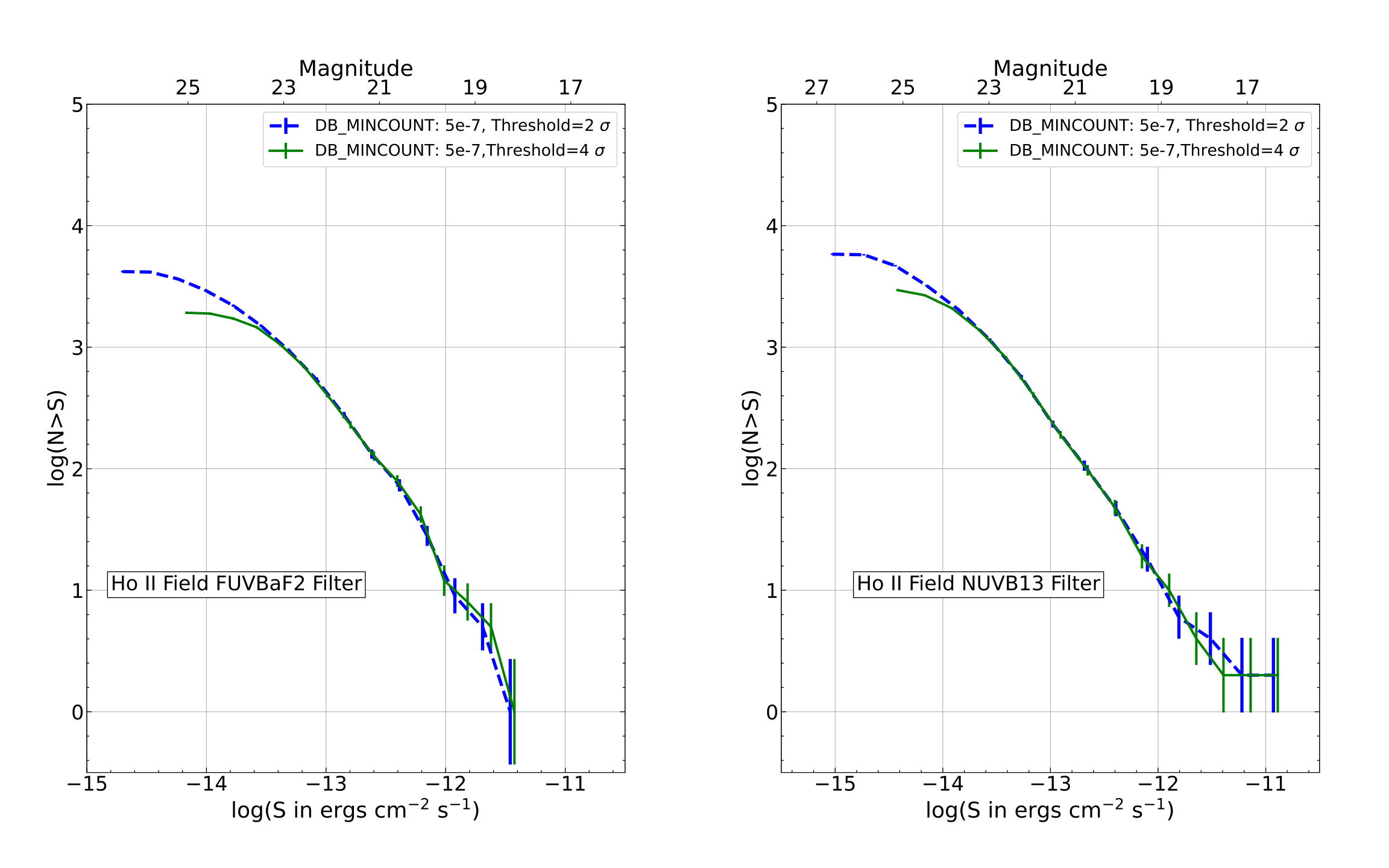}
\caption{$\log(N)$ vs. $\log(S)$ plot for sources detected in Ho~II at 4$\sigma$ and 2$\sigma$ detection thresholds. Left panel: FUV; Right panel: NUV.}
\label{fig:log_n_log_s}
\end{figure*}

The diffuse map can have a contribution from the sources that lie below the detection limit. To estimate the potential contributions from these faint sources, we reanalyzed the region by lowering the source detection threshold from 4$\sigma$ to 2$\sigma$ equivalent threshold. We observed that the contribution from faint sources to the diffuse map was approximately $5.3\%$ in the FUV and $6.2\%$ in the NUV. The $\log(N)$ vs. $\log(S)$ plot obtained at 4$\sigma$ and 2$\sigma$ thresholds is provided in Fig.~\ref{fig:log_n_log_s}, which shows the cumulative distribution of the number of sources $N$ brighter than a given flux density, $S$. The figure clearly shows the additional contribution from the undetected sources at the 4$\sigma$ threshold, in comparison with those detected at 2$\sigma$. Also, the sources detected at 4$\sigma$ and 2$\sigma$ are consistent with the bright sources, as expected.

To test the robustness of our methodology, we used simulated fields. We utilized detected source positions, magnitudes, and PSF in the field of Ho II to create the simulated sources. The sources are added using the {\it addstar} task in IRAF. For the diffuse component, we used the diffuse map generated in the field. To include the noise from UVIT, the background in each pixel is added from the random Poisson distribution with a mean value corresponding to the derived diffuse map in that pixel.  
Our analysis is able to retrieve the diffuse component within $2\%$ uncertainty, indicating that the method is robust in capturing large-scale variations (on the scale of $128\times128$ pixels or above) in the background of Ho~II.

In order to study local diffuse emission away from bright star-forming regions, we also selected 142 locations throughout the galaxy devoid of bright point sources. The UV diffuse intensities for these locations were calculated as the mean of the counts in an aperture of 5$^{\prime\prime}$ radius. The intensities range from a few hundred photon units in cavities to $\sim$10000 photons units near OB associations (where 1 photon unit equals 1 photons/cm$^{2}$/s/sr/\AA).

\subsubsection{UV Diffuse fraction}
%\label{sec:diffuse_fraction}
The ratio of the background to the total counts from Ho~II in the analysis region was calculated to obtain the diffuse fraction of Ho~II.
We obtained the diffuse fractions separately for crowded (indicated by the purple boxes in top panel of Fig.~\ref{fig:diffuse}) and non-crowded regions (white boxes). We obtain total diffuse fraction values of 70.66\% and 58.51\% in the FUV and NUV respectively (see Table~\ref{tab:Luminosity_diffuse}). Similar high values have been observed for the SMC bar \citep{pradhan2011} and Orion nebula \citep{Bohlin1982}. It is interesting to note that, the diffuse fraction obtained in the crowded regions is nearly a factor of two lower compared to the less crowded regions (Table~\ref{tab:Luminosity_diffuse}). This trend is similar to what is observed in the Magellanic Clouds \citep{pradhan2010far, pradhan2011}. One of the possible reasons could be the escaping UV photons from distant OB associations and young star clusters being scattered by dust in regions with fewer stars \citep{Cole1999}. The obtained diffuse maps for the FUV and NUV are shown in the bottom panels of Fig.~\ref{fig:diffuse}, respectively. It is evident that the diffuse emission extends beyond the bright massive star-forming regions and follows the HI profile, which is one striking similarity in the diffuse UV distribution between Ho~II and the Milky Way \citep{henry2014mystery}. Also, the diffuse intensities derived here, close to the young massive star clusters and away from them, have values comparable to what is observed in the Galactic plane and in the Galactic poles, respectively.

\begin{table}
\centering
\caption{Total and diffuse FUV and NUV luminosities and corresponding diffuse fractions for Ho~II.}
\begin{tabular}{ccllc}
\hline
        Band & Region & Total UV & Diffuse UV & Diffuse \\
        & & Luminosity & Luminosity & Fraction \\
         & & ($\times 10^{41}$ ergs/s) &($\times 10^{41}$ ergs/s)& (\%)\\
        \hline
         FUV& All&{ 8.954$\pm$0.006} &{6.327$\pm$0.005 }& {70.66}\\
         & Crowded & { 6.680$\pm$ 0.005}&{4.08$\pm$0.004} &{ 61.05} \\
         & Non- & {2.274$\pm$0.003} &{ 2.249$\pm$0.003} &{ 98.90}\\
         &crowded & & & \\
         \hline
        NUV& All&{6.165$\pm$0.003} & { 3.607 $\pm$ 0.002}&{58.51}\\
         & Crowded &{ 4.586 $\pm$0.003 }&{2.289 $\pm$ 0.002}& { 49.92}\\
         & Non- &{1.579$\pm$0.002 }&{1.318$\pm$0.001} &{83.46}\\
         &crowded & & & \\
          \hline
    \end{tabular} 
    \label{tab:Luminosity_diffuse}      
\end{table}

\subsection{IR intensities and HI column densities in regions with diffuse UV emission}

We utilized IRAF tools, specifically the {\it imalign} and {\it psfmatch} tasks, to align and convolve the IR images to a common PSF. We then extracted the IR intensities in circular apertures of radius $5''$ at the 142 selected diffuse UV locations. Among these, only 33 locations had non-zero intensities in all of the IR wavebands which we considered for the correlation (4.5, 5.8, 24, 70, 100, and 160 $\mu m$). 
Similarly, for the UV--HI correlation, we extracted the HI column densities for all 142 locations. 50 of these locations had HI column density $>$ $1 \times 10^{21}$ cm$^{2}$. The list of selected 33 locations for correlation study along with diffuse FUV and NUV intensities is given in Table~\ref{tab:fuv_intensity}. 
Interestingly, we noted the absence of 8 $\micron$ emission in all of our considered locations, consistent with the suggested scarcity of PAH in this particular galaxy \citep{Li2020}. Calculated IR intensities for these 33 locations, along with their associated errors, are given in Table~\ref{tab:IR_intensity_30}. The combined table with FUV, NUV, IR intensities, and neutral hydrogen column densities N(HI) for 33 selected locations with non-zero IR intensities, and the table with IR intensities for all 142 locations are provided online (see Sec.~\ref{sec:data}).

\subsection{Sources of Diffuse UV emission}

In the following sections, we investigate contributing factors to the diffuse UV emission in Ho~II, with a specific focus on the dust scattering contribution and the different dust populations present via their thermal emission. The different dust populations are explored via UV--IR correlations and the dust scattering contribution is estimated via radiative transfer modelling. In addition, we discuss possible contributions from other processes such as H$_{2}$ fluorescent emission (via FUV--N(HI) correlation) and two-photon continuum emission.

\subsubsection{UV--IR and UV--N({\rm HI}) Correlations}

Correlation study is an important statistical tool that signifies the relation between various quantities. Dust grains absorb stellar radiation at short wavelengths, such as UV and optical, get heated up, and subsequently emit radiation at longer wavelengths (IR and submm). The Near-IR (NIR) and Mid-IR (MIR) emission is attributed to small/very small grains at high temperatures or to PAHs. On the other hand, Far-IR (FIR) emission is attributed to colder dust grains. Therefore, due to the complimentary nature of dust scattering and thermal emission, a correlation study of UV and IR intensities can help in ascertaining the abundance and distribution of the dust populations at different temperatures. We have calculated the correlation between the UV (at mean wavelengths 1541~\AA\, for FUV and 2447~\AA \, for NUV) and the IR intensities for six IR wavelengths. Since we have assumed the 3.6~$\mu$m emission is stellar emission during the stellar contamination correction (see Sec.~\ref{sec:ir_data}), we have not considered the 3.6~$\mu$m band for our correlation study. 
To correlate with the 4.5~$\mu$m and 5.8~$\mu$m bands, the UVIT image was convolved to a common resolution (FWHM) of 2.5$''$, and for 24~$\micron$ it was convolved to 6$''$. UV intensities from these convolved images were used to calculate the correlation. For longer wavelengths (beyond 24~$\mu$m) the UVIT images were not convolved, since the IR data had a poor resolution otherwise resulting in nearby UV sources contaminating the aperture fluxes.
The 24~$\mu$m emission is attributed to the warm dust emission by the Very Small Grains (VSG), associated with locations close to hot and young UV emitting stars, such as HII regions \citep{wu2005pah}. The 70~$\mu$m emission which shows a tight linear correlation with 24~$\mu$m emission \citep{zhu2008correlations} is also considered as a warm dust tracer in galaxies \citep{walter2007dust}. In our study, we have considered 70~$\mu$m emission as tracer of warm dust emission, since it remains unaffected by the presence of point sources \citep{walter2007dust}. 

\begin{figure*}
\centering
    \includegraphics[scale=0.38]{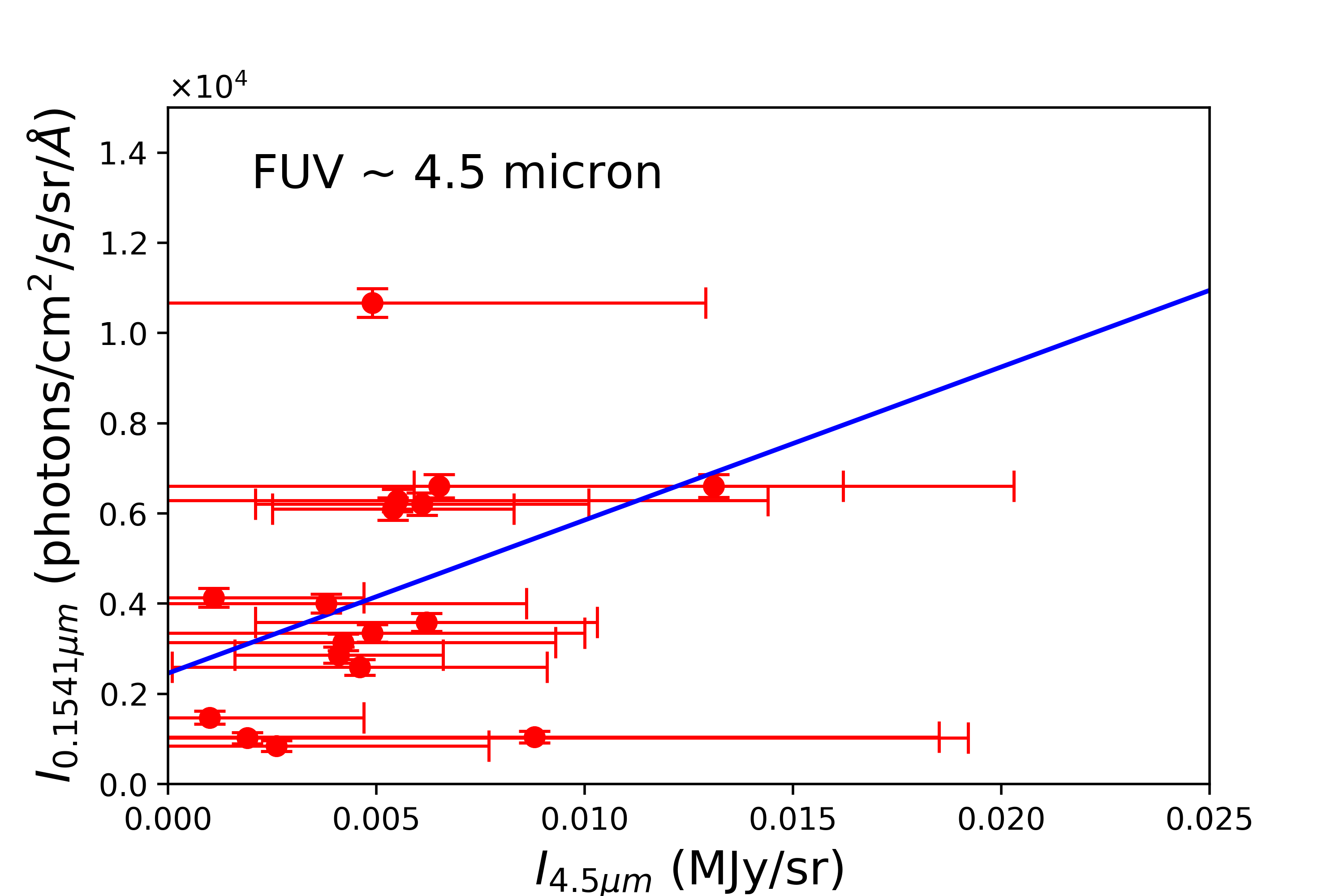}
    \includegraphics[scale=0.38]{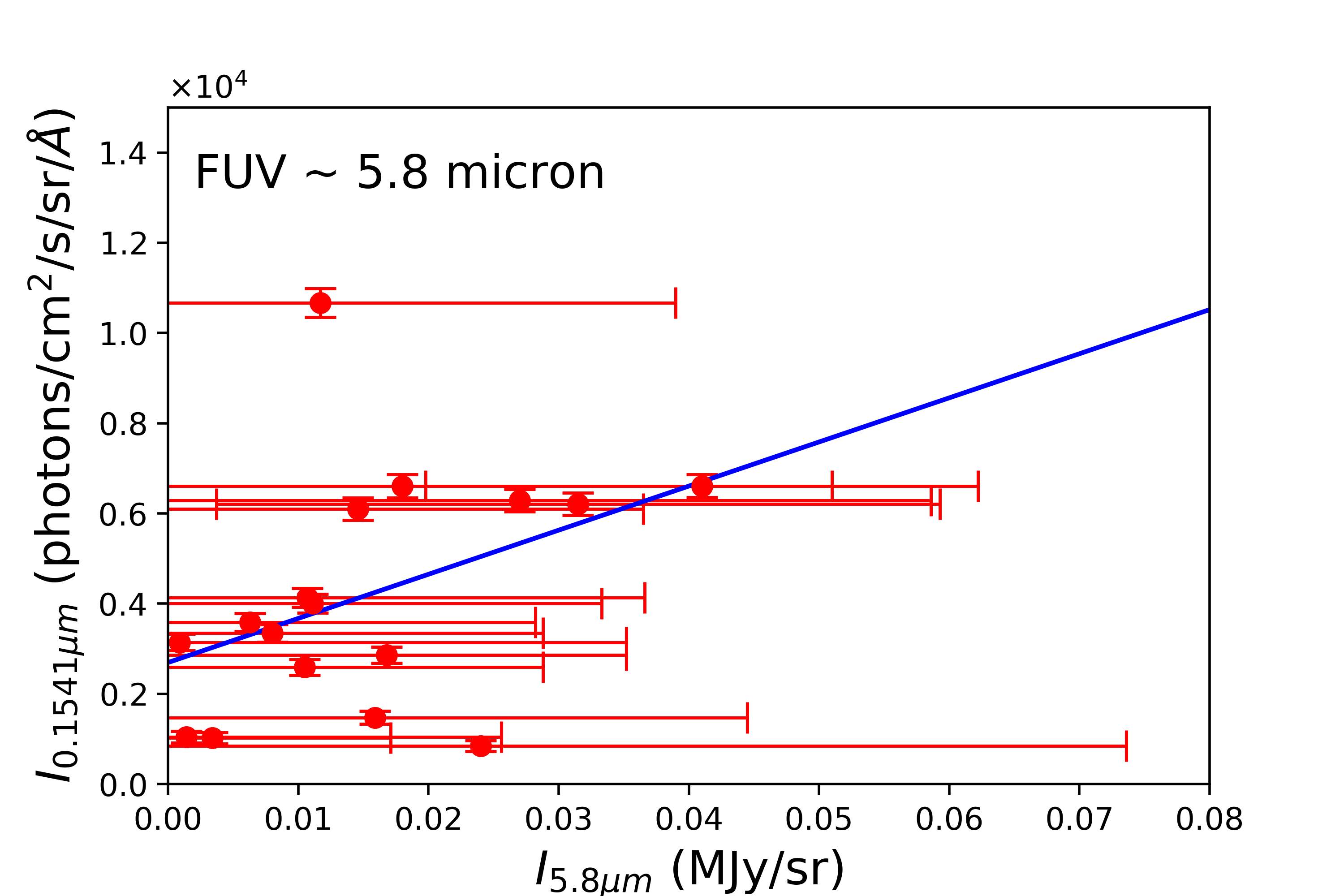}
    \includegraphics[scale=0.38]{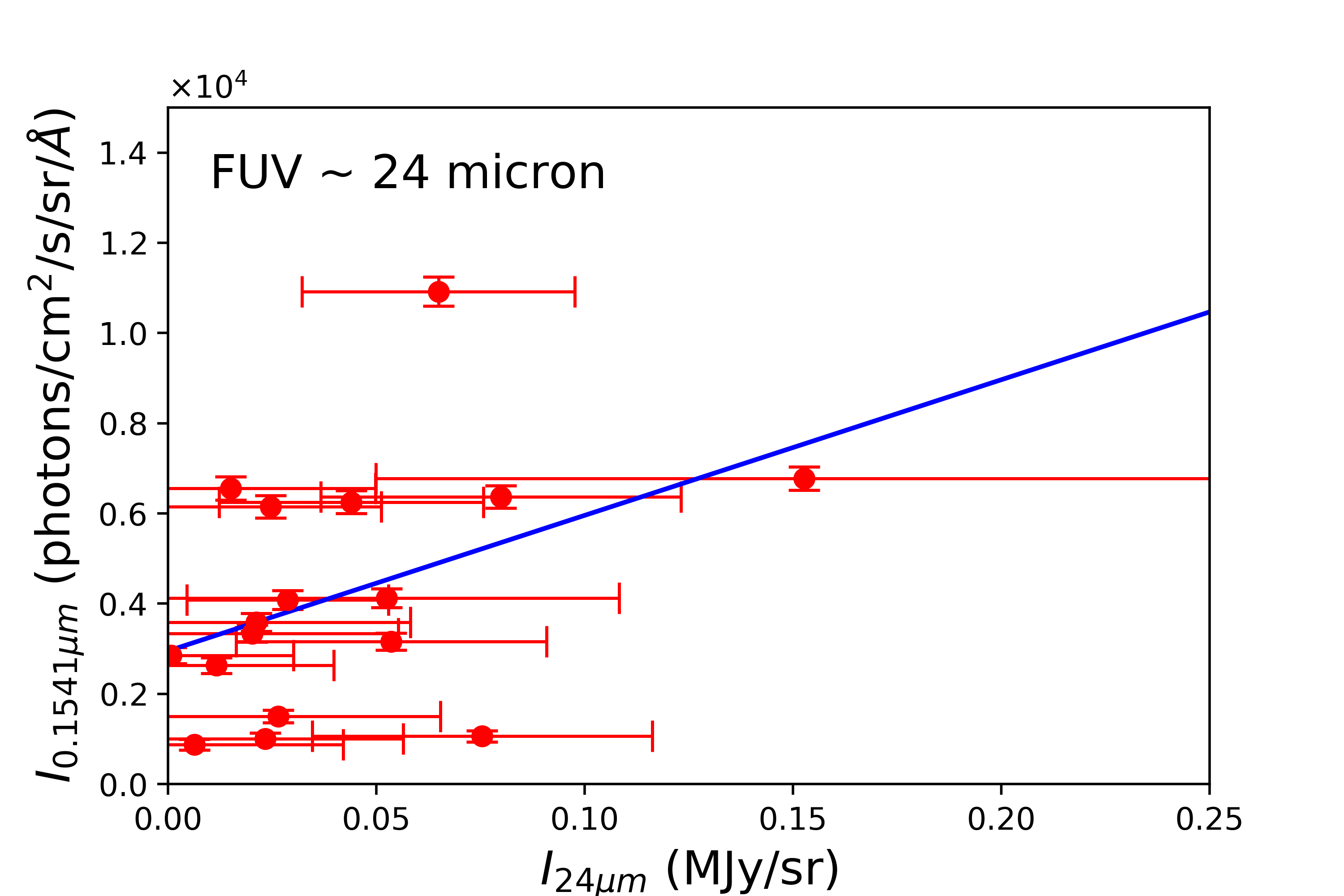}
    \includegraphics[scale=0.38]{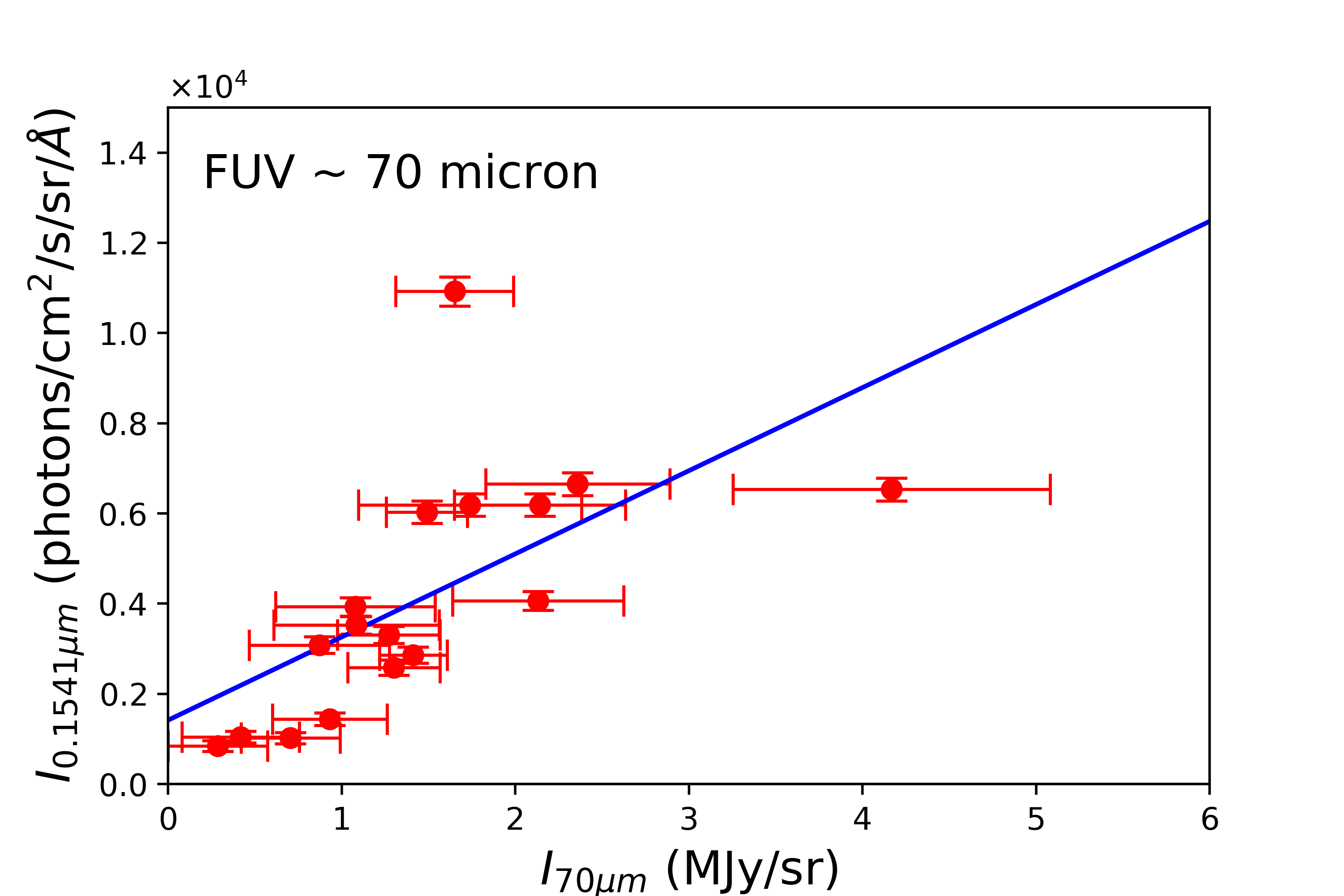}
    \includegraphics[scale=0.38]{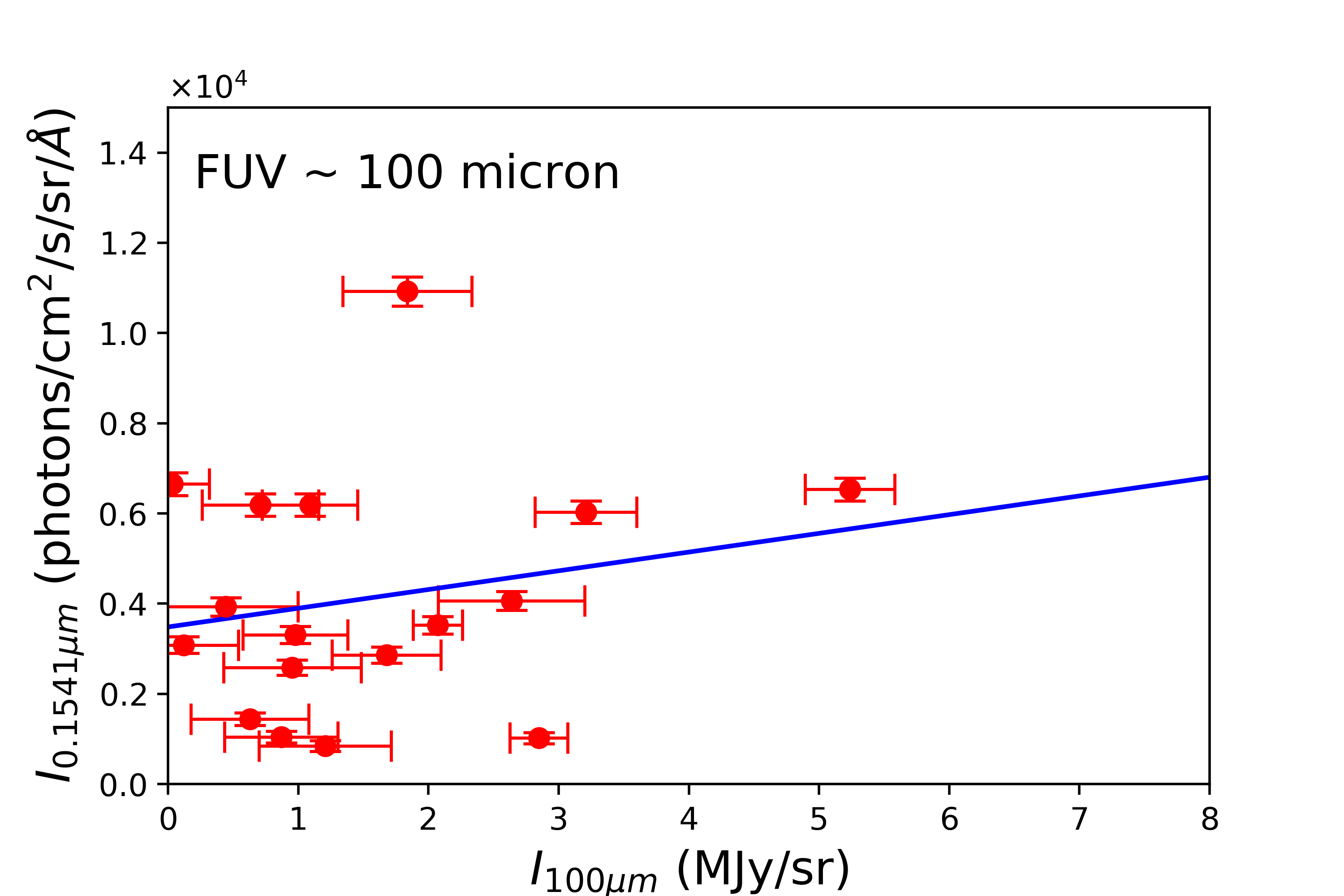}
    \includegraphics[scale=0.38]{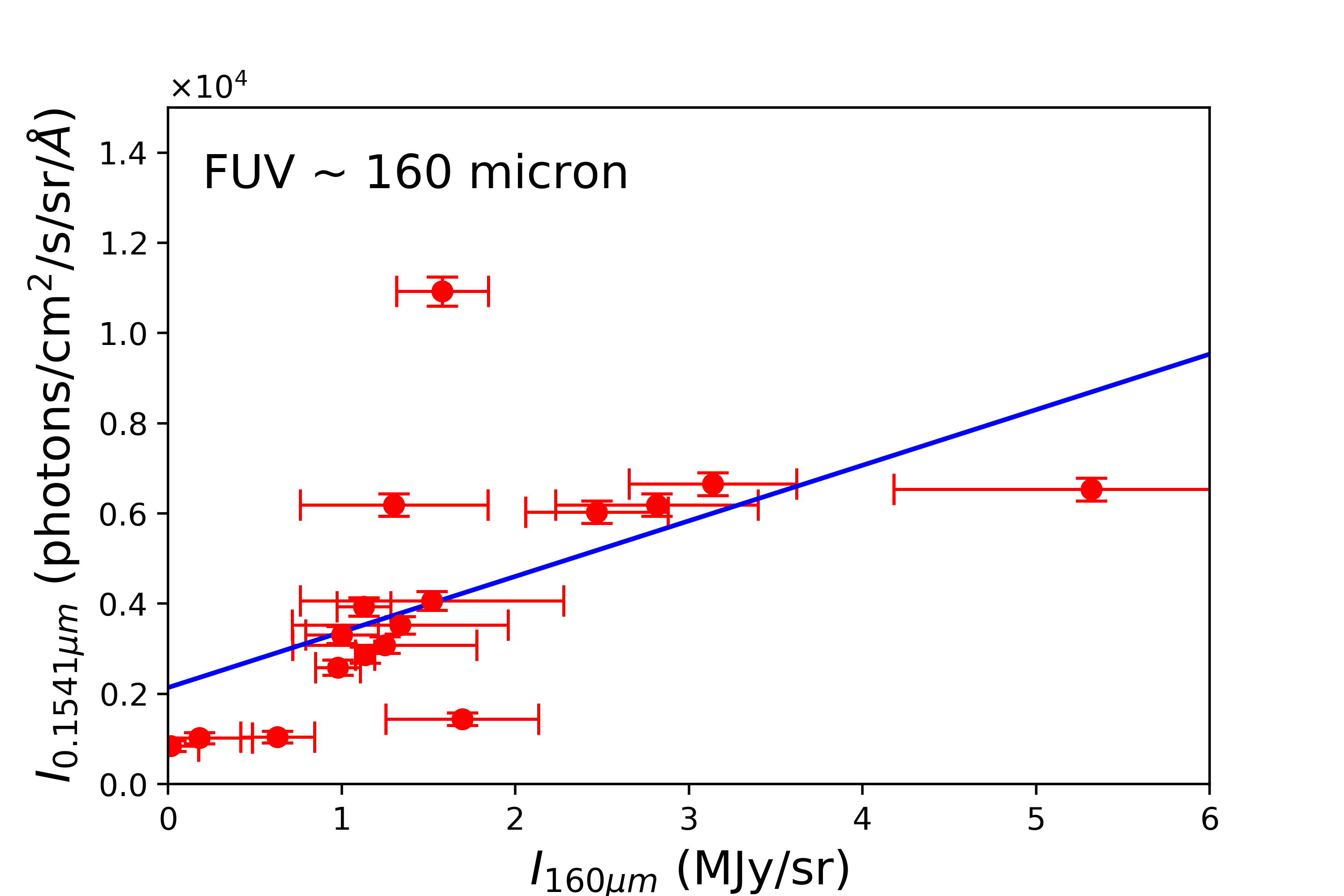}
    \includegraphics[scale=0.38]{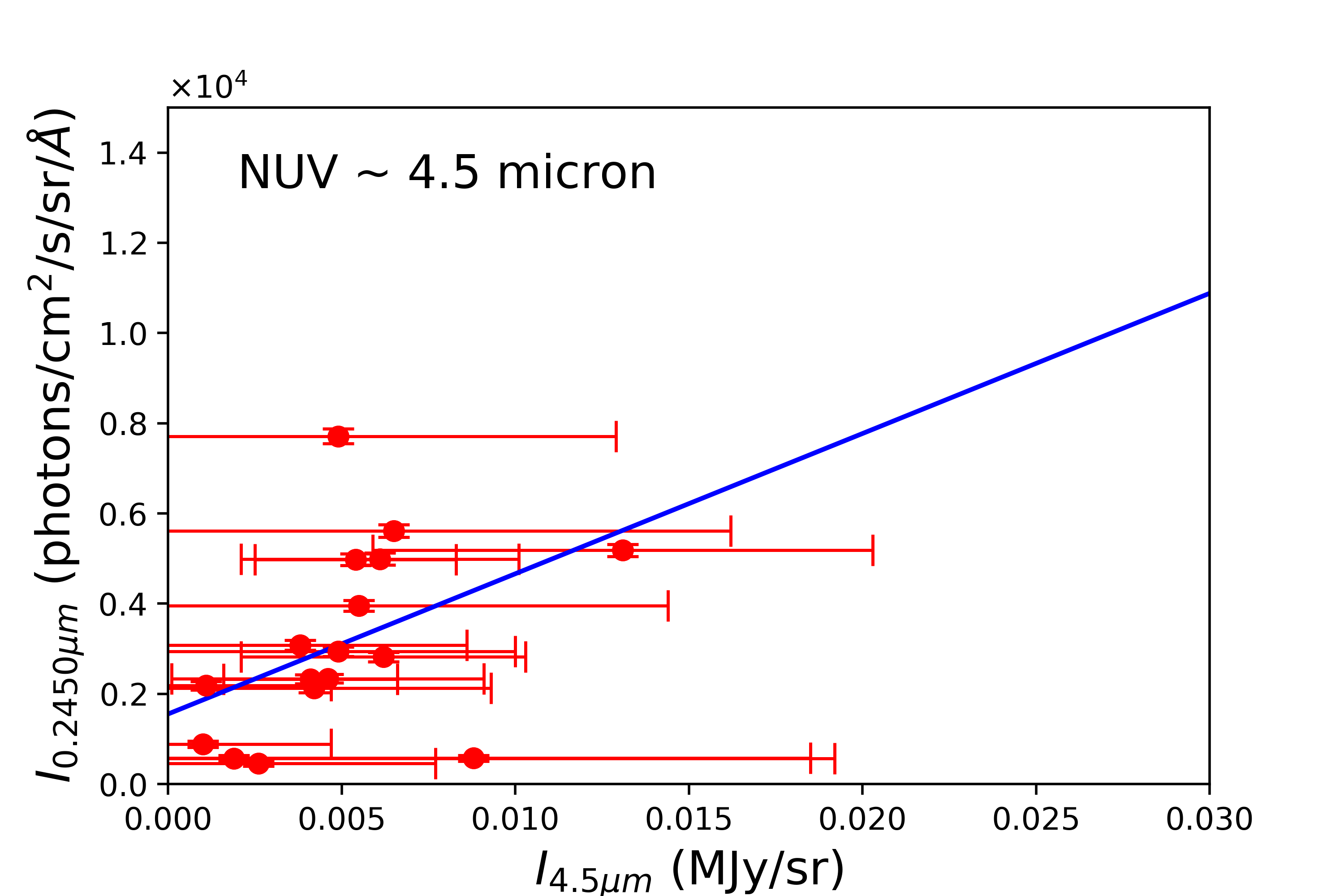}
    \includegraphics[scale=0.38]{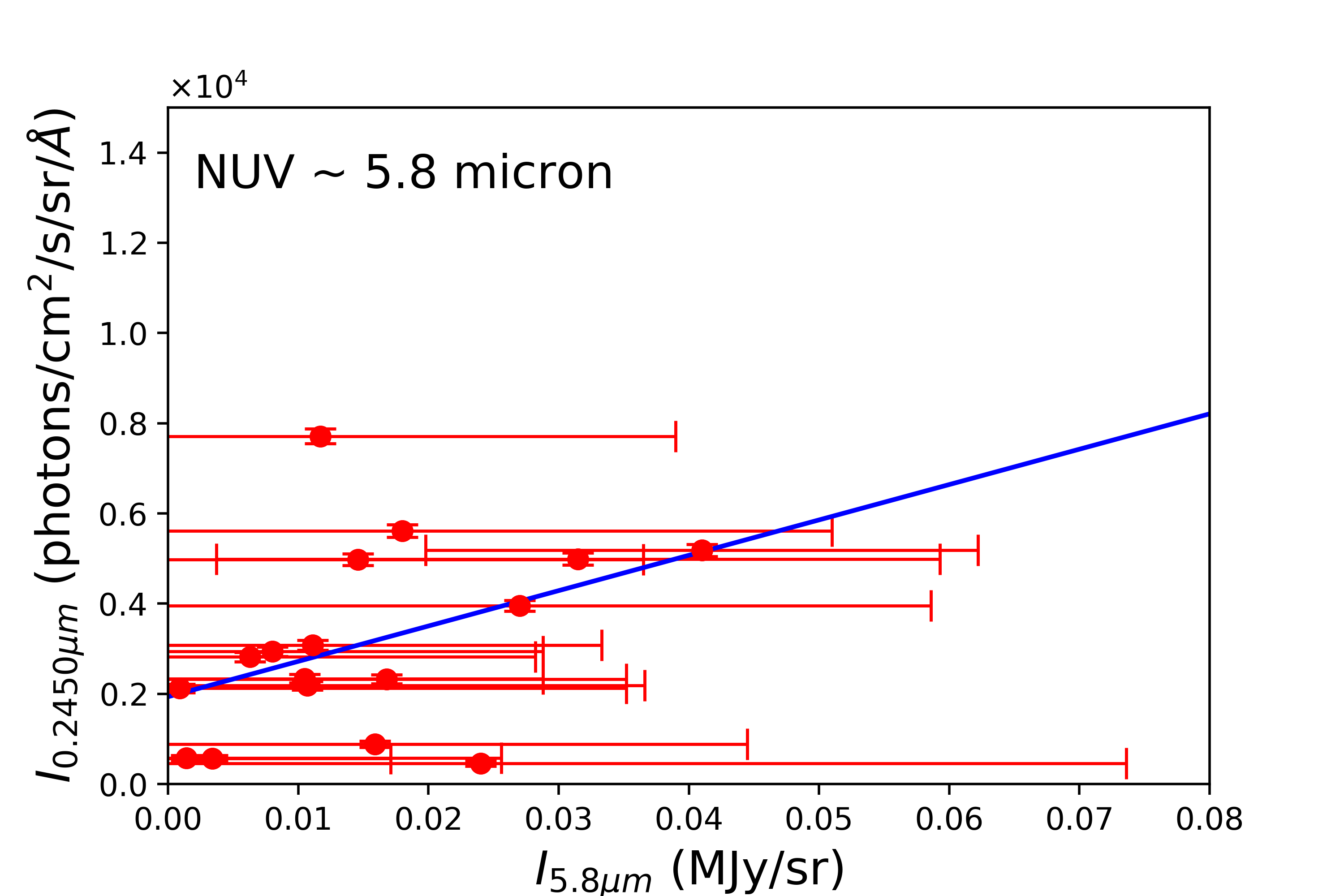}
    \includegraphics[scale=0.38]{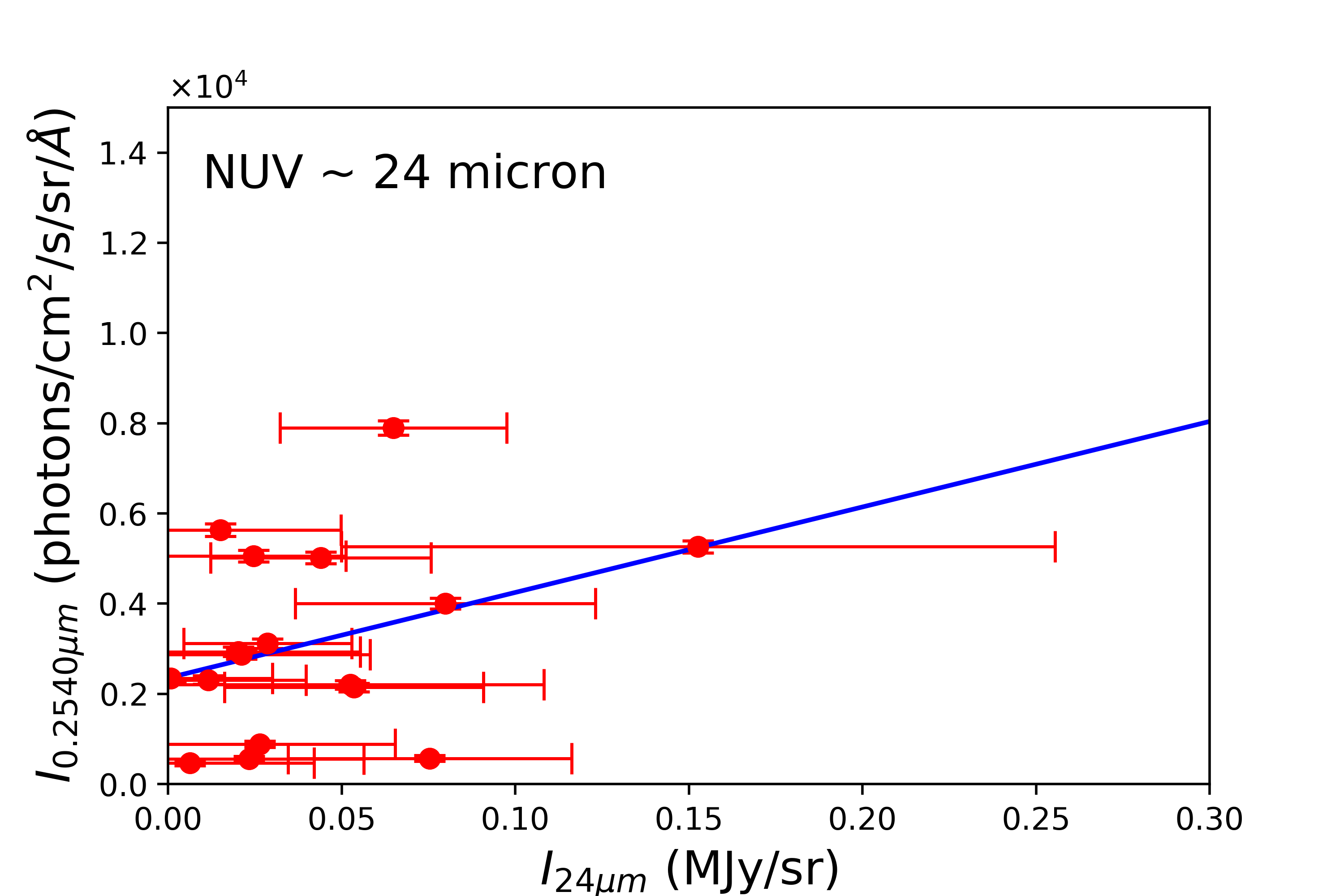}
    \includegraphics[scale=0.38]{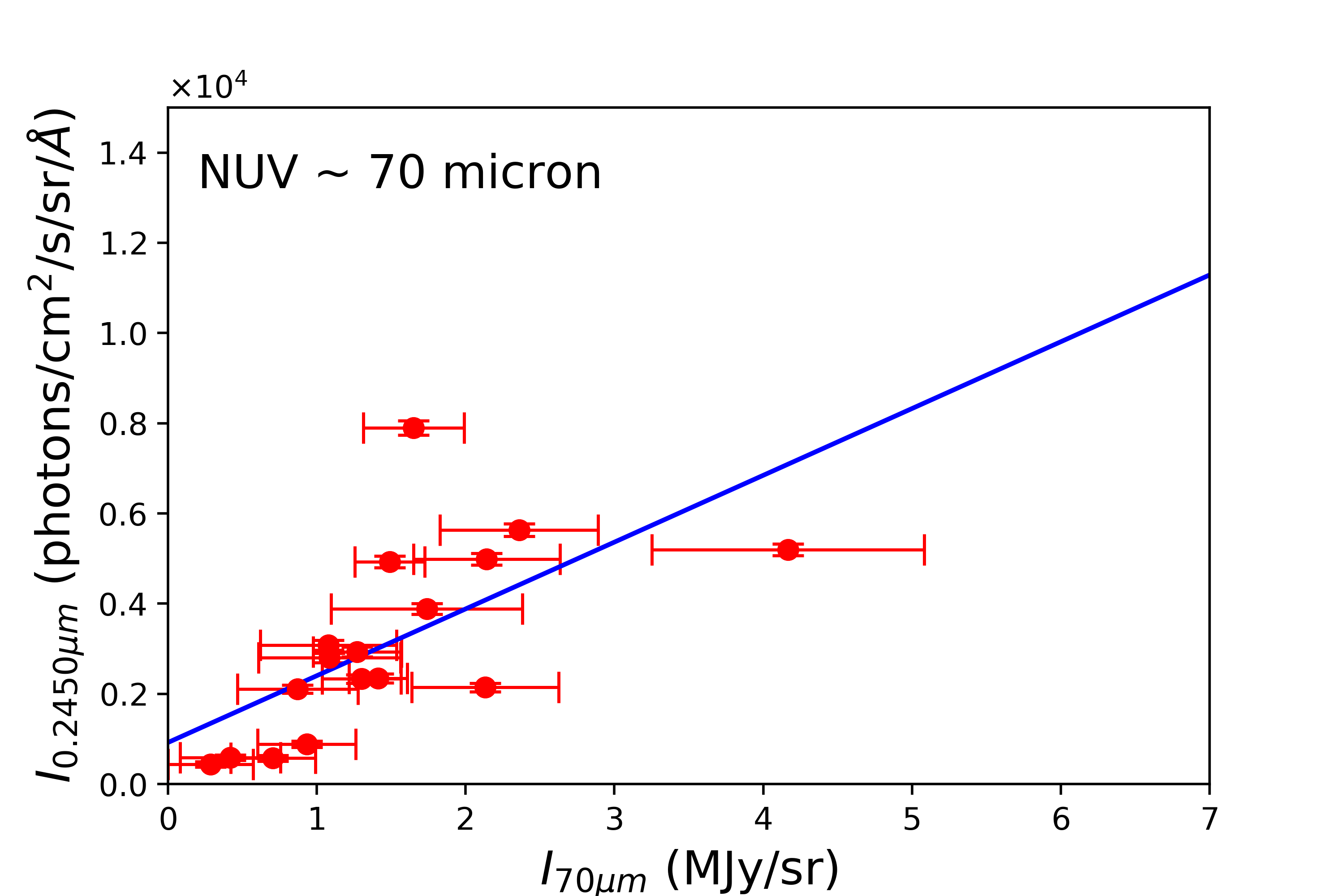}
    \includegraphics[scale=0.38]{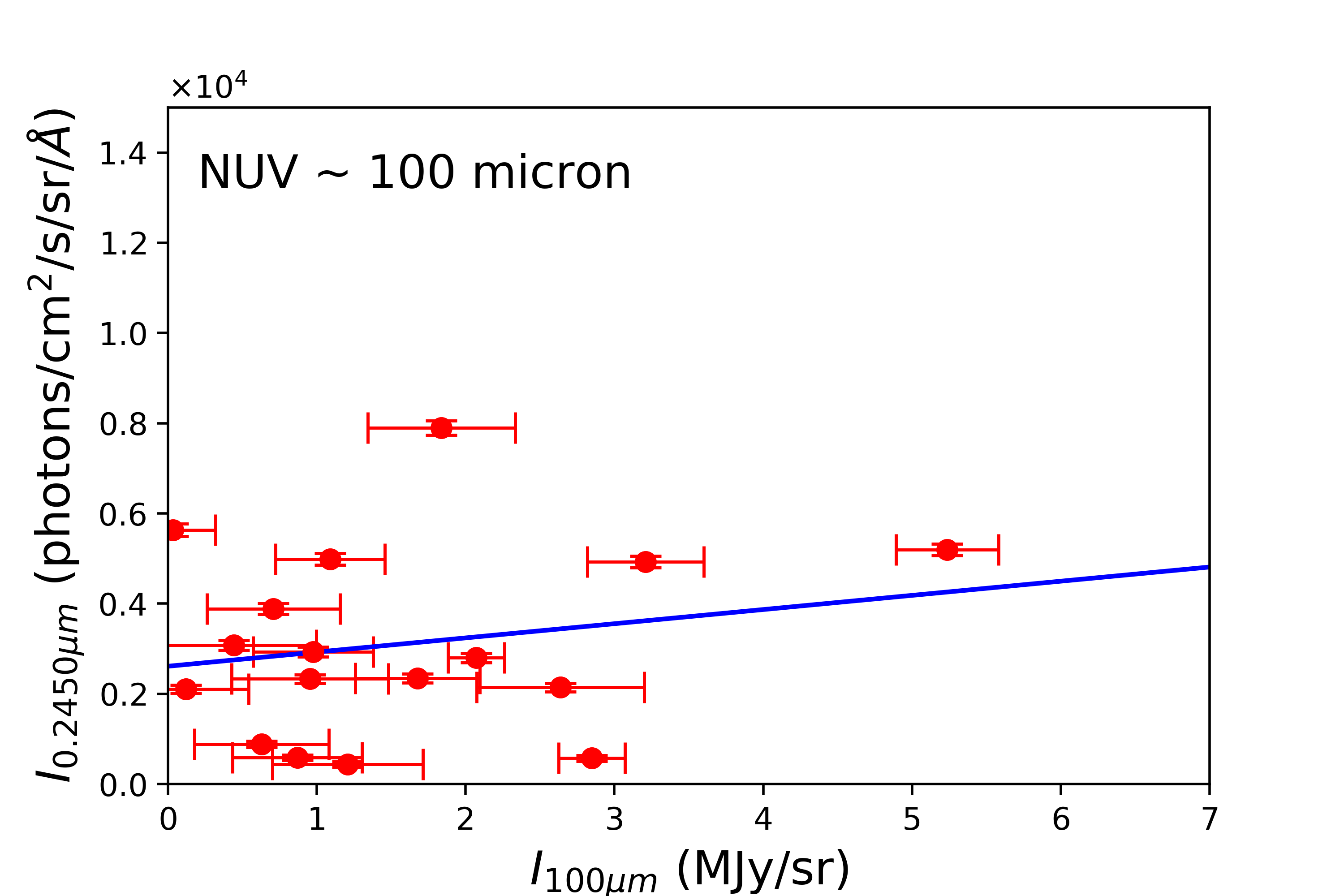}
    \includegraphics[scale=0.38]{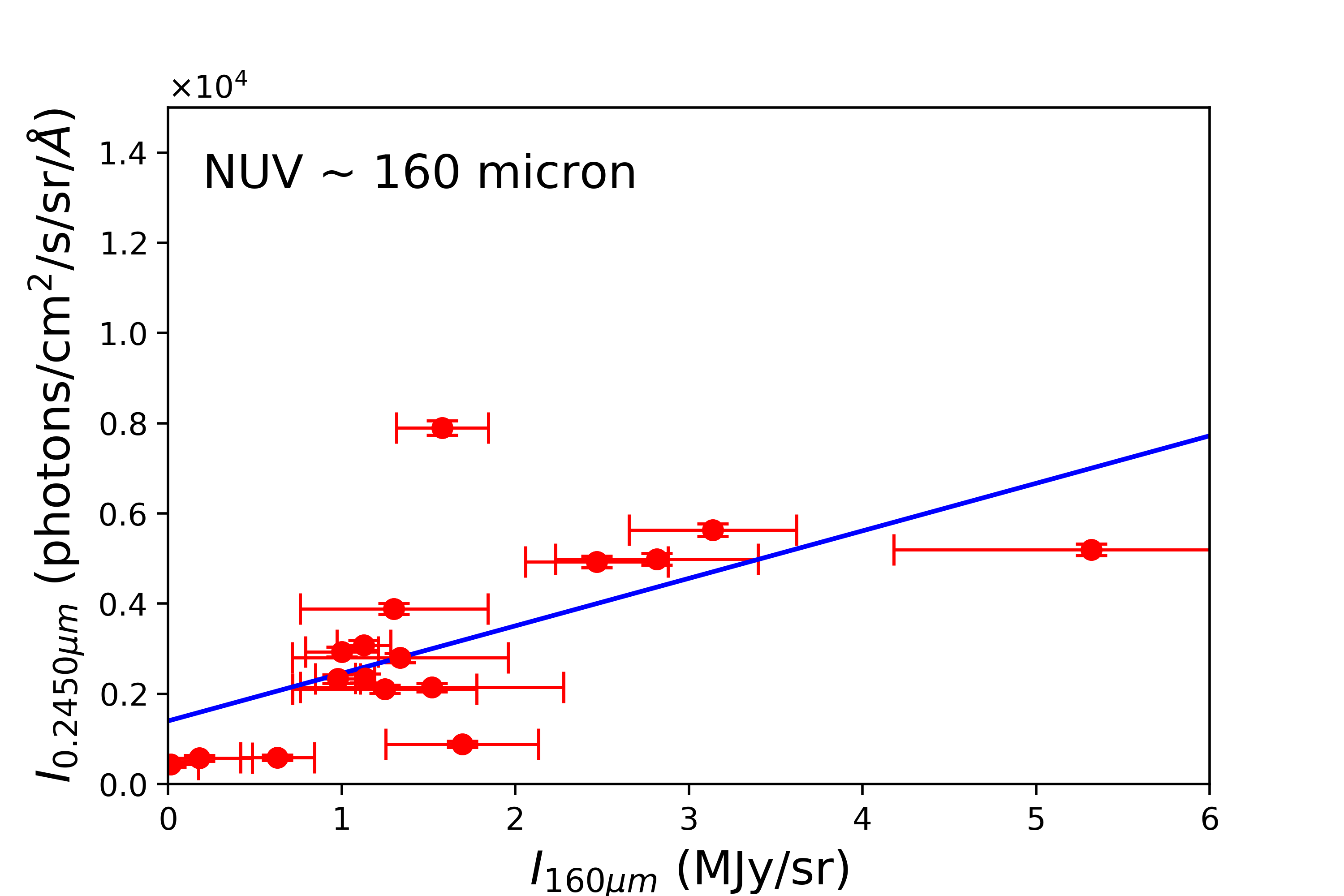}
    \caption{FUV--IR and NUV--IR correlation plots for 17 selected locations with N(HI) $> 1 \times 10^{21}$ cm $^{-2}$. 
    }
\label{fig:correlation_plots}
\end{figure*}

For the correlation study, we have considered the 33 locations for which non-zero IR intensities are observed at all considered wavelengths. Since most of the dust emission in Ho~II is associated with the regions having HI column density greater than $1 \times 10^{21}$ cm$^{-2}$ \citep{walter2007dust}, we have derived the HI column densities from THINGS integrated HI map \citep{walter2008things} to divide our observed locations into two groups: {\it a}) locations with HI column density greater than $1 \times 10^{21}$ cm$^{-2}$, comprising 17 locations; and {\it b}) locations with HI column density less than $1 \times 10^{21}$ cm$^{-2}$, comprising 16 locations. The HI column densities of the locations are listed in Table \ref{tab:fuv_intensity}. 
Both FUV and NUV vs IR intensities for the 17 locations with N(HI) $>$ $1 \times 10^{21}$ cm$^{-2}$ are plotted in Fig.~\ref{fig:correlation_plots}. From the plots, an overall trend of increasing IR intensities with UV can be recognized. 

Such a correlated increase of IR and UV along with its inherent spread may indicate spatial variations of the dust mass in illuminated clouds and the illuminating UV intensity, as well. The former demonstrates proportionality to the dust mass $I_{ir}\propto M_d$, while the latter manifests via the relation $I_{ir}\propto \exp(-h\nu/kT_d)$ with dust temperature $T_d\propto I_{ uv}^{1/4+\beta}$, $\beta$ being the dust spectral index (see also below). 
The spread of IR intensities might be also attributed to random variations of the mass-to-size ratio of IR emitting clouds, because clouds with smaller radii and cross-sections absorb a smaller amount of the heating UV photons. Local variations in the dust-size distribution function cannot be excluded either, though they don't  seem likely given that the areas under consideration are located far from high-density star formation regions with frequent SNe shocks harmful for dust. However, as we will see below in Table~\ref{tab:derivative_high_hI}, the overall trend of the derivative $dI_{ir}/dI_{uv}$ versus $\lambda_{ir}$ indicates that UV heating is the primary cause that determines dust emission intensity along with its variations. The FUV and NUV are thought to trace regions of recent massive star formation within a wide time scale: from 10 to 100 Myr for the first, and 10 to 200 Myr for the latter \citep[][]{Kennicutt2012}. In these conditions, one might attribute the observed large spread of IR intensities to possible spatial variations in the FUV vs. NUV interrelations and their contribution to dust heating. However, as we will see below (Sec. \ref{odfuv}), the FUV/NUV ratio is invariant over the whole set of locations under study. This circumstance may indicate that a large spread of IR intensities reflects a considerable spread in properties of dust clouds.   

\begin{figure*}
\centering
\includegraphics[scale=0.38]{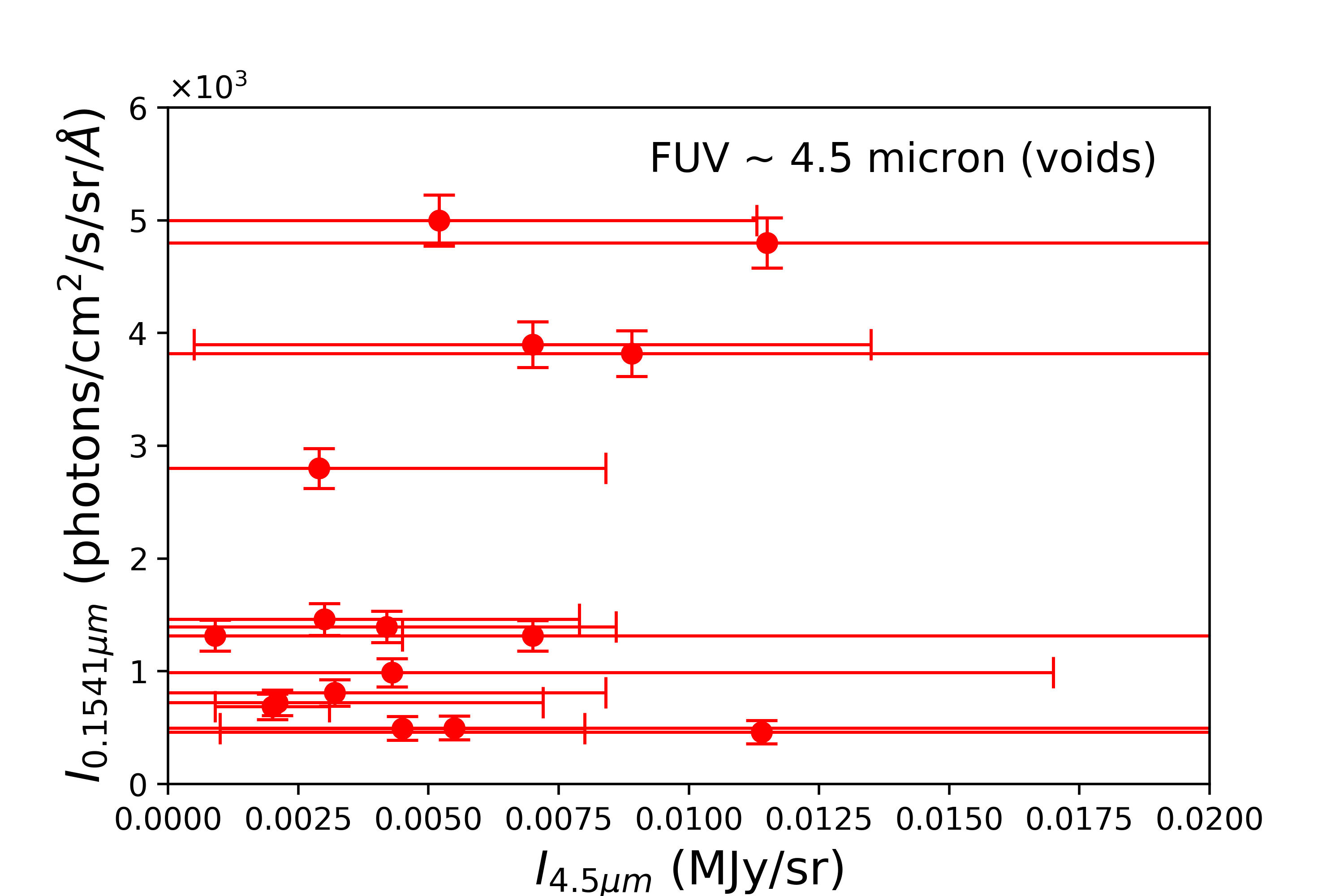}
\includegraphics[scale=0.38]{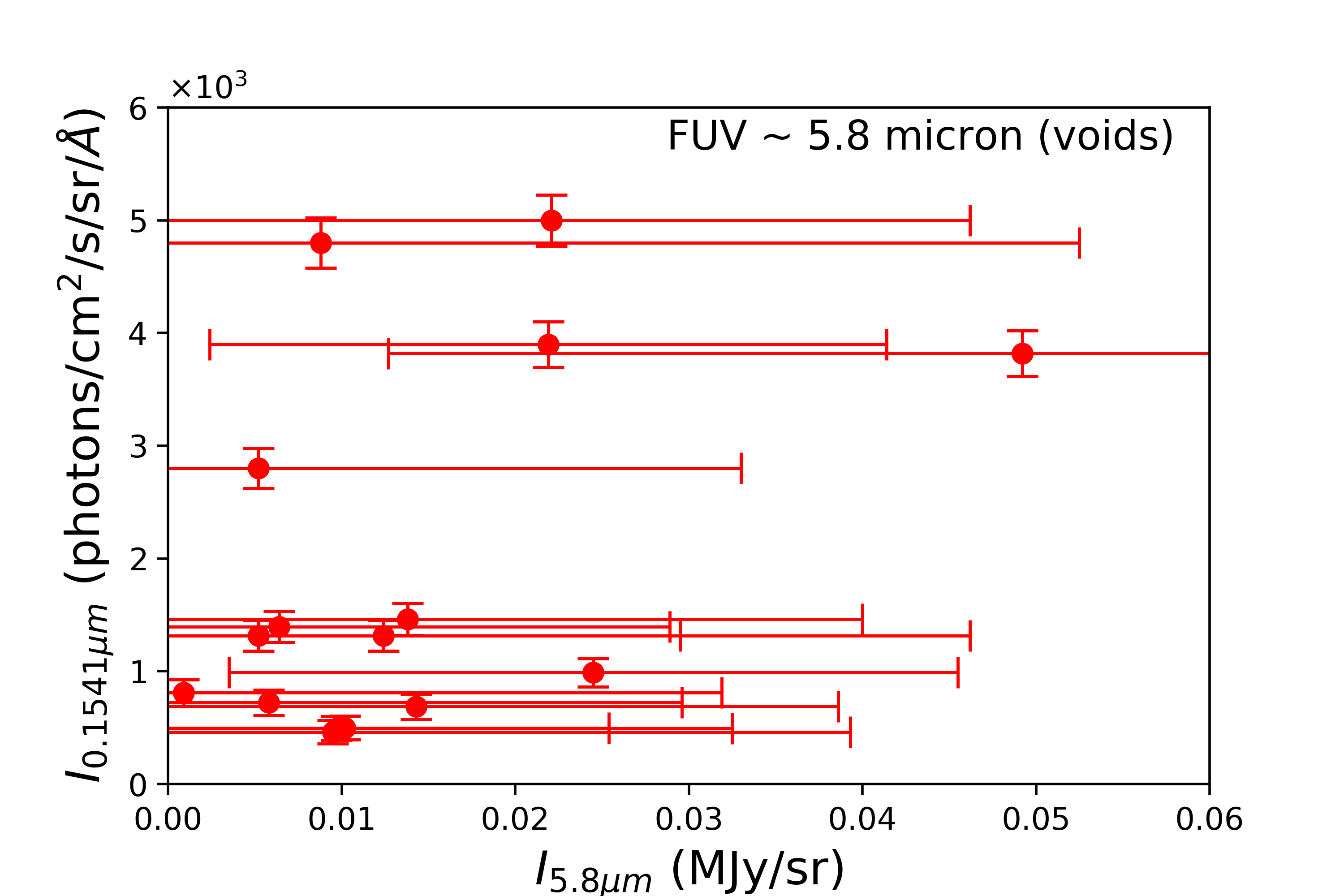}
\includegraphics[scale=0.38]{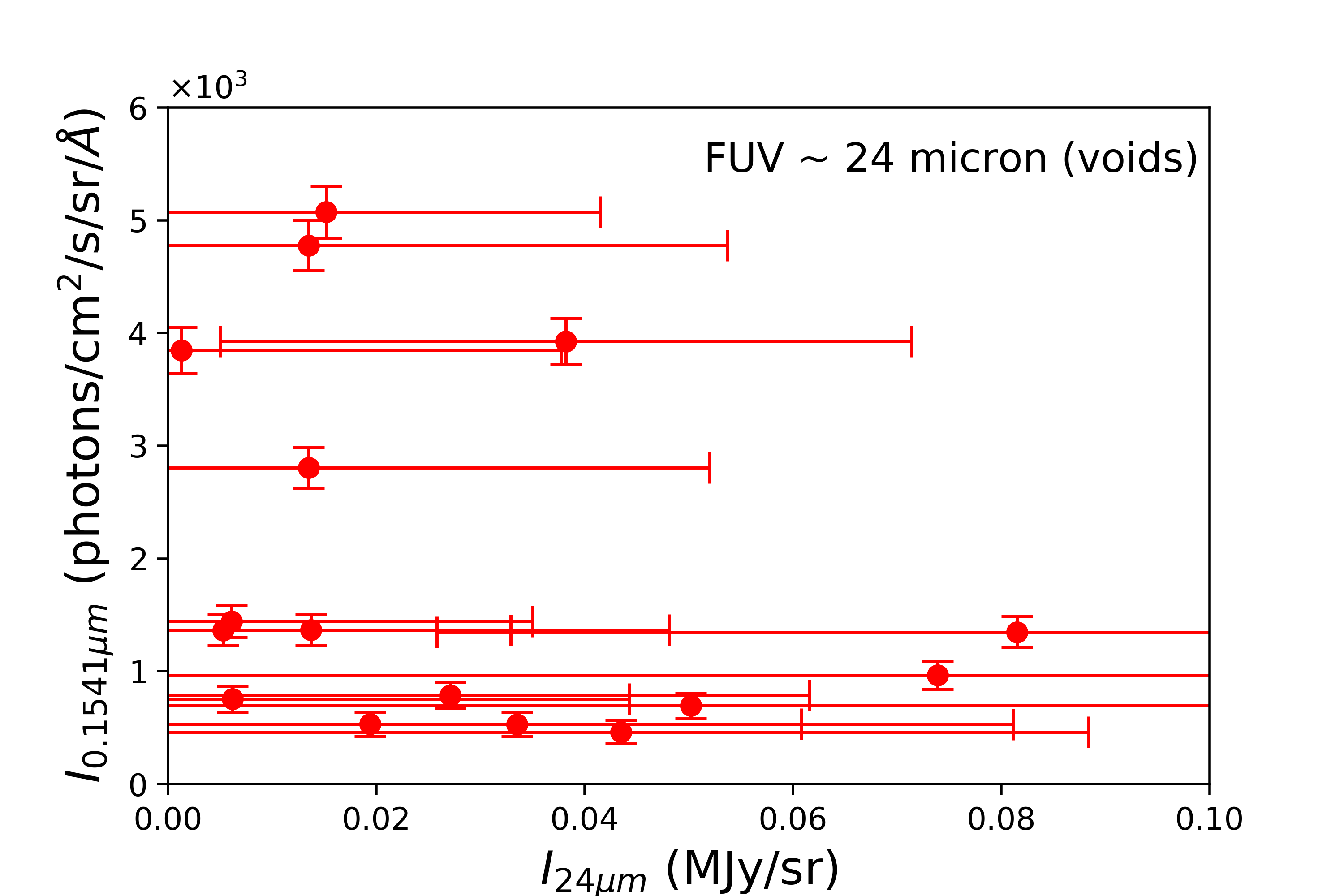}
\includegraphics[scale=0.38]{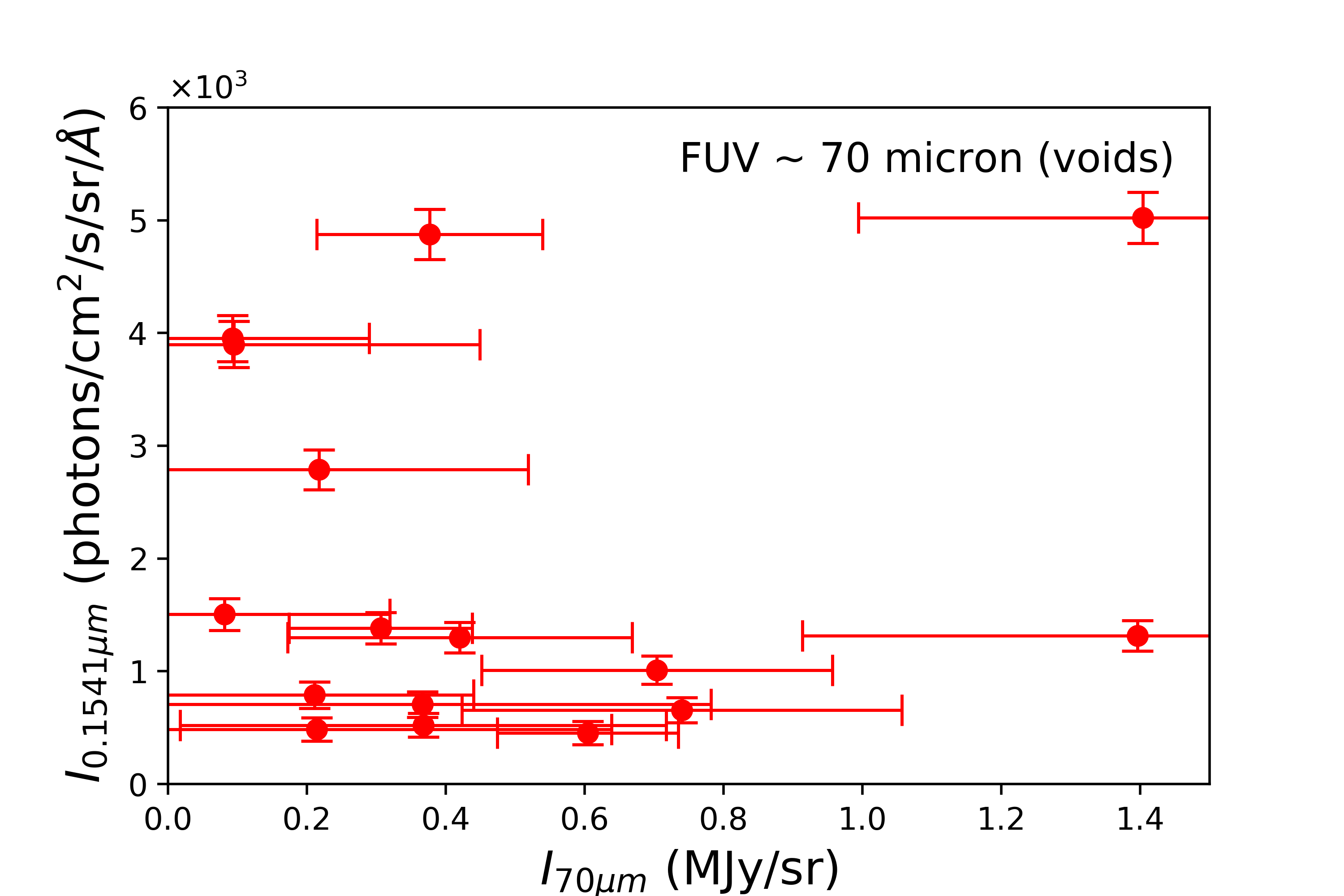}
\includegraphics[scale=0.38]{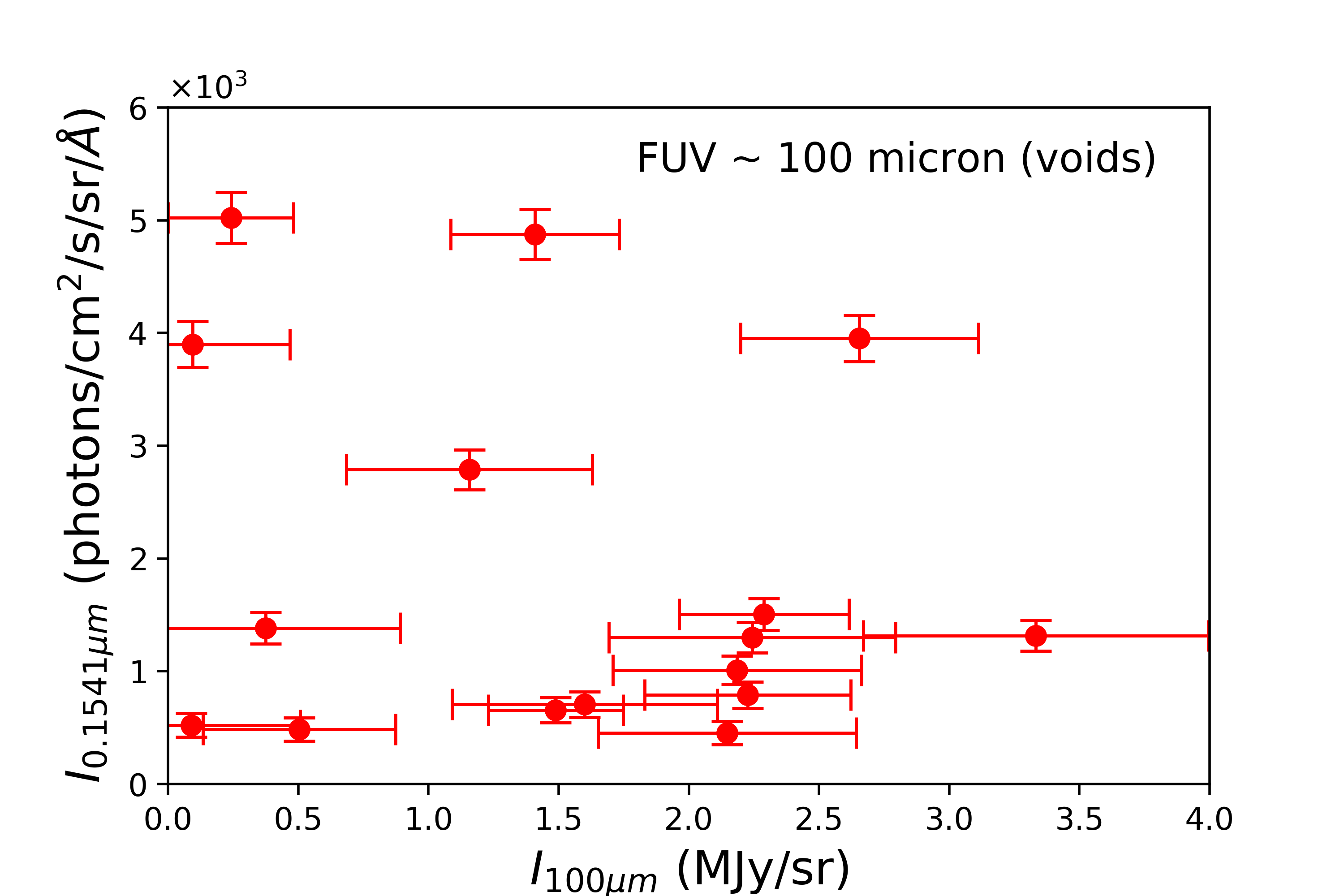}
\includegraphics[scale=0.38]{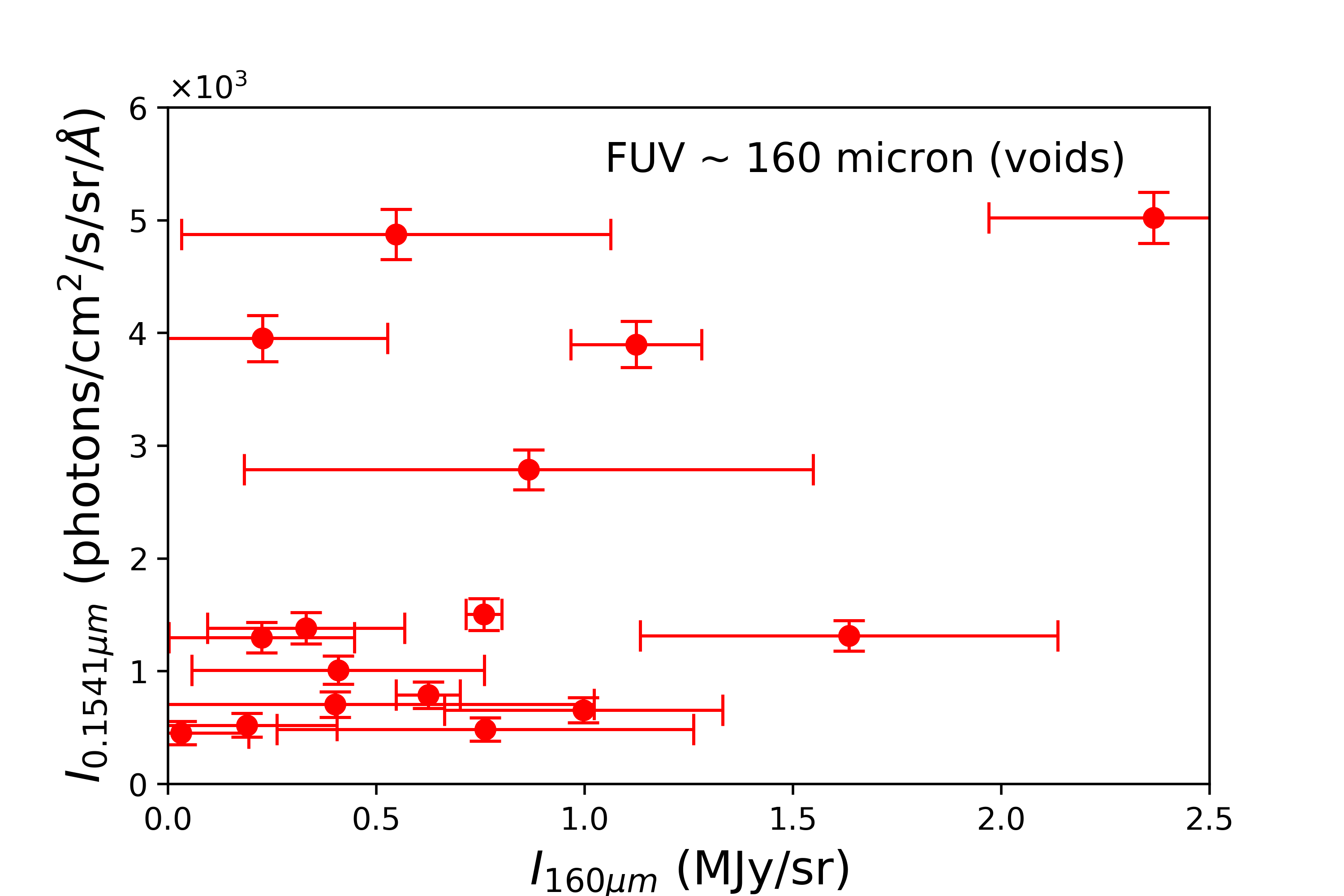}
\includegraphics[scale=0.38]{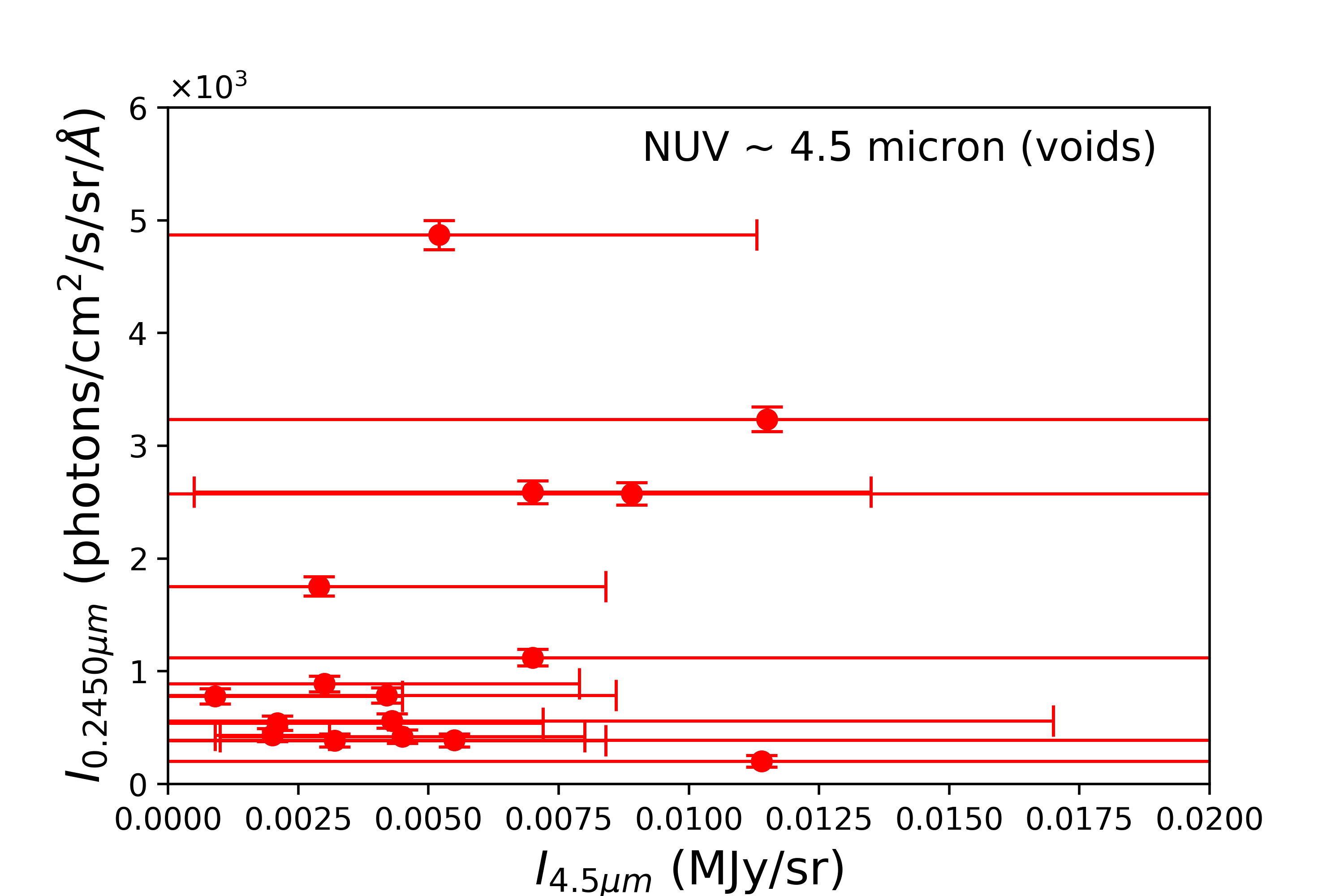}
\includegraphics[scale=0.38]{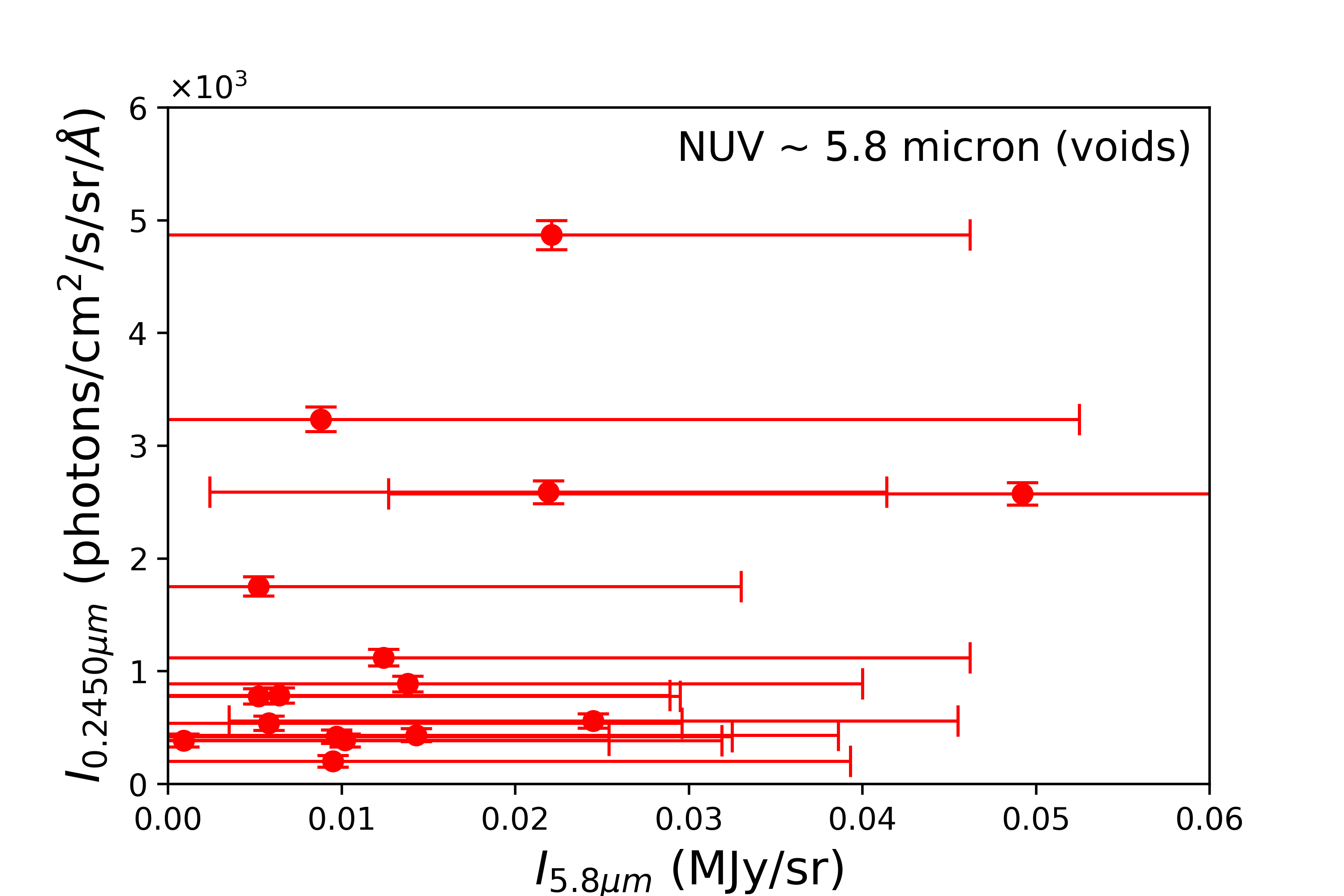}
\includegraphics[scale=0.38]{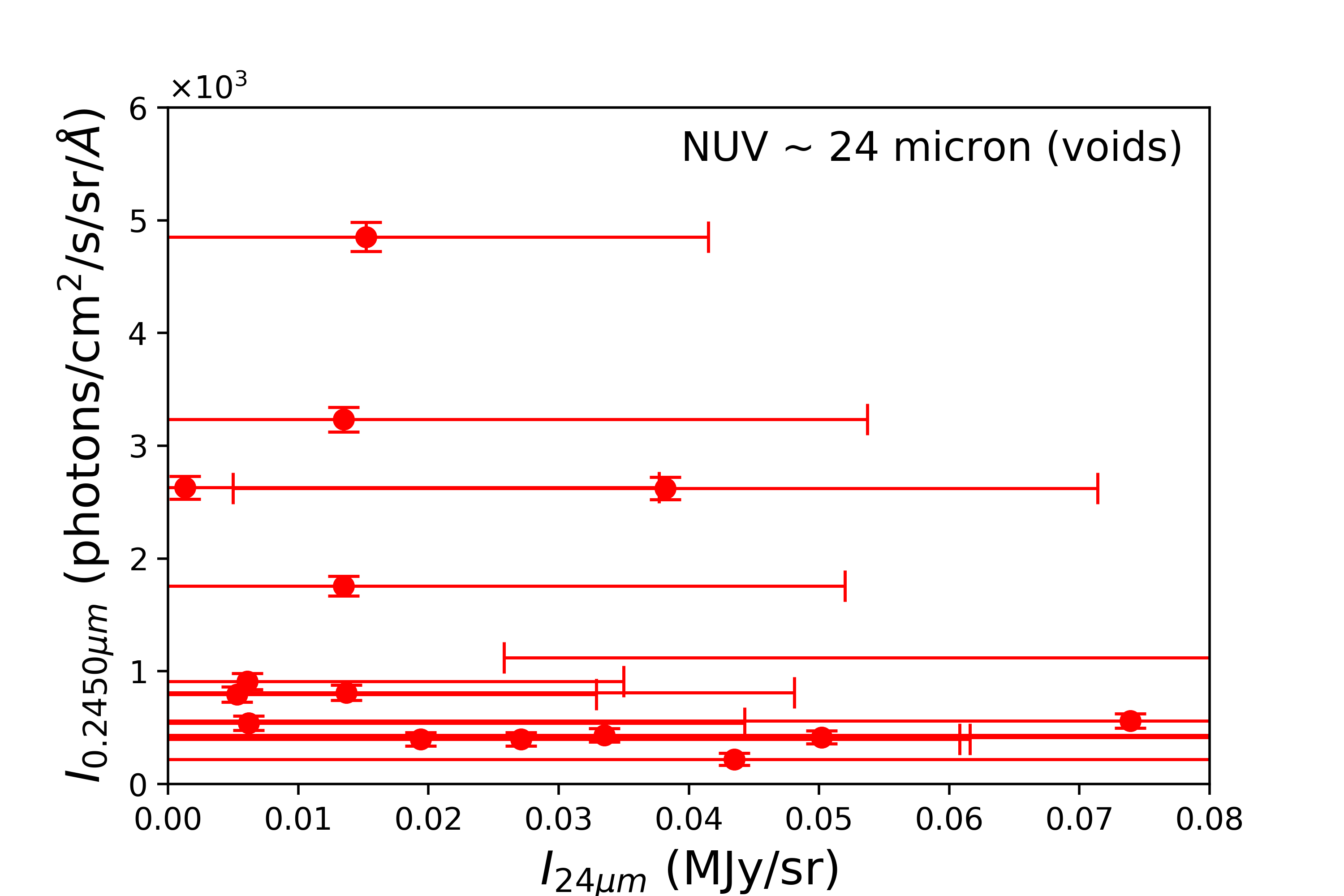}
\includegraphics[scale=0.38]{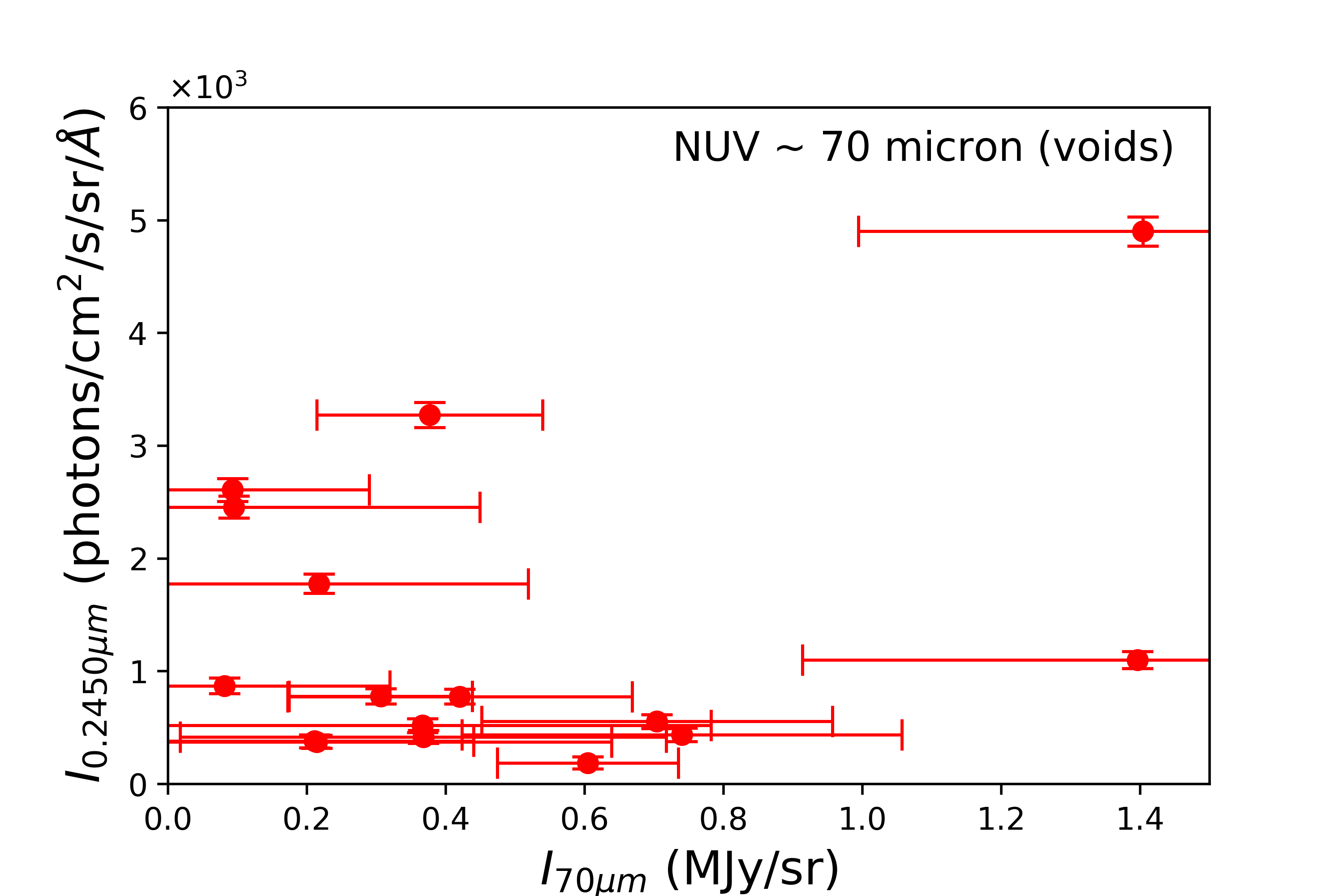}
\includegraphics[scale=0.38]{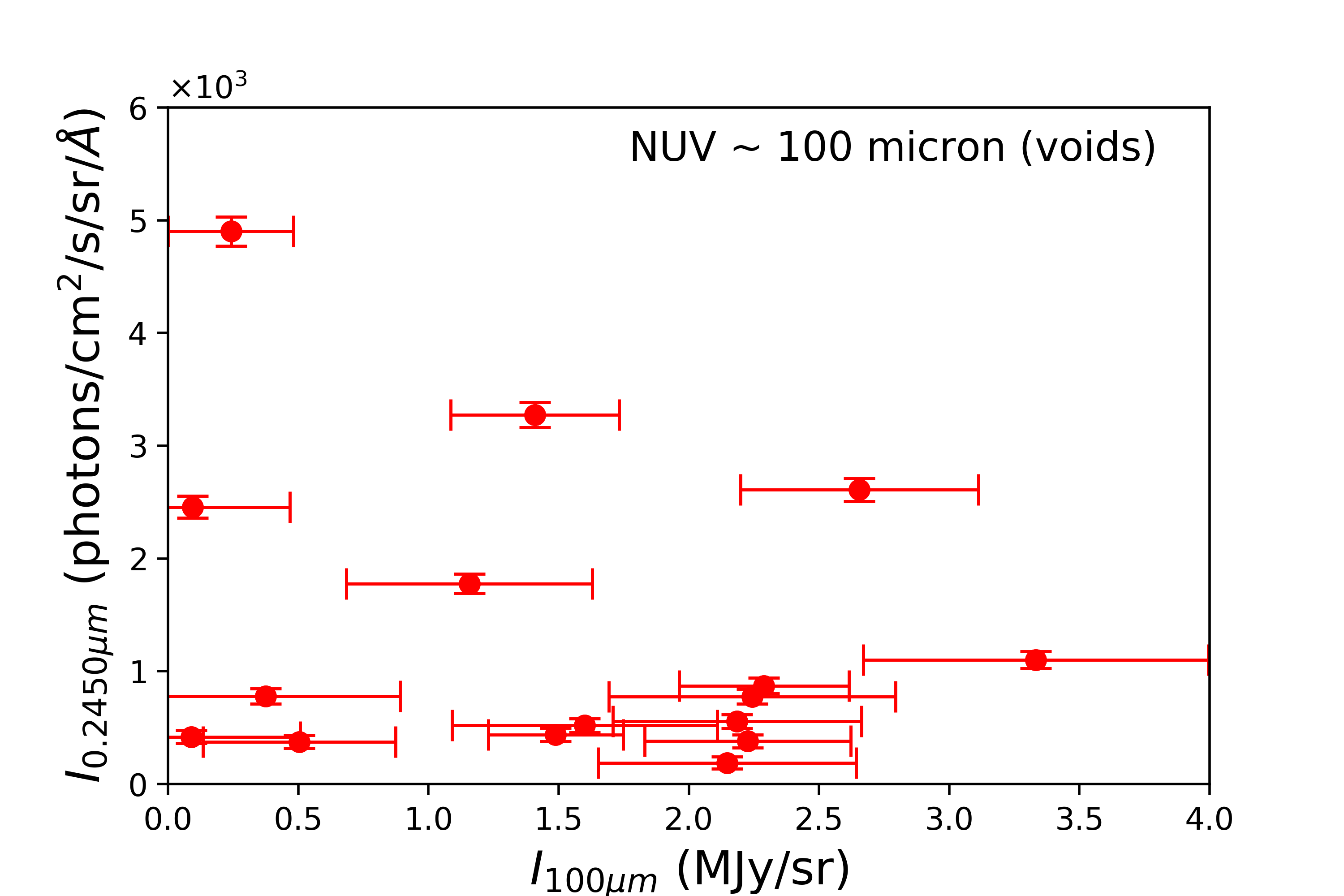}
\includegraphics[scale=0.38]{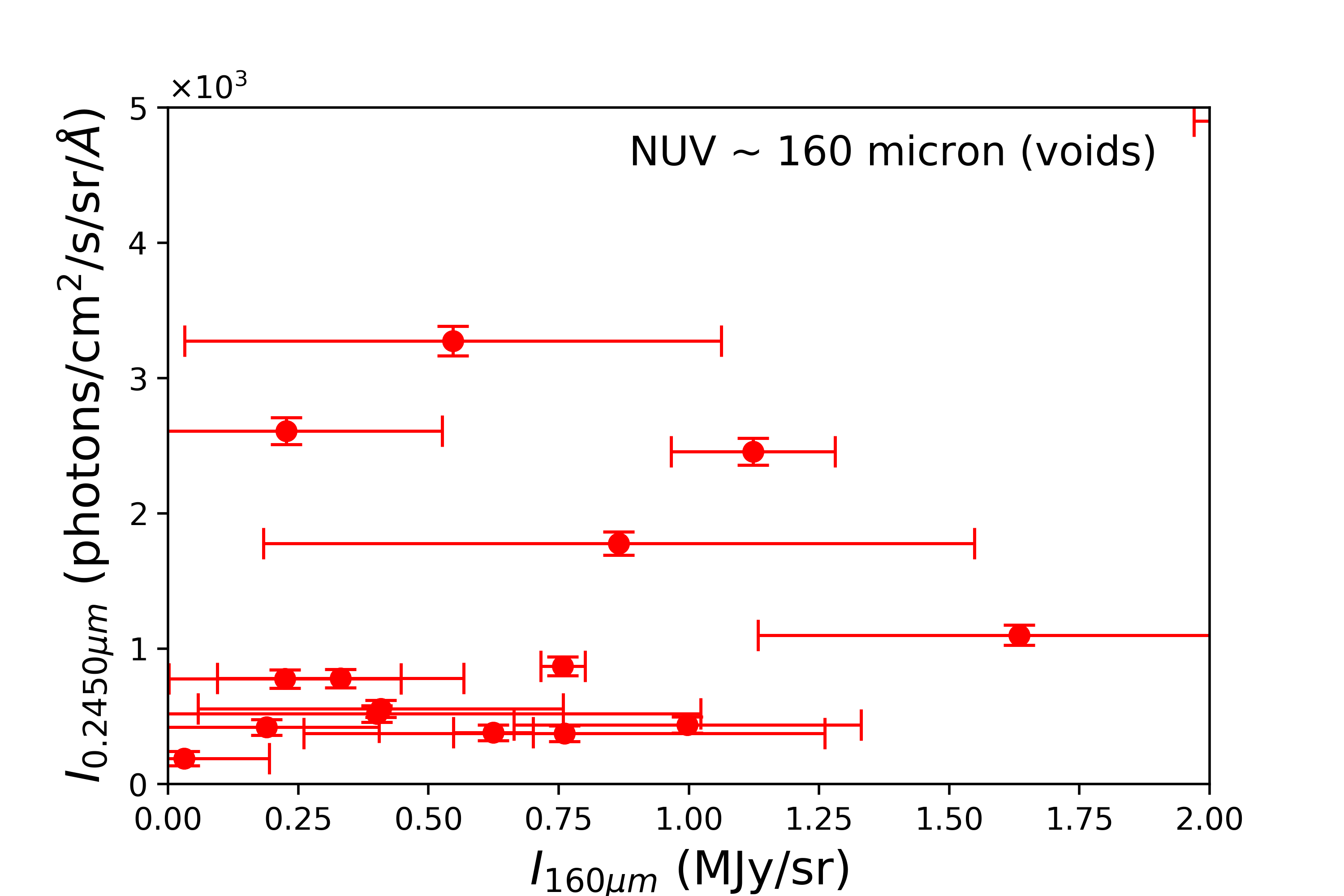}
\caption{FUV--IR and NUV--IR correlation plots for {16} locations with N(HI) $< 1 \times 10^{21}$ cm$^{-2}$. 
} 
\label{fig:one_void}
\end{figure*}

Most of the locations with N(HI) $<1 \times 10^{21}$ cm$^{-2}$, except a few (at shorter wavelengths), tend to have lower dust surface density, manifested in low IR fluxes -- nearly half of those typical for regions with higher N(HI), %$I_\lambda<1$MJy/sr, 
as expected in a sparse environment (see Fig.~\ref{fig:one_void}). 

\begin{table*}
   \centering
    \begin{tabular}{lcclcc}
     \hline
     FUV--IR  & Pearson coefficient ($r$) & $p$-value & NUV--IR & Pearson coefficient ($r$) & $p$-value \\
    \hline
       FUV $\sim I_{4.5 \micron}$  & {0.14 $\pm$ 0.27} & {0.13} & NUV $\sim I_{4.5 \micron}$ & {0.17 $\pm$ 0.28} & {0.07}    \\
       FUV $\sim I_{5.8 \micron}$ &  {0.17 $\pm$ 0.24} & {0.10}  & NUV $\sim I_{5.8 \micron}$ & {0.16 $\pm$ 0.25} & {0.09} \\
       FUV $\sim I_{24 \micron}$ & {0.28 $\pm$ 0.17} & {0.09}  & NUV $\sim I_{24 \micron}$ & {0.21 $\pm$ 0.18} & {0.18} \\
       FUV $\sim I_{70 \micron}$ &  {0.57 $\pm$ 0.08} & {0.007} & NUV $\sim I_{70 \micron}$ & {0.59 $\pm$ 0.08} & {0.005}  \\
       FUV $\sim I_{100 \micron}$ & {0.19 $\pm$ 0.08} & {0.42} & NUV $\sim I_{100 \micron}$ & {0.19 $\pm$ 0.07} & {0.44} \\
       FUV $\sim I_{160 \micron}$ & {0.54 $\pm$ 0.06} & {0.01} & NUV $\sim I_{160 \micron}$ & {0.61 $\pm$ 0.06} & {0.005}   \\
       \hline
\end{tabular}

\caption{FUV--IR and NUV--IR correlation values for locations with N(HI) $>$ $1 \times 10^{21}$ cm$^{-2}$ {(Fig.~\ref{fig:correlation_plots}).}}
    \label{tab:correlation_high_hI}
\end{table*} 

\begin{table*}
    \centering
    \begin{tabular}{lcclcc}
     \hline
     FUV--IR  & Pearson coefficient ($r$) & $p$-value & NUV--IR  & Pearson coefficient ($r$) & $p$-value \\
     \hline
       FUV $\sim I_{4.5 \micron}$  &{0.06 $\pm$ 0.30} & {0.10} & NUV $\sim I_{4.5 \micron}$ & {0.05 $\pm$ 0.24} & {0.17}  \\
       FUV $\sim I_{5.8 \micron}$ & {0.18 $\pm$ 0.25} & {0.07}  & NUV $\sim I_{5.8 \micron}$ & {0.17 $\pm$ 0.29} & {0.08}  \\
      
       FUV $\sim I_{24 \micron}$ & {$-0.14 \pm 0.20$} & {0.22} & NUV $\sim I_{24 \micron}$ & {$-0.14 \pm 0.19$} &{0.29}  \\
       FUV $\sim I_{70 \micron}$ & {0.07 $\pm$ 0.16} & {0.77} & NUV $\sim I_{70 \micron}$ & {0.23 $\pm$ 0.16} & {0.29}  \\
       FUV $\sim I_{100 \micron}$ & {$-0.20 \pm 0.09$} & {0.42}  & NUV $\sim I_{100 \micron}$ & {$-0.25 \pm 0.08$} & {0.32} \\
       FUV $\sim I_{160 \micron}$ & {0.37 $\pm$ 0.14} & {0.08} & NUV $\sim I_{160 \micron}$ & {0.51 $\pm$ 0.13} & {0.01} \\
       \hline
\end{tabular}
\caption{FUV--IR and NUV--IR correlation values for locations with N(HI) $<$ $1 \times 10^{21}$ cm$^{-2}$ {(Fig.~\ref{fig:one_void}).}} 
%and \ref{fig:remaining_voids}).}
    \label{tab:correlation_low_hI}
\end{table*}

We study the UV--IR (both FUV--IR and NUV--IR) correlations for our locations by calculating the Pearson and Spearman correlation coefficients. We find that the Pearson coefficient represents better our results, suggesting that the UV--IR correlation follows a linear relationship rather than a monotonic relationship. 

The Pearson correlation coefficient does not incorporate the data errors in the calculations. To include the errors, we generated 1000 simulated data sets from the Gaussian distribution considering the uncertainty in the data as the standard deviation for each data point. The correlation coefficient is then calculated as the mean of the correlation coefficients of these 1000 simulated data sets and its standard deviation was considered as the error in the correlation coefficient. The observed Pearson correlation coefficients between FUV(NUV) and IR data for the two groups of locations are shown in Tables~\ref{tab:correlation_high_hI} and \ref{tab:correlation_low_hI}. By comparing the results given in the two tables, it is evident that for locations in Table~\ref{tab:correlation_high_hI}, there is a reasonable correlation between the FUV and IR intensities at 70 and 160 $\micron$, and the coefficients are statistically significant.
{For high column density locations, the correlation coefficients between FUV and IR are highest at 70 $\mu$m compared to other IR wavelengths.} This result indicates that the MIR emission is mostly dominated by warm dust emission, which is in agreement with the MIR spectra of low-metallicity dwarf galaxies. 

On the other hand, locations with low N(HI) show poor or even a weak negative correlation between the UV and IR intensities, except at 160 $\micron$ (see Table~\ref{tab:correlation_low_hI}) which shows a reasonable correlation with NUV. This may reflect the fact that in HI deficient regions only a rather small  fraction of UV photons can transfer their energy to dust heating: with the LMC dust model $A_V\sim 2.5\times 10^{-22}N({\rm HI})<0.25$ mag which gives $A_{uv}<0.5$ mag. However, assuming the Milky Way-like extinction law $A_V\sim 5\times 10^{-22}N({\rm HI})$ and accounting for an order of magnitude lower metallicity and dust abundance in Ho II one can arrive at $A_{uv}<0.1$ mag.

\begin{figure*}
\centering
    \includegraphics[scale=0.5]{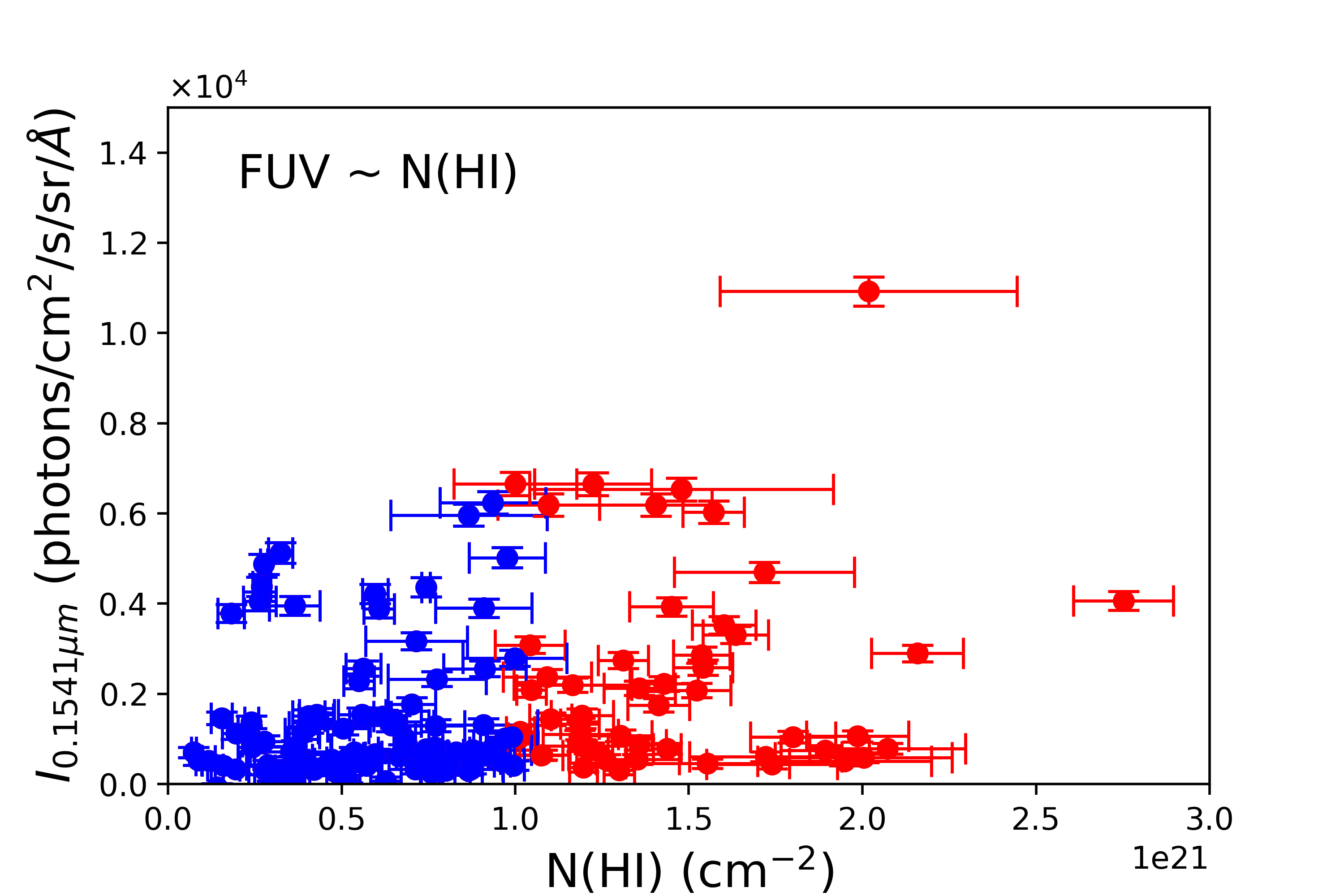}
    \includegraphics[scale=0.5]{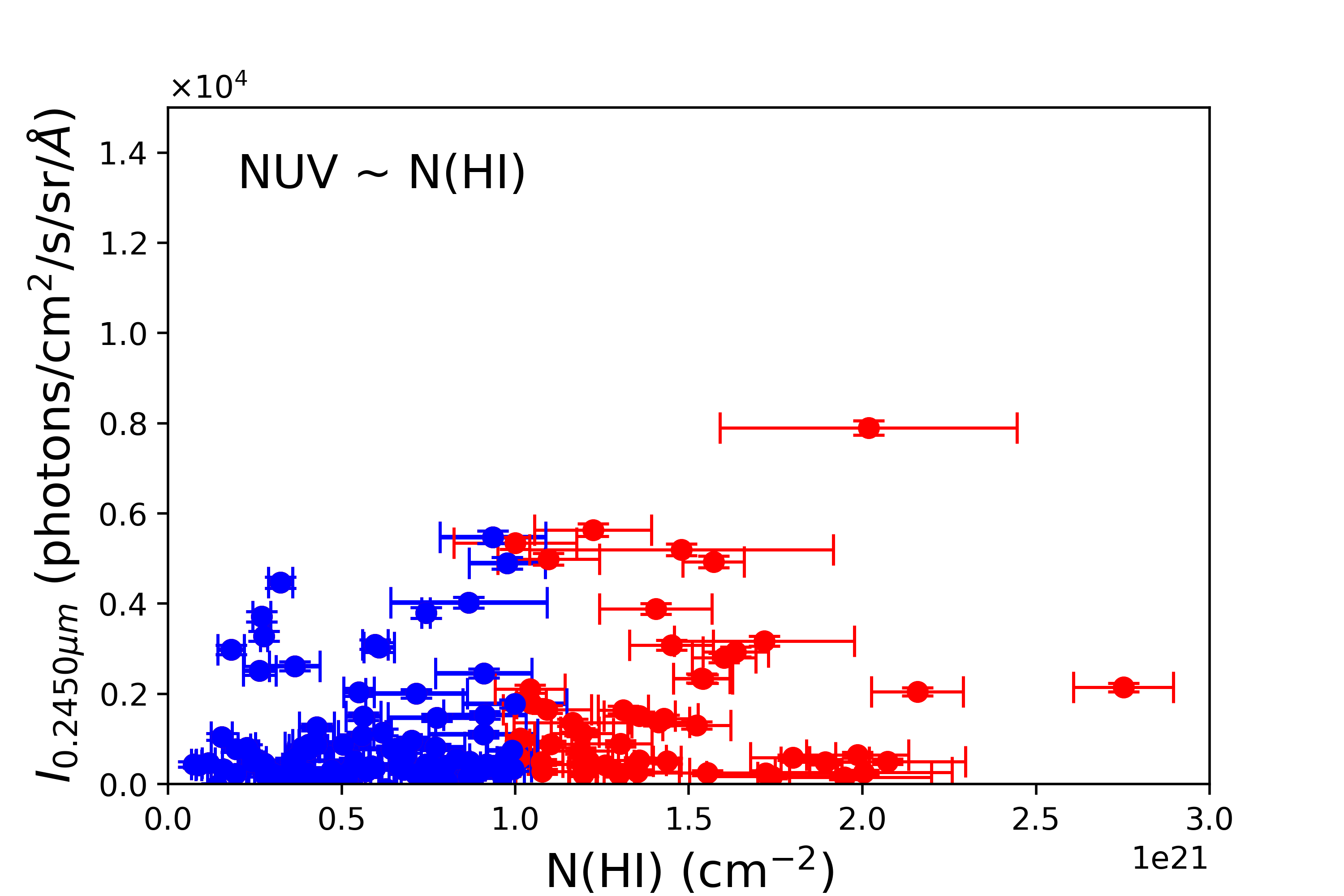}
    \caption{{UV--N(HI) correlation plots for selected 142 locations. \textit{Left}: FUV, \textit{Right}: NUV. The blue symbols represent locations with N(HI)  $< 1 \times 10^{21}$ cm$^{-2}$, and the red symbols represent locations with N(HI)  $> 1 \times 10^{21}$ cm$^{-2}$.}}
    \label{fig:nh_correlation}
 \end{figure*}

It is evident from the MIR spectra of Ho~II that 8~$\micron$ emission by PAH molecules is suppressed in this galaxy \citep[see discussion in][]{walter2007dust}. From photometric measurements, we too did not find 8 $\micron$ emission from any of our considered locations. As this emission is suppressed, the correlation for 8 $\micron$ was not calculated. 

The interrelations between the FUV and IR, and between the NUV and IR flux densities are similar. Such a similarity can take place if the dominant source of dust heating is due to UV radiation, with the interrelation between FUV and NUV intensities being spatially invariant, indicating  their common  origin.

It is worth noting that the slopes of the correlations between the IR and {UV} intensities vary at different IR wavelengths as seen in Fig.~\ref{fig:correlation_plots}: they increase from the shorter towards longer wavelengths, except at 160 $\micron$ (see Table~\ref{tab:derivative_high_hI}). This trend is consistent with the expected one, if the dust heating is determined by diffuse {UV}: the derivative  $dI_{ir}/dI_{uv}$ at shorter wavelengths decreases approximately as $\nu^{4+\beta}\exp(-h\nu/kT_d)$, with $\beta$ being the dust spectral index; it is assumed that dust temperature $T_d\propto I_{uv}^{1/(4+\beta)}$.

\begin{table*}
\centering
\caption{Best-fit ratios of IR and UV intensities in units of 10$^{-5}$ MJy/(phot/cm$^2$/s/\AA) for locations with N(HI) $>1\times 10^{21}$ cm$^{-2}$.}
    \begin{tabular}{ccccccc}
     \hline
     & $dI_{4.5}/dI_{uv}$ & $dI_{5.8}/dI_{uv}$ & $dI_{24}/dI_{uv}$ & $dI_{70}/dI_{uv}$ & $dI_{100}/dI_{uv}$ & $dI_{160}/dI_{uv}$ \\
     \hline
       FUV & {0.29}  & {1.02} & {3.32} & {54.2} & {240.9} & {81.1} \\
       NUV & {0.32} & {1.27} & {5.27} & {67.5} & {318.1} & {94.9}  \\
       \hline
\end{tabular}
    \label{tab:derivative_high_hI}
\end{table*}

\citet[][see panels (a--c) in their Fig.~21]{Seon2011} have reported a correlation between the FUV, HI, 100 $\mu$m and H$\alpha$ in our Galaxy, similar to what we show in Fig.~\ref{fig:correlation_plots} for IR. Such a similarity in their case is naturally explained by reflection of light from hot stars in the disk by nearby dusty clumps and clouds. This suggests a non-negligible optical depth of the clumps in order to cause a considerable amount of FUV photons to be reflected. We have also looked at the relation between UV and HI (both FUV-HI and NUV-HI) in all considered {142} locations (Fig.~\ref{fig:nh_correlation}). We see that the correlation between these two quantities is poor with a coefficient of 0.27 for FUV-HI and 0.24 for NUV-HI. It is remarkable that HI deficient regions (``cavities'', blue symbols) demonstrate UV intensities very close to those from higher HI column density locations (red symbols). This may indicate that the origin of diffuse UV light is not tightly associated with HI gas. One could therefore assume that the HII component is more diffuse, while HI gas is immersed in it in the form of clumps and filaments. At the same time, a weak correlation (about $\simeq 0.25$) is worth mentioning, with slopes of $\simeq 0.1$ for FUV vs $N({\rm HI})$ and $\simeq 0.07$ for NUV vs $N({\rm HI})$. This difference in the slope is consistent with the difference of a factor of 2 between the scattering cross-sections in FUV and in NUV, and does apparently reflect a minor contribution from dust scattering.

\subsubsection{3D radiative transfer model of dust scattering}

One of the known sources of diffuse UV emission is the scattering of starlight by dust grains in the ISM. To find out how much the scattering contributes to the diffuse emission we have performed radiative transfer modelling in the FUV for selected locations.
 
We have extracted the diffuse UV intensities at {33} locations (see Table~\ref{tab:fuv_intensity}) as described in the earlier section. In Fig.~\ref{fig:rgb} ({\it Right}), locations with HI column density greater than $1 \times 10^{21}$ cm$^{-2}$ are marked as purple circles, whereas locations with HI column density less than $1 \times 10^{21}$ cm$^{-2}$, or locations with cavities, are marked by white squares. The diffuse intensities vary from {$\sim$$450-10000$} photon units, with the brightest values corresponding to regions close to regions with recent massive star formation. In radiative transfer modelling, the scattered intensity is quantified in terms of two important wavelength-dependent parameters -- the single scattering albedo $\alpha$, and the scattering phase function asymmetry parameter $g$. 

In order to predict the dust scattered intensities in the FUV for Holmberg~II, we have used a single scattering radiative transfer model successfully used for the Orion region by \citet{shalima2006far}. This model constrains the albedo $\alpha$ and the asymmetry factor \textit{g} of the dust grains \citep{shalima2006far, saikia2018modelling} in the galaxy. Since Galactic diffuse UV emission has been known to originate mainly from the forward scattering by optically thin clouds in front of hot UV emitting stars \citep{sujatha2007measurement}, a single scattering model with dust distributed in a sheet in front of the clusters is considered here. However, in reality, scattering in Ho~II could also be due to back scattering from clouds behind the stars which is not considered as part of this work. The scattered intensity is a function of dust optical properties, such as albedo $\alpha$, phase function $\phi(\theta)$, and optical depth values $\tau_{1}$ and $\tau_{2}$ through the following equation,
\begin{equation}
I_{sca} = \frac{L_{star} \times \alpha \times \phi(\theta)\times \tau_{1} \times e^{-\tau_{2}}}{4 \pi d^{2}}\,.
\label{eq:3}
\end{equation}
Here, $\tau_{1}$ is the optical depth corresponding to the scattering layer $\tau_{1} = n \sigma \delta z$, where $\sigma$ is the extinction cross-section, $n$ is the dust number density, and $\delta z$ is the thickness of the layer. $\tau_{2}$ is the optical depth of the remaining material responsible for line of sight (LOS) extinction, where $\tau$ = $\tau_{1}$ + $\tau_{2}$ corresponds to the total optical depth for a given location. The other parameters in Eq.~\ref{eq:3} are the source luminosity $L_{star}$, and $d$ is the distance between the source and the scattering layer.

The model uses the Henyey-Greenstein scattering phase function \citep{henyey1941diffuse},
\begin{equation}
    \phi(\theta)=\frac{1-g^{2}}{4\pi[1+g^{2}-2g \cos(\theta)]^{3/2}}\,,
\end{equation}
where $\phi(\theta)$ signifies the amount of energy scattered per unit solid angle in a direction $\theta$. The value of the asymmetry factor \textit{g} lies in the interval $[-1,1]$. A value of \textit{g} close to $0$ implies isotropic scattering, a value close to $-1$ implies strong backward scattering, and a value close to 1 implies strong forward scattering.

The star-forming complexes, described in \citet{egorov2017complexes}, are the main sources of UV radiation in the galaxy as each of them contains several young star clusters. For the observed fluxes of the clusters in the FUV, we have used the values at $\lambda = 1521$\AA\, from \citet{stewart2000star}. These values were already corrected for galactic foreground extinction. In order to get the intrinsic luminosity of these complexes, we also corrected for the internal extinction towards these complexes from their $E(B-V)$ values \citep{stewart2000star}. In deriving the internal extinction in Ho~II, \citet{stewart2000star} assumed an LMC reddening law because of the similar metallicity and abundance \citep{puche1992holmberg, hunter1985star}, where they assumed the metallicity of Ho~II to be $Z=0.4Z_\odot$. 

We then used these $E(B-V)$ values to calculate the optical depth $\tau$, which allows us to correct for interstellar absorption: $\tau = \frac{E(B-V)R_V}{1.0863}$, where we have adopted $R_{V}=3.41$ for average LMC reddening \citep{gordon2003quantitative}. As the observed reddening is proportional to $\tau_1$ and $\tau_2$ approximately as $\propto\tau_1+\tau_2$, (assuming forward scattering grains ($g>0$) as observed in the Magellanic clouds), one can expect that diffuse FUV originating from stellar light scattered by clumpy dust should be correlated with the observed reddening. 

The flux densities are then multiplied by $e^{\tau}$ to account for the internal extinction. The luminosities of the complexes are then calculated assuming the star clusters are at a distance of 3.39 Mpc, which is the distance to the Ho~II galaxy (see Table~\ref{tab:luminosity}).
   
The total hydrogen column density $N(H)$ towards the complexes was calculated using the following {reddening relation per H atom in the LMC} \citep{draine2003interstellar},
\begin{equation}
\frac{E(B-V)}{N(H)}=4.5 \times 10^{-23} \,\text{cm}^{2}/\text{H}\,.
\end{equation}

\begin{table}
    \centering
    \begin{tabular}{lcc}
    \hline
    Complexes  &  Luminosity & N(H)\\
         & (erg s$^{-1}$ \AA$^{-1}$) & (cm$^{-2}$) \\
         \hline
     NE & $6.28 \times 10^{37}$ & $1.27 \times 10^{21}$\\
     N & $11.15 \times 10^{37}$ & $1.62 \times 10^{21}$\\
     NW & $10.56 \times 10^{37}$ & $1.08 \times 10^{21}$\\        
     ExtN & $5.49 \times 10^{37}$ & $1.77 \times 10^{21}$\\       
     SE & $5.508 \times 10^{37}$ & $0.77 \times 10^{21}$\\
     ExtNE & $1.054 \times 10^{37}$ & $2.88 \times 10^{21}$ \\
     Int.shell & $11.35 \times 10^{37}$ & $1.99 \times 10^{21}$\\
     \hline
    \end{tabular}
\caption{Properties of the contributing star-forming complexes. The complexes have been adopted from \citet{egorov2017complexes}.}
\label{tab:luminosity}
\end{table}

\begin{figure}
\centering
\includegraphics[width=7cm,height=6cm]{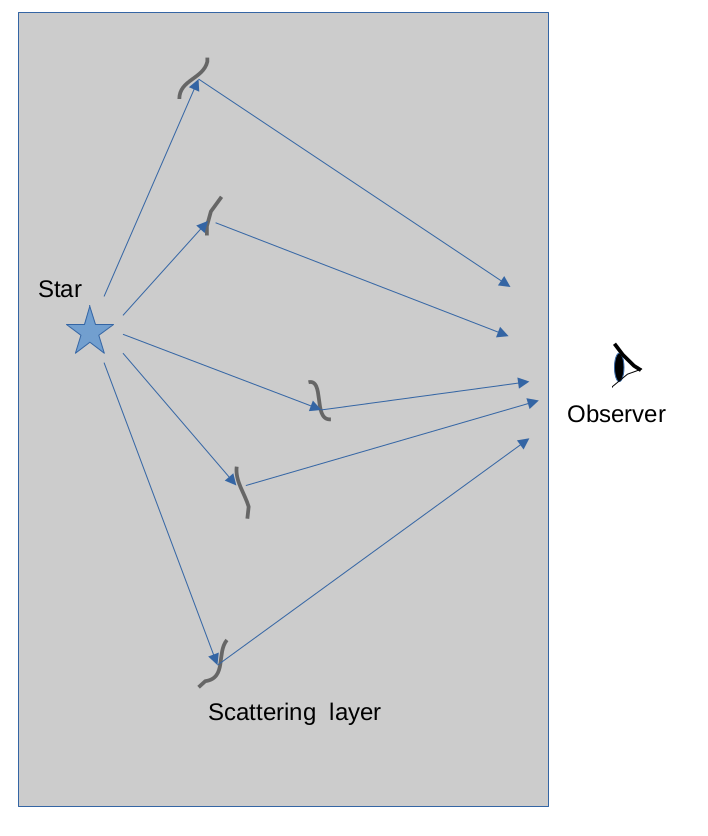}
\caption{Schematic diagram showing the relative geometry of the source (star) and the scattering of starlight by optically thin clouds towards the observer. The star-cloud distance $d$ and the optical depth of the scattering layer $\tau_{1}$ are variable parameters in the model.}
\label{fig:mod}
\end{figure}

Since the relative distribution and geometry of stars and dust in Ho~II are unknown, we have considered the scattering dust grains to be distributed in the form of optically thin filaments located at different distances between the observer and the star clusters for different locations {(see Fig.\ref{fig:mod})}.
We calculate the total optical depth $\tau$ in the LOS by multiplying the total $N(H)$ with $\sigma$, the extinction cross-section of dust. If a location lies in the LOS of any of the star-forming complexes, we consider the $N(H)$ towards that complex as the $N(H)$ for the location, and subsequently multiply by $\sigma$ to obtain the optical depth. On the other hand, if a location lies away from the LOS of the star-forming complexes, we obtain the $N(H)$ towards that location from the THINGS integrated HI map to obtain the optical depth $\tau$. While calculating the scattered intensity at any location, we have considered the contribution from each of the star forming complexes separately and then added them together to get the total scattered intensity at that location. Our model gives the scattered intensities for each combination of albedo $\alpha$, asymmetry factor \textit{g}, and optical depth $\tau$. {We vary the albedo from 0.1 to 0.9 and $g$ from 0 to 0.9 in steps of 0.1, and the star to dust cloud distance, $d$, from 47 pc to 186 pc in steps of 47 pc, to extract the best-fit optical constants and the 3D distribution of dust at the diffuse locations.} We derive the best-fit values based on the minimum $\chi^2$-statistic and a range for the parameter values within a 90\% confidence level {(Table~\ref{tab:bestfit})}.

\begin{table*}
\centering
\begin{tabular}{cccccccc}
\hline
 Locations & \textit{l} & \textit{b} & $\alpha$ & \textit{g} & $\tau_{1}$ & $\tau_{2}$ & distance \\
      &  \text{(degrees)} & \text{(degrees)} &  &  &   &  & \text{(pc)}\\
        \hline
1 & 144.2995 & 32.6691 & 0.3 (0.1-0.3) & 0.6 (0.1-0.6) & 0.041 (0.041-0.123) & 0.369 (0.369-0.287) & 140 (47-140) \\
3 & 144.2835 & 32.7029 & 0.4 (0.1-0.4) & 0.3 (0.2-0.4) & 0.122 (0.122-0.366) & 1.098 (1.098-0.854) & 93 (47-140) \\
      6 &  144.2682 & 32.7303 & 0.4 (0.1-0.4) & 0.4 (0.0-0.4) & 0.041 (0.041-0.082) & 0.369 (0.369-0.328) & 140 (93-186) \\
       9 & 144.3063 & 32.6755 & 0.2 (0.1-0.3) & 0.4 (0.2-0.7) & 0.124 (0.062-0.124) & 0.496 (0.558-0.496) & 47 (47-93) \\
       10 & 144.2544 & 32.6734 & 0.1 (0.1-0.2) & 0.4 (0.1-0.4) & 0.088 (0.044-0.088) & 0.352 (0.396-0.352) & 140 (93-186) \\
    14 &   144.2816 & 32.6657 & 0.1 (0.1-0.2) & 0.5 (0.4-0.6) & 0.060 (0.060-0.120) & 0.540 (0.540-0.480) & 140 (47-186) \\
   15 &  144.2885 & 32.7289 & 0.2 (0.2-0.6) & 0.6 (0.2-0.6) & 0.058 (0.029-0.058) & 0.232 (0.261-0.232) & 140 (93-186) \\
       16 & 144.2847 & 32.6633 & 0.2 (0.1-0.2) & 0.5 (0.3-0.7) & 0.062 (0.062-0.124) & 0.558 (0.558-0.496) & 140 (93-186) \\
      19 &  144.2226 & 32.6552 & 0.1 (0.1-0.2) & 0.5 (0.2-0.6) & 0.046 (0.046-0.092) & 0.414 (0.414-0.368) & 186 (93-186) \\
      21 &  144.3336 & 32.6575 & 0.2 (0.1-0.2) & 0.5 (0.3-0.7) & 0.082 (0.082-0.123) & 0.328 (0.328-0.287) & 186 (140-186) \\
      22 &  144.2679 & 32.7278 & 0.3 (0.2-0.6) & 0.5 (0.2-0.7) & 0.049 (0.049-0.098) & 0.441 (0.441-0.392) & 93 (93-140) \\
      23 &  144.2625 & 32.6781 & 0.1 (0.1-0.3) & 0.7 (0.3-0.7) & 0.041 (0.041-0.082) & 0.369 (0.369-0.328) & 93 (93-186) \\
     25 &  144.2843 & 32.7338 & 0.1 (0.1-0.3) & 0.4 (0.3-0.6) & 0.049 (0.049-0.098) & 0.441 (0.441-0.392) & 140 (140-186) \\
    26 & 144.2823 & 32.6637 & 0.1 (0.1-0.2) & 0.5 (0.3-0.6) & 0.124 (0.062-0.124) & 0.496 (0.558-0.496) & 140 (140-186) \\
    30 & 144.2723 & 32.6760 & 0.3 (0.1-0.3) & 0.4 (0.3-0.6) & 0.062 (0.062-0.124) & 0.558 (0.558-0.496) & 93 (93-140) \\
    32 & 144.2721 & 32.7334 & 0.5 (0.3-0.6) & 0.2 (0.1-0.3) & 0.029 (0.029-0.058) & 0.261 (0.261-0.232) & 140 (140-186) \\
    33 & 144.2765 & 32.6735 & 0.2 (0.1-0.3) & 0.6 (0.3-0.6) & 0.046 (0.046-0.092) & 0.414 (0.414-0.368) & 93 (47-140) \\
        \hline
\end{tabular}
\caption{The values of the best-fit {model} parameters (based on minimum $\chi^2$-statistic) for 17 locations with N(HI) $>1 \times 10^{21}$ cm$^{-2}$. The range of values within 90\% confidence level for each parameter are shown inside parentheses. The location numbers are the same as in Table~\ref{tab:fuv_intensity}.}
\label{tab:bestfit}
\end{table*}

For modelling of the dust scattered emissions, we have considered only those locations where the HI column density is greater than $1 \times 10^{21}$ cm$^{-2}$, as these are the regions where dust emission was detected. The radial surface brightness profile shows the detection of dust up to $\sim$4 kpc \citep{walter2007dust}. The input parameters of our model are mainly the luminosities of the star-forming complexes and the extinction cross-section $\sigma$, described previously. Since we are assuming an LMC reddening law, we adopted the $\sigma$ value for the LMC of 3.867 $\times$ 10$^{-22}$ cm$^2$ at our mean FUV wavelength of 1541~\AA. We then compare our model intensities with {\it AstroSat} FUV values in order to find the best-fit values of $\alpha$, \textit{g}, and distance to different dust locations.  

From our calculations, we have obtained a median value of $\alpha = 0.2$ and $g = 0.5$ at 1541~\AA\, for the dust grains in Ho~II. This is in reasonable agreement with the theoretically predicted value of $\alpha = 0.3$ and $g=0.6$ at a similar wavelength for an average LMC dust \citep{draine2003interstellar}. Our model-derived optical depths correspond to an optically-thin scattering medium ($\tau \sim$ 0.04 to 0.13) at distances of $\sim$47 pc to $\sim$186 pc in the foreground of the stars.
The slight discrepancy between the observed and theoretical values could be due to large uncertainties in the star-dust geometry and the use of the Henyey-Greenstein phase function at $\lambda < 0.16\,\micron$ \citep{draine2003scattering}. This phase function provides a good approximation to the calculated scattering phase function at wavelengths between $\sim 4 \,\micron$ and $1 \,\micron$, but does not provide a good fit at shorter wavelengths, mainly in the UV. If we take the metallicity of Ho~II to be 0.1 Z$_{\odot}$ \citep{egorov2013emission}, the extinction law will be similar to that of the SMC. The extinction cross-section $\sigma_{ext}$ would get reduced by a factor of $\sim$3, as the dust-to-gas ratio of SMC is almost one-third compared to LMC \citep{roman2022metal}. This would require an increase in the {FUV} albedo of the dust grains by a factor of $\sim$3 {($\sim$0.6)}, in the model which is higher than the theoretical predictions for {SMC bar ($\sim$0.42), a region with reasonably high albedo values in the FUV}.

From Table~\ref{tab:bestfit}, we can see that the optical depths of the scattering layer $\tau_1$ are small, {in the range 0.02 - 0.12} compared to the optical depths of the layer responsible for LOS extinction implying forward scattering by optically thin clouds, such as seen in our Galaxy \citep{sujatha2007measurement}.

\subsubsection{Diffuse fraction for isolated regions}

Another method we have used to identify the source of the diffuse UV emission is to extract the aperture intensities in relatively non-crowded regions that have a single UV source: either a star or a cluster of stars. We {selected 8 such locations and derived diffuse emission by removing the source contribution in a box of $32\times32$ pixels centered at the source position.} We estimated the ratio of diffuse flux from Ho~II and total flux for these {8} regions ({Table~\ref{tab:fuv_intensity_isolated}}).
If dust scattering is a source of diffuse emission, the diffuse fraction is expected to be close to the value of the single scattering albedo of dust grains in the vicinity of a star for an optically thin medium. The observed median diffuse fraction was {{$\sim$81.61\%}} in FUV and {{$\sim$76.93\%}} in NUV. The obtained FUV and NUV diffuse fractions in individual aperture regions are provided in Table~\ref{tab:fuv_intensity_isolated}. The value in the FUV is {much higher than} the theoretical predictions of dust albedo for average LMC dust, dust in the SMC bar \citep{weingartner2001dust, draine2003scattering} as well as to our model-derived albedo values, implying the presence of {other components of diffuse UV emission apart from dust scattering.}

\begin{table}[h!]
\centering
\caption{Diffuse UV fractions in {8} isolated regions having single star/stellar cluster.}
\begin{tabular}{cccc}
\hline
 $l$  &  $b$ & FUV Diffuse & NUV Diffuse \\
 (degrees)  & (degrees) & Fraction (\%) & Fraction (\%) \\
\hline
144.3018  & 32.6582 & 62.10 & 44.29 \\
144.2962  & 32.6612 & 65.61 & 72.65 \\
144.2731  & 32.6721 & 84.79 & 79.39 \\
144.3107  & 32.6634 & 84.82 & 74.66 \\
144.3333  & 32.6447 & 78.42 & 79.20 \\
144.3260  &32.6412 & 87.85 & 81.56 \\
144.2154  & 32.6638 & 93.99 & 82.50 \\
144.2045  & 32.6814 & 44.72 & 38.91 \\
\hline
Median& & 81.61 & 76.93 \\
\hline
\end{tabular}
\label{tab:fuv_intensity_isolated}
\end{table}

\subsubsection{Possible origins of the diffuse UV emission}
\label{odfuv}

Although our model is able to predict the observed UV intensities as being associated with dust scattering, the dust-to-HI mass ratio for Ho II is an order of magnitude lower than in Milky Way ($M_d/M_{HI}\simeq 10^{-3}$) as estimated by \citet{walter2007dust,Draine2007}. This results in the extinction of $A_V\sim 0.5\times 10^{-22}N({\rm HI})$, if we do a linear extrapolation from the Milky Way dust-to-gas mass ratio of $\sim 0.01$; which is consistent with \citet{kahre2018extinction}. In the UV ($\lambda=1000-1500$~\AA), $A_\lambda$ can be as high as $\sim (2-5)\times 10^{-23}N({\rm HI})\sim 0.1-0.15$ for $N({\rm HI})\sim 10^{21}$ cm$^{-2}$, provided the extinction law is the same as in the LMC. The reflected fraction is correspondingly $f_{r}\sim (0.1-0.15)$, which with the albedo $\alpha \simeq 0.6$ as estimated by \cite{draine2003scattering} for Galactic diffuse light at FUV--NUV, gives $f_r\simeq 0.06-0.09$. Therefore, with our model estimates of albedo, $\alpha = 0.1-0.4$, it results in reflected fraction of $f_{r}\sim (0.01-0.04)$. This seems to be an upper limit for the fraction of diffuse FUV that can be associated with dust scattering. 

Another evidence for the dust contribution to the diffuse FUV emission in regions of high HI column densities is clearly seen from Figs.~\ref{fig:correlation_plots} \& \ref{fig:one_void}, where the slopes of the correlations between the IR and FUV(NUV) fluxes vary at different IR wavelengths: they grow from the shorter towards longer wavelengths, as given in Table~\ref{tab:derivative_high_hI}. This trend is consistent with the expected one if the dust heating (which is complimentary to scattering) is determined by incident diffuse FUV(NUV) radiation. This means that in high N(HI) density regions of Ho~II, a fraction of the diffuse FUV can be attributed to originate from the scattering of FUV photons off the dust grains. But, the FUV contour plot in Fig.~\ref{fig:rgb} (\textit{Left}) clearly highlights the presence of faint FUV emission from the HI cavities having very low HI column density. For these regions,  including the HI cavities, it is very hard to explain the origin of diffuse FUV from dust scattering. In this case, other conventional sources of production of diffuse FUV need to be considered. In the context of Ho~II, one such possible source can be the two-photon continuum from warm ionized medium and low-velocity shocks. In the ISM of our galaxy, these low-velocity shocks are abundant. In the H$\alpha$ image of Ho~II (see Figs.~1 \& 2 in \citet{egorov2017complexes}), we see many bubble kind of structures, which can originate from supernova remnants as high-velocity shocks, eventually cascading into low-velocity shocks. These low-velocity shocks mainly cool through Ly$\alpha$ emission, two-photon continuum, and H$\alpha$ emission. Ly$\alpha$ photons are trapped inside the gas and get absorbed by the dust grains. But the two-photon continuum emission, which peaks at 1400~\AA \, (close to the mean wavelength of FUV observations), can be a possible source for the diffuse FUV emission in the cavity regions (for further discussions on a two-photon continuum, see \citet{Kulkarni2023}).

Contributions to the diffuse FUV emission could be from recombinations in diffuse HII regions similar to those observed in the Milky Way \citep{Haffner2009}, as well as from the warm diffuse HI gas with $T\sim 10^4$ K due to collisions {followed by two-photon decays of 2s-1s transition of atomic hydrogen} \citep{kulkarni2022far}.
In collisionally-dominated HI gas at $T\sim 10^4$ K, the photon production rate from two-photon decays can be as efficient as 50\% of Ly$\alpha$ \cite[Fig.~4 in][]{Kulkarni2023}. Following \citet{Kulkarni2023} (see their Fig.~2), a very rough estimate of the two-photon emissivity for Ho~II at FUV frequencies is 
\be 
L_{2s}\sim (0.6-2)\times 10^{44}f_{2s}(T)n_en_{\rm HI}~{\rm erg~s^{-1}}\,. 
\ee 
If $f_{2s}\simlt 0.3$ at $T\lesssim 10^4$K, where radius of {the central part of} Ho~II ISM disk is assumed to be $R_{\rm HoII}\sim 1$ kpc, {the HI scale height $h_{\rm HI}\simeq 0.4$ kpc \citep{Banerjee2011},} $n_e\sim 0.1$ cm$^{-3}$ and $n_{\rm HI}\sim 0.3$ cm$^{-3}$ in the diffuse gas, it gives $L_{2s}\sim 10^{42}$ erg s$^{-1}$, {which is within factor 3, close to the value shown in Table \ref{tab:Luminosity_diffuse}.} This accounts for around 5\% of stellar UV emission in Ho II, if we convert the observed SF rate $\dot M_{\ast,UV}\simeq 0.12~\msun$ yr$^{-1}$ \citep{kahre2018extinction} to the total UV luminosity calibrated in \citep[][ their Table  1]{Kennicutt2012} to obtain $L_{UV}\simeq 2.4\times 10^{43}$ erg s$^{-1}$.

One of the most efficient sources of two-photon decays is connected with ionized gas behind shock waves with intermediate ($v_{sh} \simeq 40-50$ km s$^{-1}$) velocities \citep{Kulkarni2023}. Assuming SN remnants at a Sedov-Taylor phase, the  rough estimates result in the rate of shock waves with a given velocity $u_0$ at a given point of the ISM \citep{Draine1979} to be
\be 
N_{sh}(u_0)\simeq 8\times 10^{-7}\frac{E_{51}}{n}\frac{S_{1/100}}{u_{100}^2}R_{G,1}^{-2}h_{200}^{-1}~{\rm yr^{-1}}\,, 
\ee
where $E_{51}$ is the explosion energy in $10^{51}$ erg, $n$ -- the number density, $u_{100}=u_0/100~$km s$^{-1}$, $S_{1/100}$ is the integral SN rate in units of 1 SN per 100 years per galaxy having radius $R_G=1$ kpc and scale height $h=200$ pc; $R_{G,1}=R_G/1~{\rm kpc}$, $h_{200}=h/200~{\rm pc}$. For the SFR $\simeq 0.15~\msun$ yr$^{-1}$, we arrive roughly at $S_{1/100}\sim 0.1$, and for $u_{100}\sim 0.5$ at $N_{sh}\sim 8\times 10^{-7}n^{-1}$ yr$^{-1}$. With the recombination coefficient $\alpha_B\simeq 2.7\times 10^{-13}$~cm$^3$~s$^{-1}$ at $T\simeq 10^4$ K, the ambient gas recombines with the rate $A_r=\alpha_r n\simeq 8\times 10^{-6}n$ yr$^{-1}$. At $n\sim 1$ cm$^{-3}$, the recombination rate is much faster than the rate of shocks impinging on any given ISM gas parcel, thus resulting in sporadic two-photon decays. Important, however, is that the interrelation between the $N_{sh}$ and $A_r$ scales differently with gas density. In those regions with $n\sim 0.18$, recombination and {shock} heating rates are nearly equal. Previous estimates of the average gas density in Ho~II ISM vary from 0.3 to 0.5 cm$^{-3}$ \citep{puche1992holmberg,Bagetakos2011,egorov2017complexes}. If one assumes that HI deficient regions in Ho~II are less dense than the average disk density, the diffuse FUV emission in these areas can be attributed to two-photon decays connected to weak ($v_{sh}\lesssim 100$ km s$^{-1}$) shocks. Apart from that there could also be some contribution from an unresolved stellar population.

\begin{figure}
    \centering
    \includegraphics[scale=0.5]{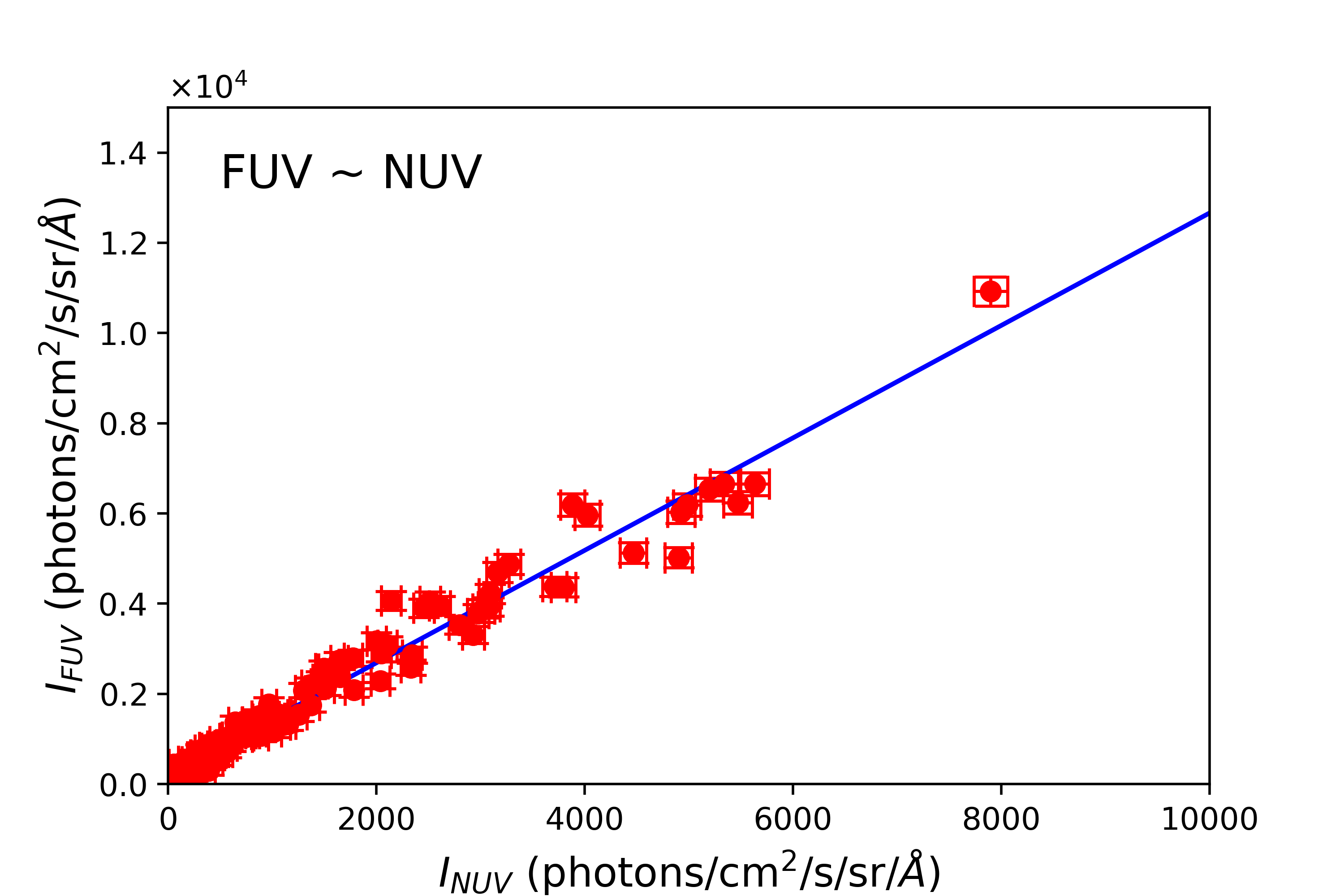}
    \caption{{FUV-NUV correlation plot for selected 142 diffuse locations (slope $f_r\simeq 1.25$, Pearson correlation coefficient $r$ = 0.98).} }
    \label{fig:fuv_nuv_corr}
\end{figure}

{Regardless of whether two-photon decay emission is connected with recombination or with collisional excitations of the 2s HI states, one can expect the FUV ($1200-1800$~\AA) and NUV ($1800-4000$~\AA) fluxes to be connected by an approximate proportionality $f_r=$FUV$_{1300-1800}$/NUV$_{1200-4000}$ $\sim 1$, as can be judged from the spectrum \citep[][]{Osterbrock2006,Draine2011}. Such a proportionality can be recognized while comparing the interrelations between FUV--IR and NUV--IR in Fig.~\ref{fig:correlation_plots}, and is confirmed in Fig.~\ref{fig:fuv_nuv_corr} with FUV/NUV $\simeq 1.25$. Note that for the Milky Way this ratio is nearly half of this value, as measured by GALEX \citep[see Fig.~1 and Sec.~3 in][]{kulkarni2022far}. This difference can be attributed to dust-scattered light \citep{akshaya2018diffuse}. However, the contribution of dust-scattered light in Ho II galaxy is apparently an order of magnitude lower than in the Milky Way case, and this can explain a higher value of FUV to NUV ratio inferred for Ho~II. } 

Another conventional source of diffuse FUV is connected with fluorescent emission of H$_2$ in the Lyman band \citep[$1435-1630$~\AA,\ see][]{Jo2017} originated in molecular clouds. In principle, in our case, one can expect that molecular gas is present there in the form of CO-dark molecular gas, such that molecular emission can be seen {\it only} from H$_2$. However, for this to be possible, the optical depth in the Lyman-Werner band of H$_2$ has to be sufficiently high in order to provide their self-shielding. In the Milky Way, the self-shielding requires a minimum optical depth $A_V\simeq 0.2$ \citep[][]{Draine1996}. In Ho II it can differ because of: i) at least an order of magnitude lower dust content, ii) lower dust amount inhibits formation of H$_2$, iii) uncertain interstellar UV flux in the Lyman-Werner band. In CO-dark clouds, star formation takes place from the fragmentation of molecular clouds in a relatively dispersed manner \citep[see discussion in][]{Planck2011,Shchekinov2017,Madden2020,Chiang2023}. The FUV photons produced by these hot and young stars can escape the immediate vicinity of the star-forming regions, get subsequently scattered by the dust grains, and contribute to the diffuse FUV emission.

If a correlation between the FUV H$_2$ fluorescence emission and N(HI) is similar to that in the Milky Way, presented by \citet[][see panel (b) in their Fig.~7]{Jo2017}, it would indicate that some of the regions with $N({\rm HI})>10^{21}$ cm$^{-2}$ could represent CO-dark molecular gas in Ho II. {In case of the Milky Way, where molecular gas is of $\simeq 20$\% of the ISM mass, the FUV H$_2$ fluorescence accounts for an average $\simeq 8.7$\% of the total diffuse UV emission \citep{Jo2017}. If we assume that H$_2$ mass fraction is proportional to the dust mass (or equivalently to the metallicity), one can infer $\simeq 2$\% for the H$_2$ mass in the ISM of Ho II galaxy, and the corresponding contribution of the H$_2$ fluorescence of $\sim 1$\% to the FUV diffuse emission.} 

\section{Conclusions} 
 
In this work, we have used the highest resolution observations of Ho~II obtained with the UVIT instrument of {\it AstroSat} in order to construct the diffuse UV map of Ho~II and understand the nature and origin of this diffuse emission. Using these observations, we derive {the total diffuse fractions in the NUV and FUV for the entire galaxy and median diffuse fractions for isolated locations in the galaxy.} We also performed UV--IR correlation studies for selected locations and derived the Pearson correlation coefficients for six IR wavelengths, i.e. 4.5, 5.8, 24, 70, 100, and 160 $\mu$m.
\begin{enumerate}
\item
The FUV and NUV diffuse maps, presented here, show the intensity distribution and overall morphology of the diffuse UV emission in Ho~II. These maps, when combined with data at other wavelengths, can help to understand the origin of the diffuse UV background in Ho~II and also in low-metallicity dwarf irregular galaxies in general that are thought to mimic the galaxies in the early universe.

\item The total diffuse UV fraction in Ho~II is {found to be higher than the LMC value} and closer to the observed value for the {SMC bar, in agreement with observations of low metallicity in this galaxy (similar to SMC)}. For a few of the selected locations in regions with N(HI) $> 1 \times 10^{21}$ cm$^{-2}$, the agreement of the observed aperture values of the diffuse UV fraction with the theoretical as well as our model-derived albedo values for similar environments implies dust scattering to be one of the contributors to the diffuse UV radiation. {A weak correlation ($r\simeq 0.25$), revealed between UV and $N({\rm HI})$, may also indicate the contribution from dust scattering. Slightly different slopes between the FUV vs $N({\rm HI})$ ($\simeq 0.1$) and NUV vs $N({\rm HI})$ ($\simeq 0.07$) interrelations can reflect the difference by factor of 2 between the scattering cross-sections in FUV and in NUV. However, for 8 selected isolated regions, the diffuse fractions are much larger than the theoretically predicted albedos for the Magellanic clouds, implying a larger contribution to the diffuse emission from sources other than dust scattering. } 
\item From our FUV modelling, we conclude that the diffuse UV emission in high HI density regions contains a dust scattering component. We also find a low value of optical depth for the layer responsible for the scattering, and a high $g$ value, similar to earlier results (for our Galaxy) that this component could be due to forward scattering by optically thin clouds. However, estimates, based on the low albedo values and optical depths derived from our model, show that only a small fraction of the total diffuse FUV emission can be from dust scattering. {This is further supported by the high diffuse fraction obtained here, which is similar to what is observed in the SMC bar, as well as the scattering model, based on SMC dust, which requires nearly 50\% higher FUV albedos than the theoretical predictions.}
\item We find that 70 $\mu$m IR emission, followed by 160 $\mu$m is better correlated with the UV compared to the other wavelengths for high HI density regions. Since 70 $\mu$m emission is usually attributed to dust heated by the UV photons in regions close to hot and young stars \citep{zhu2008correlations}, it shows that most of the UV emission is absorbed and re-radiated by the warm dust component, while there is also some contribution from the colder dust grains heated by the general radiation field. Cavities did not show any significant UV--IR correlation, except at 160 $\mu$m, which shows better correlation with the NUV rather than the FUV. Therefore, the dust emission in the high HI density regions of this galaxy can be considered to be mostly dominated by the warm dust grains heated by FUV photons, while HI cavities contain colder grains irradiated by the general interstellar radiation field. 
\item Although the origin of diffuse UV close to the OB associations can be partly attributed to the scattering of UV photons from the dust grains, we do see UV emission from the HI cavities, as well as from regions with low column density, which are mostly devoid of dust. The diffuse intensities in these regions match the Galactic polar intensities which contain an offset component \citep{akshaya2018diffuse}, with the FUV diffuse fraction as high as $\sim$99\%. In these regions, the origin of diffuse UV could be due to two-photon continuum emission from low-velocity shocks which peaks near 1400~\AA, very close to our wavelength of observation. A rather tight correlation between FUV and NUV with the ratio $I_{\rm FUV}/I_{\rm NUV}\simeq 1.25$ shown in Fig.~\ref{fig:fuv_nuv_corr}, which is typical for two-photon continuum, strenghtens this preliminary conclusion. In the case of Ho~II, the diffuse UV emission seems to be spread widely across the disk, and only a few particular locations can have similar origin as described by \citet{Holberg1990} and \citet{Witt1989}, where the illuminating sources (stars) are located very close to the reflecting nebulae and the nebulae themselves have a non-negligible optical depth. However, in order to conclusively confirm the individual contributions to the diffuse UV emission including the emission from H$_{2}$ fluorescence, further precise analysis of a larger unbiased sample from star-forming regions of Ho~II, as well as other similar dwarf irregular galaxies, is crucial and will be addressed in future work. 
\end{enumerate}

\section{Acknowledgements}

This work uses the data from the AstroSat mission of the Indian Space Research Organisation (ISRO), archived at the Indian Space Science Data Centre (ISSDC). This research has used observations made with the Spitzer Space Telescope, which is operated by the Jet Propulsion Laboratory, California Institute of Technology under a contract with NASA. Herschel is an ESA space observatory with science instruments provided by European-led Principal Investigator consortia and with important participation from NASA. OPB and RG are thankful to Science \& Engineering Research Board (SERB), Department of Science \& Technology (DST), Government of India for financial support (EMR/2017/003092). OPB gratefully acknowledges the help received from his colleagues Anshuman Borgohain and Hritwik Bora. BA acknowledges the financial support by DST, Government of India, under the DST-INSPIRE Fellowship {{(Application Reference Number: DST/INSPIRE/03/2018/000689; INSPIRE Code: IF190146)} program. {BA, SP \& DB thank Prof. P. Sreekumar, MCNS, MAHE, Manipal for the fruitful discussions.} BA, SP, and DB acknowledge Manipal Centre for Natural Sciences, Manipal Academy of Higher Education (MAHE) for facilities and support. {MS acknowledges the financial support by the DST, Government of India,
under the Women Scientist Scheme (PH) project reference number SR/WOS-A/PM-17/2019 and the hospitality of the M.P. Birla Institute of Fundamental Research (MPBIFR), Bangalore, India, where part of the work has been carried out. YS acknowledges the hospitality of Raman Research Institute, Bangalore, India.}
 
\section{Data availability}
\label{sec:data}

The diffuse UV maps generated as part of this study are available upon request from the authors. {The combined table with FUV, NUV, IR intensities, and neutral hydrogen column densities N(HI) for 33 selected locations with non-zero IR intensities in all IR considered bands, and the table with IR intensities for all 142 locations are provided online on the following GitHub link: \url{https://github.com/olagpratim/Supplementary_tables_HoII_PASP}.
}

%% To help institutions obtain information on the effectiveness of their telescopes the AAS Journals has created a group of keywords for telescope facilities. Following the acknowledgments section, use the following syntax and the \facility{} or \facilities{} macros to list the keywords of facilities used in the research for the paper. Each keyword is check against the master list during copy editing. Individual instruments can be provided in parentheses, after the keyword, but they are not verified.

\vspace{5mm}
\noindent
\facilities{AstroSat (UVIT), GALEX, Spitzer (IRAC and MIPS), Herschel, VLA.}
\noindent
\software{IRAF \citep{iraf1986},   
          Source Extractor \citep{bertin1996}, 
          CCDLAB \citep{Postma2017, Postma2021}, 
          SAOImageDS9 \citep{ds92003}, 
          TOPCAT \citep{Taylor2005},
          Astropy \citep{Astropy2013,Astropy2018},
          Photutils \citep{photutil}.
          }

\newpage
\appendix

\section{Correlation coefficients}

\subsection{Spearman rank Correlation Coefficient}

Spearman's rank correlation coefficient ($\rho$) \citep{spearman1904} is a reliable method to test the monotonic relationship between two quantities rather than a linear relationship \citep{maurice1990rank}. In order to calculate the coefficient between two quantities with \textit{n} number of data pairs, first the observed values ($X_{i}$, $Y_{i}$) for each data pair are converted to dimensionless ranks $x_{i}$, $y_{i}$: the highest data value for each quantity will be assigned rank '1', the second highest data value will be assigned rank '2', and so on. Afterwards, $\rho$ is calculated as:
\begin{equation}\label{corr}
    \rho = 1-\frac{6 \Sigma d_{i}^2}{n(n^2-1)}\,,
\end{equation}
where $d_{i}=x_{i}-y_{i}$ is the difference between the ranks of corresponding observed values, determined as mentioned above, and \textit{n} = number of data pairs.
The Spearman's rank correlation coefficient lies in the interval $[-1,1]$. Higher value of rank correlation coefficient implies a better correlation between the two quantities. The coefficient value of $1$ implies a perfect association between the ranks (as one quantity increases, the other also increases), $0$ implies the association between the rankings are completely independent (the quantities are independent of one another), and $-1$ implies a perfect negative association between the rankings (as one quantity increases, other decreases).

\subsection{Pearson Correlation Coefficient}

The Pearson correlation coefficient ($r$) \citep{pearson1895}, is a widely used statistical measure that quantifies the strength and direction of a linear relationship between two continuous variables. It assesses how well the relationship between these variables can be described by a straight line. 

Pearson coefficient is calculated as:
\begin{equation}
    r = \frac{\Sigma((X_{i} - X_{mean})(Y_{i} - Y_{mean}))}{\sqrt{\Sigma((X_{i} - X_{mean})^2) \Sigma((Y_{i} - Y_{mean})^2)}}\,,
\end{equation}
where $X_i$ and $Y_i$ are the data points from the two variables, $X_{mean}$ and $Y_{mean}$ are the mean values of the two variables. Pearson coefficient falls within the range of $[-1, 1]$, with $-1$ indicating a perfect negative linear relationship, $1$ indicating a perfect positive linear relationship, and $0$ suggesting no linear relationship at all.

\subsection{Probability, or $p$-value}

The $p$-value is a measure of the evidence against a null hypothesis (there is no correlation between the two variables). It lies between 0 and 1, where a high $p$-value (close to 1) suggests no correlation other than due to chance and one must accept the null hypothesis. A lower $p$-value (close to 0) signifies the correlation is unlikely to be due to chance and there is a high probability that the null hypothesis is wrong. Therefore, one must accept the alternative hypothesis that a correlation exists between the two quantities. The $p$-value is calculated using a two tailed Student's $t$-distribution \citep{student1908}. The calculated correlation coefficient is transformed to a $t$-statistic using the formula
\begin{equation}
    t = \frac{r \sqrt{N-2}}{1 - r^2}\,,
\end{equation}
where $r$ is the Pearson correlation coefficient and $N$ is the number of data points. The $p$-value is then calculated using a $t$-distribution table or a statistical software \citep{R2023}.
\newpage

%\section{Author publication charges} \label{sec:pubcharge}

%PASP is a hybrid open access journal. Authors have the option to pay article publication charge (APC) to publish their article on an open access basis under a Creative Commons Attribution (CC BY) license. Additionally, PASP is supported in part by page charges. The current cost for publication charges is available at \url{https://iopscience.iop.org/journal/1538-3873/page/publication-charges}. 

% \highlightReference{spearman1904}
% \highlightReference{pearson1895}
% \highlightReference{student1908}
% \highlightReference{R2023}
%\highlightReference{Cole1999} 
%\highlightReference{Draine2011}
%\highlightReference{Osterbrock2006}
%\highlightReference{draine2003scattering}
%\highlightReference{Bohlin1982}
%\highlightReference{Postma2020}
%\highlightReference{Nilson1973}
%\highlightReference{Ananthamoorthy2024}
\bibliography{PASP101695}

\end{document}